\documentclass[journal]{IEEEtran}
\ifCLASSINFOpdf
\else
\fi
\usepackage{amsmath}
\usepackage{mathtools}
\usepackage{amssymb}
\usepackage{enumitem}
\usepackage{xspace}
\usepackage{pifont}
\usepackage{dsfont}
\usepackage{algpseudocode,algorithm}
\usepackage{graphicx}
\usepackage{subfig}
\usepackage{grffile}
\usepackage{pdflscape}
\usepackage{booktabs}
\usepackage{amsthm}
\usepackage{bm}
\allowdisplaybreaks
\graphicspath{{./figs/}}

\newtheorem{prop}{Proposition}

\newcommand{\vect}[1]{\boldsymbol{#1}}
\graphicspath{{./figs/}}
\newcommand{\pbar}{\overline{p}}
\newcommand{\indic}{\mathds{1}}

\begin{document}
\title{On the Stochastic Analysis of a Quantum Entanglement Switch}
\author{Gayane~Vardoyan,~Saikat~Guha,~Philippe~Nain,~and~Don~Towsley%
%\author{Michael~Shell,~\IEEEmembership{Member,~IEEE,}
%        John~Doe,~\IEEEmembership{Fellow,~OSA,}
%        and~Jane~Doe,~\IEEEmembership{Life~Fellow,~IEEE}% <-this % stops a space
\thanks{G. Vardoyan and D. Towsley are with the College of Information and Computer Sciences, University of Massachusetts, Amherst, MA, 01003 USA (e-mails: gvardoyan@cs.umass.edu, towsley@cs.umass.edu).}%
%M. Shell was with the Department of Electrical and Computer Engineering, Georgia Institute of Technology, Atlanta,
\thanks{Saikat Guha is with the College of Optical Sciences, University of Arizona, Tucson, AZ, 85721 USA (e-mail: saikat@optics.arizona.edu).}%
\thanks{Philippe Nain is with Inria, France (e-mail: philippe.nain@inria.fr).}%
%\thanks{Manuscript received July 1, 2019.%; revised August 26, 2015.}
}

\maketitle

\begin{abstract}
We study a quantum entanglement switch that serves $k$ users in a star topology. We model variants of the system using Markov chains and standard queueing theory and obtain expressions for switch capacity and the expected number of qubits stored in memory at the switch. While it is more accurate to use a discrete-time Markov chain (DTMC) to model such systems, we quickly encounter practical constraints of using this technique and switch to using continuous-time Markov chains (CTMCs). Using CTMCs allows us to obtain a number of analytic results for systems in which the links are homogeneous or heterogeneous and for switches that have infinite or finite buffer sizes. In addition, we can model the effects of decoherence of quantum states fairly easily using CTMCs. We also compare the results we obtain from the DTMC against the CTMC in the case of homogeneous links and infinite buffer, and learn that the CTMC is a reasonable approximation of the DTMC. 
From numerical observations, we discover that decoherence has little effect on capacity and expected number of stored qubits for homogeneous systems. For heterogeneous systems, especially those operating close to stability constraints, buffer size and decoherence can have significant effects on performance metrics. We also learn that in general, increasing the buffer size from one to two qubits per link is advantageous to most systems, while increasing the buffer size further yields diminishing returns.
%From numerical observations, we discover that buffer size has little effect on capacity and expected number of stored qubits. We also learn that while decoherence does not significantly affect these performance metrics in homogeneous systems, it can have drastic consequences when heterogeneity is introduced.
\end{abstract}


% \fi
\IEEEpeerreviewmaketitle

\section{Introduction}
%\IEEEPARstart{T}{his} demo file is intended to serve as a ``starter file''
%for IEEE journal papers produced under \LaTeX\ using
%IEEEtran.cls version 1.8b and later.
%%!TEX PS-program = pdflatexmk
%%!TEX root = quantum_jrnl.tex
\IEEEPARstart{E}{ntanglement} is an essential component of quantum computation, information, and communication. Its applications range from cryptography (\emph{e.g.} quantum key distribution \cite{ekert1991quantum}, quantum error correction \cite{bennett1996mixed}) to ensemble sensing (\emph{e.g.} multipartite entanglement for quantum metrology \cite{giovannetti2011advances} and spectroscopy \cite{leibfried2004toward}). These applications drive the increasing need for a quantum switching network that can supply end-to-end entanglements to groups of endpoints that request them \cite{pant2019,Schoute2016-ir,Van_Meter2014-vz}. To realize such quantum systems, architectures have been proposed to support high entanglement generation rates, high fidelity, and long coherence times \cite{li2018crossbar,armstrong2012programmable,herbauts2013demonstration,hall2011ultrafast}.

In this paper, we study a single quantum switch that serves $k$ users in a star topology. Each user has a dedicated link to the switch, and all sets of users of size $n\leq k$, for a fixed $n$, wish to share an entangled state. To achieve this, link-level entanglements are generated at a constant rate across each link, resulting in two-qubit maximally-entangled states (\emph{i.e.} Bell pairs or EPR states). These qubits are stored at local quantum memories: one half of a Bell pair at the user and the other half at the switch. When enough of these entanglements are accrued (at least $n$ of them), the switch performs multi-qubit measurements to provide end-to-end entanglements to user groups of size $n$. When $n=2$, the switch uses Bell-state measurements (BSMs) and when $n\geq 3$, it uses $n$-qubit Greenberger-Horne-Zeilinger (GHZ) basis measurements \cite{nielsenchuang}.

We consider systems in which links may generate entanglements at different rates and where the switch can store one or more qubits (each entangled with another qubit held by a user) per link.
%We consider a number of variants of this system: for instance, a scenario in which all links generate entanglements at the same rate, and another in which they generate link-level entanglements at different rates. We also consider the effects of extra memory (or buffer) at the switch, which allows it to store qubits (each entangled with another qubit held by a user) for future use. 
Throughout this paper, we refer to these pairs of stored qubits as \emph{stored entanglements}. 
Another factor that impacts performance is decoherence of quantum states; we model it and study its effect. The main metric of interest for this network is its capacity $C$, \emph{i.e.}, the number of end-to-end entanglements served by the switch per time unit. Another metric of interest is the expected number of qubits $Q$ in memory at the switch, $E[Q]$. Both $C$ and $E[Q]$ depend on the values of $k$, $n$, entanglement generation and decoherence rates, number of quantum memories, and the switching mechanism, including the scheduling policy used by the switch.

Contributions of this work are as follows: using continuous-time Markov chains (CTMCs), we derive $C$ and $E[Q]$ for $n=2$ for a particular scheduling policy and study how they vary as functions of $k$, buffer size, and decoherence rate. For $n\geq 3$, we derive $C$ and $E[Q]$ for the case where all links are identical, the switch has infinite memory, and there is no decoherence.
From our analysis, we gain valuable insight into which factors influence capacity the most, and which ones are of lesser consequence. For instance, we find that when $n=2$ and links are identical, the number of links and their entanglement generation rate are the most impactful, while decoherence and buffer size have little effect. However, the same is not true in the case of non-identical links, where the distribution of entanglement generation rates, combined with finite coherence time, can drastically affect both $C$ and $E[Q]$. Last, we compare our results for the $n=2$, identical-link, infinite buffer case against a logically more accurate discrete-time Markov chain (DTMC) model and find that {\em (i)} they predict the same capacity, and {\em(ii)} the difference in predictions of $E[Q]$ is small although relative errors can be large for small values of $k$. Consequently, we rely on CTMC models as we relax assumptions.

The remainder of this paper is organized as follows: in Section \ref{sec:background}, we discuss relevant background and related work. In Section \ref{sec:modelandObj}, we cover modeling techniques, assumptions, and objectives. In Section \ref{sec:CTMCBipartite}, we introduce our CTMC models for $n=2$ and present their analyses. In Section \ref{sec:DTMCBipartite}, we introduce and analyze our DTMC model for $n=2$ infinite-buffer, homogeneous-link scenario and compare the results to its CTMC analogue. In Section \ref{sec:nPartite}, we derive $C$ and $E[Q]$ for the simplest case of $n\geq3$ using a CTMC model. Numerical observations are discussed in Section \ref{sec:numerObs}. We conclude in Section \ref{sec:concl}.

\section{Background}
%%!TEX PS-program = pdflatexmk
%%!TEX root = quantum_jrnl.tex
\label{sec:background}
In \cite{herbauts2013demonstration}, Herbauts \emph{et al.} implement an entanglement distribution network intended for quantum communication applications. The fidelities of entanglements generated in this network were $93\%$ post-distribution, and fidelities of $99\%$ were shown to be achievable. The demonstration entails distributing bipartite entanglements to any pair of users wishing to share entanglement in a multi-user network (there were eight users in the experimental setup). Delivering multiple bipartite entanglements was shown to be possible virtually simultaneously. The authors specifically cite a possible application of the network in a scenario where a single central switch dynamically allocates two-party entanglements to any pair of users in a static network.
In this paper, we study variants of this system, where we assume that the switch has the ability to store entangled qubits for future use, and that successfully-generated entanglements have fidelity one (a reasonable assumption based on the results in \cite{herbauts2013demonstration}).

In \cite{shchukin2017waiting}, the authors use Markov chains to compute the expected waiting time in quantum repeaters with probabilistic entanglement swapping. While this work considers an arbitrary number of links, explicit expressions are only provided for up to four links. In contrast, the scheduling policy used in our work simplifies the construction of the Markov chains, allowing us to derive closed-form expressions for a diverse set of systems. Further, in some cases we also consider extra memory at the switch and study its effect on performance.% metrics.

In \cite{vardoyan2019capacity}, we analyze the capacity region of a quantum entanglement switch that serves users in a star topology and is constrained to store one or two qubits per link. The problem setup is quite similar to that of this work, with the exception that the switch has the ability to serve bipartite and tripartite end-to-end entanglements. We examine a set of randomized switching policies and find policies that perform better than time-division multiplexing between bipartite and tripartite entanglement switching.
As new quantum architectures and technologies emerge, we expect quantum networks to be more prevalent and suitable for practical use. With link-level and especially end-to-end entanglements being a valuable commodity in these networks, proper resource management will be imperative for reliable and efficient operation. 
%To our knowledge, \cite{vardoyan2019capacity} and this work are the the first to shed light on this problem using classical queuing theory.
\section{Model and Objectives}
%%!TEX PS-program = pdflatexmk
%%!TEX root = quantum_jrnl.tex
\label{sec:modelandObj}
Consider first a fairly general setting of the proposed problem: $k$ users are attached to a quantum entanglement switch via $k$ dedicated links. At any given time step, any set of $n$ users (with $n\leq k$) wish to share an end-to-end entangled state. 
The creation of an end-to-end entanglement involves two steps. 
First, users generate pairwise entanglements with the switch, which we call link-level entanglements. Each of these results in a two-qubit Bell state, with one qubit stored at the switch and the other stored at a user. Once at least $n$ links have created entanglements, the process enters step two: the creation of an end-to-end entanglement. The switch chooses a set of $n$ locally-held qubits (that are entangled with $n$ qubits held by $n$ distinct users) and performs an entangling measurement. If such a measurement is successful, the result is an $n$-qubit maximally-entangled state between the corresponding $n$ users. If after this step more link-level entanglements are available, the switch repeats the second step until there are fewer than $n$ local qubits left.

The reason that we focus on the case where any $n$ users wish to share an entangled state is that in this work, we would like to obtain a bound on the aggregate capacity of the switch for any workload. 
Hence, it is helpful if the switch has no restrictions on which measurements to perform whenever $n$ distinct link-level entanglements are available. The results of this analysis can then be used as a comparison basis for other types of scenarios, in which, for example, each $n$-group of users may specify a desired rate of communication with each other through the switch. Another utility of this analysis is that by examining a switch that operates at or near maximum capacity, one may gain insight on the practical memory requirements of a switch.

Both link-level entanglement generation and entangling measurements can be modeled as probabilistic phenomena \cite{Guha2015-qo}. In this work, we model the former as follows: at each time step of length $\tau$ seconds, all $k$ users attempt to generate link-level entanglements. In general, link $l$ succeeds in generating an entanglement with probability $p_l\approx e^{-\beta L}$, where $L$ is the length of the link (\emph{e.g.} optical fiber) and $\beta$ its attenuation coefficient. We refer to the special case of $p_l=p_m,~\forall l,m\in\{1,\dots,k\}$ as a \emph{homogeneous} system, and when they are not necessarily equal, as a \emph{heterogeneous} system. We assume that whenever a link-level entanglement is generated successfully, it always has fidelity one, but in certain cases we will consider decoherence post-generation. 
We also assume that measurements performed by the switch succeed with probability $q$\footnote{With a linear optical circuit, four unentangled ancilla single photons and photon number resolving detectors, with all the devices being lossless, $q = 25/32 = 0.78$ can be achieved for BSMs \cite{ewert20143}. With other technologies $q$ close to 1 can be achieved \cite{grice2011arbitrarily}.}.

If at any time there are fewer than $n$ link-level entanglements, the switch may choose to store the available entangled qubits and wait until there are enough new ones generated to create an end-to-end entanglement. We assume that the switch can store $B\geq 1$ qubits in its buffer, per link. If on the other hand, there are more than $n$ link-level entanglements, the switch must decide which set(s) of them to use in measurement(s). Such decisions can be made according to a pre-specified scheduling policy: for example, a user or a set of users may be given higher priority for being involved in an end-to-end entanglement. Other scheduling policies may be adaptive, random, or any number of hybrid policies.

In this work, we utilize a hybrid scheduling policy. First, we assume that the switch adheres to the \emph{Oldest Link Entanglement First (OLEF)} rule, wherein the oldest link-level entanglements have priority to be used in entangling measurements. A practical reason for this rule is that quantum states are subject to decoherence, which is a function of time; hence, our goal is to make use of link-level entanglements as soon as possible. Second, we assume that as long as the switch follows the OLEF rule, sets of link-level entanglements are chosen at random for measurements, provided that each set consists of $n$ entanglements belonging to $n$ distinct links. The state space of this system can be represented by a vector $\vect{Q}(t)\in\{0,1,\dots,B\}^k$, where the $l$th element corresponds to the number of stored entanglements at link $l$ at time $t$. 
Note that one consequence of the OLEF rule and the assumption that any set of $n$ users always wish to share an entangled state is that at most $n-1$ distinct users will store entanglements at any time.

One way to model a system as described above is to construct a DTMC on the appropriate state space. Unfortunately, this method is not the most scalable (in terms of $k$ or $n$) and is not the easiest to analyze even in the simple setting of homogeneous links and infinite switch buffer size. Further obstacles arise when one considers, for example, accounting for decoherence in a DTMC model. Another possibility is to use a CTMC: instead of viewing a link-level entanglement as a Bernoulli trial, view it as an exponential random variable (r.v.) with successful generation rate equal to $\mu_l=p_l/\tau$. The analysis is significantly less challenging, and we can easily incorporate decoherence by modeling coherence time as an exponential r.v. with mean $1/\alpha$. Under this assumption, an entanglement's fidelity goes from one directly to zero upon decoherence, \emph{i.e.}, the fidelity does not degrade while the entanglement is in storage. A disadvantage is that a discrete model describes the operation of our systems more accurately. However, in Section \ref{sec:ctmcVsdtmc}, we argue that using CTMCs as an approximation is quite reasonable: at least in the homogeneous case with infinite buffer, not much is lost in terms of accuracy.

Our goal in this work is to derive the system capacity $C$ (\emph{i.e.}, the number of end-to-end entanglements produced per time unit) and the expected number of stored qubits $E[Q]$. 
%For brevity, we analyze capacity with the assumption that all measurements performed by the switch succeed, but note that removing this assumption is simple: if $q$ is the measurement success probability, then the effective capacity becomes $qC$.
A note on mathematical notation: in this paper, we will use the convention that for any $y>x$, the term ${x \choose y}=0$.
Throughout the paper, we also use the result that if the balance equations of an irreducible CTMC have a unique and strictly positive solution, then this solution represents the stationary distribution of the chain.
\section{CTMC for Bipartite Entanglements}
%%!TEX PS-program = pdflatexmk
%%!TEX root = quantum_jrnl.tex
\label{sec:CTMCBipartite}
In this section, we introduce and analyze a CTMC model of a bipartite entanglement switch serving $k$ users. 
We first assume that memories do not decohere and 
obtain expressions for capacity and the expected number of qubits stored at the switch. 
%In the homogeneous links case, the expressions are closed-form, while in the heterogeneous case they are still fairly simple and interpretable. 
We then modify the models to incorporate decoherence and analyze it. 
Last, we derive an upper bound for the capacity of the switch.
%Numerical observations and comparisons are presented in Section \ref{sec:numerObs}.
%%!TEX PS-program = pdflatexmk
%%!TEX root = quantum_jrnl_short.tex
\label{sec:CTMCanalysis}
\subsection{The Heterogeneous Case}
Assume $\mu_l$ depends on $l$, \emph{i.e.} the links are heterogeneous. For subsequent analysis, it is useful to define
\begin{align*}
\gamma &\coloneqq \sum\limits_{l=1}^k \mu_l,
\end{align*}
the aggregate entanglement generation rate over all links. 
Also, let $\vect{e}_l$ be a size $k$ vector with all zeros except for the $l$th component, which is 1, and let $\vect{0}$ be a vector of size $k$ with all entries equal to 0.

We are interested in the stationary distribution and stability conditions for a heterogeneous system with infinite and finite buffers. As discussed in Section \ref{sec:modelandObj}, in bipartite entanglement switching, only one link will store entanglements, but since links generate entanglements at different rates, we must keep track of which link is associated with the stored entanglement(s). Let $\vect{Q}(t)=(Q_1(t), \dots, Q_k(t))\in\{0,1,2,\dots\}^k$ represent the 
state of the system at time $t$, where $Q_l(t)$ is the
number of entanglements stored at link $l$, $l\in\{1,\dots,k\}$, at time $t$. 
As a consequence of the scheduling policy described in Section \ref{sec:modelandObj}, if $Q_i(t)>0$ for some $i$, then $Q_j(t)=0$, $j\neq i$.
In other words, $\vect{Q}(t)$ only takes on values $\vect{0}$ or $j\vect{e}_l$, $l\in\{1,\dots,k\}$, $j\in\{1,2,\dots\}$. 
Here, $\vect{0}$ represents the state where no entanglements are stored, and
 $j\vect{e}_l$ represents the state where the $l$th link has $j$  stored entanglements.

Define the following limits when they exist:
\begin{align*}
\pi_0 &= \lim\limits_{t\to\infty}P(\vect{Q}(t)=\vect{0}),\\
\pi_l^{(j)} &= \lim\limits_{t\to\infty}P(\vect{Q}(t)=j\vect{e}_l).
\end{align*}
Once we obtain expressions for $\pi_0$ and $\pi_l^{(j)}$, we can derive expressions for capacity and the expected number of stored qubits $E[Q]$.
\subsubsection{Infinite Buffer}
\label{sec:infBufHeterog}
\begin{figure}
\centering
\includegraphics[width=0.45\textwidth]{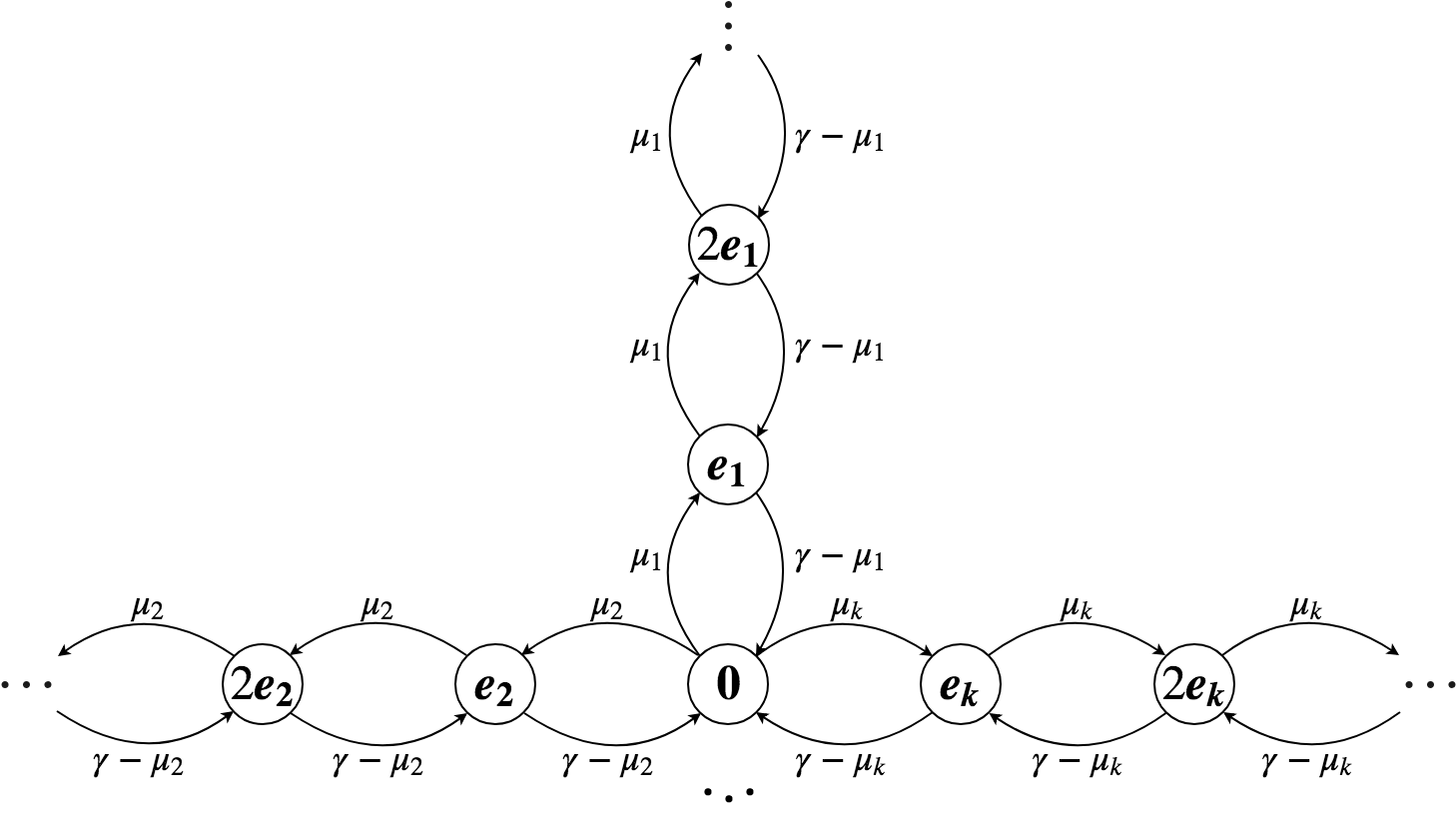}
\caption{A CTMC for a $k$-user, infinite buffer, heterogeneous-link switch. $\mu_l$ is the entanglement generation rate of link $l$, while $\gamma$ is the aggregate entanglement generation rate of all links. $\vect{e}_l$ is a vector of all zeros except for the $l$th position, which is equal to one.}
\label{fig:ctmcInfBufHeterog}
\end{figure}

Figure \ref{fig:ctmcInfBufHeterog} presents the CTMC for an infinite buffer. 
%The system is assumed to initiate in 
Consider state $\vect{0}$ (no stored entanglements). 
From there, a transition along one of the $k$ ``arms'' of the CTMC occurs with rate $\mu_l$, when the $l$th link successfully generates an entanglement. For a BSM to occur, any of the $k-1$ other links must successfully generate an entanglement: this occurs with rate $\gamma-\mu_l$.
The balance equations are 
\begin{align*}
&\pi_0\mu_l = \pi_l^{(1)}(\gamma-\mu_l),~l\in\{1,\dots,k\},\\
&\pi_l^{(j-1)}\mu_l = \pi_l^{(j)}(\gamma-\mu_l),~l\in\{1,\dots,k\},~j\in\{2,3,\dots\},\\
&\pi_0+\sum\limits_{l=1}^k\sum\limits_{j=1}^{\infty}\pi_l^{(j)} =1.
\end{align*}
From above, we see that for $j=1,2,\dots$,
\begin{align*}
\pi_l^{(j)} &= \rho_l^j\pi_0, \text{ where }\rho_l \equiv \frac{\mu_l}{\gamma-\mu_l}, \forall ~l.
\end{align*}
It remains to obtain $\pi_0$; we can use the normalizing condition:
\begin{align*}
\pi_0+\pi_0\sum\limits_{l=1}^k\sum\limits_{j=1}^{\infty} \rho_l^j %&=1,\\
=
\pi_0\left(1+\sum\limits_{l=1}^k\left(\sum\limits_{j=0}^{\infty} \rho_l^j-1\right)\right) &=1.
\end{align*}
Now, assume that for all $l\in\{1,\dots,k\}$, $\rho_l<1$. This implies that for all $l$, $\mu_l< \gamma/2$. This is the stability condition for this chain. 
%Then,
%\begin{align*}
%\pi_0\left(1+\sum\limits_{l=1}^k\left(\frac{1}{1-\rho_l}-1\right)\right) &=1,\\
%\pi_0\left(1+\sum\limits_{l=1}^k\frac{\rho_l}{1-\rho_l}\right) &=1,
%\end{align*}
%and therefore,
\begin{flalign}
&\text{Then,}\quad
\pi_0 = \left(1+\sum\limits_{l=1}^k\frac{\rho_l}{1-\rho_l}\right)^{-1}
\quad\text{and the capacity is}&\nonumber\\
%\end{flalign*}
%The capacity is 
%\begin{align*}
C &= q\sum\limits_{l=1}^k\sum\limits_{j=1}^{\infty}\pi_l^{(j)}(\gamma-\mu_l)% = 
%q\sum\limits_{l=1}^k\sum\limits_{j=1}^{\infty}\pi_0\rho_l^{j}(\gamma-\mu_l)%\\ 
%&=\pi_0\sum\limits_{l=1}^k(\gamma-\mu_l)\sum\limits_{j=1}^{\infty}\rho_l^{j}
%= \pi_0\sum\limits_{l=1}^k(\gamma-\mu_l)\left(\sum\limits_{j=0}^{\infty}\rho_l^{j}-1\right)\\
%&= \pi_0\sum\limits_{l=1}^k(\gamma-\mu_l)\frac{\rho_l}{1-\rho_l}\\
%&= \frac{\sum\limits_{l=1}^k(\gamma-\mu_l)\frac{\rho_l}{1-\rho_l}}{1+\sum\limits_{l=1}^k\frac{\rho_l}{1-\rho_l}}
%=\frac{\sum\limits_{l=1}^k\frac{\mu_l}{\rho_l}\frac{\rho_l}{1-\rho_l}}{1+\sum\limits_{l=1}^k\frac{\rho_l}{1-\rho_l}}
=\frac{q\sum\limits_{l=1}^k\frac{\mu_l}{1-\rho_l}}{1+\sum\limits_{l=1}^k\frac{\rho_l}{1-\rho_l}}
=\frac{q\gamma}{2}.
\label{eq:heterogCapInfBuf}
\end{flalign}
See Appendix \ref{heterogCapApp} for a proof of the last equality.
The distribution of the number of stored entanglements is 
\begin{align*}
P(Q=j) &= \begin{cases}
    \pi_0, & \text{if $j=0$},\\
    \sum\limits_{l=1}^k\pi_l^{(j)} = \pi_0\sum\limits_{l=1}^k\rho_l^{j}, & \text{if $j>0$}.
  \end{cases}
\end{align*}
The expected number of stored entanglements is
\begin{align*}
E[Q] &= \sum\limits_{j=1}^{\infty}jP(Q=j) = \sum\limits_{j=1}^{\infty}j\pi_0\sum\limits_{l=1}^k\rho_l^{j}
%=\pi_0\sum\limits_{l=1}^k\sum\limits_{j=1}^{\infty}j\rho_l^{j}\\
%&=\pi_0\sum\limits_{l=1}^k \rho_l \sum\limits_{j=1}^{\infty}j\rho_l^{j-1}
%=\pi_0\sum\limits_{l=1}^k  \frac{\rho_l}{(1-\rho_l)^2}
=\frac{\sum\limits_{l=1}^k  \frac{\rho_l}{(1-\rho_l)^2}}{1+\sum\limits_{l=1}^k\frac{\rho_l}{1-\rho_l}}.
\end{align*}
\subsubsection{Finite Buffer}
In the case of heterogeneous links and a finite buffer of size $B$, the CTMC has the same structure as in Figure \ref{fig:ctmcInfBufHeterog}, except that each ``arm'' of the chain terminates at $B\vect{e}_l$, $\forall l\in\{1,\dots,k\}$. The balance equations are
\begin{align*}
&\pi_0\mu_l = \pi_l^{(1)}(\gamma-\mu_l),~l\in\{1,\dots,k\},\\
&\pi_l^{(j-1)}\mu_l = \pi_l^{(j)}(\gamma-\mu_l),~l\in\{1,\dots,k\},~j\in\{2,\dots,B\},\\
&\pi_0+\sum\limits_{l=1}^k\sum\limits_{j=1}^{B}\pi_l^{(j)} =1
\end{align*}
and have solution
\begin{align*}
\pi_l^{(j)} &= \rho_l^j\pi_0,~l\in\{1,\dots,k\},~j\in\{1,\dots,B\},
\end{align*}
where $\rho_l$ is defined as in the infinite buffer case. %For $\pi_0$, we have
Then,
\begin{flalign*}
&\pi_0\left(1+\sum\limits_{l=1}^k\sum\limits_{j=1}^{B}\rho_l^{j}\right) =1, \text{ hence }
%\end{align*}
%\begin{align*}
\pi_0 = \left(1+\sum\limits_{l=1}^k\sum\limits_{j=1}^{B}\rho_l^{j}\right)^{-1}\hspace{-0.5em},&\\
%\end{align*}
%\begin{flalign*}
&\text{and the capacity is }\\
&C = q\sum\limits_{l=1}^k\sum\limits_{j=1}^{B}(\gamma-\mu_l)\pi_l^{(j)} %&\\
%&\qquad\qquad\qquad\qquad=q\pi_0\sum\limits_{l=1}^k\sum\limits_{j=1}^{B}(\gamma-\mu_l)\rho_l^{j}%\\
%&= \frac{\sum\limits_{l=1}^k\sum\limits_{j=1}^{B}(\gamma-\mu_l)\rho_l^{j}}{1+\sum\limits_{l=1}^k\sum\limits_{j=1}^{B}\rho_l^{j}}
%=\frac{\sum\limits_{l=1}^k(\gamma-\mu_l)\frac{\rho_l(1-\rho_l^B)}{1-\rho_l}}{1+\sum\limits_{l=1}^k\frac{\rho_l(1-\rho_l^B)}{1-\rho_l}}
=\frac{q\sum\limits_{l=1}^k\frac{\mu_l(1-\rho_l^B)}{1-\rho_l}}{1+\sum\limits_{l=1}^k\frac{\rho_l(1-\rho_l^B)}{1-\rho_l}}.
\end{flalign*}
The distribution of the number of stored qubits is given by
\begin{align*}
P(Q=j) &= \begin{cases}
    \pi_0, & \text{if $j=0$},\\
    \sum\limits_{l=1}^k\pi_l^{(j)} = \pi_0\sum\limits_{l=1}^k\rho_l^{j}, & \text{if $0<j\leq B$}.
  \end{cases}
\end{align*}
The expected number of stored qubits is
\begin{align*}
&E[Q] = \sum\limits_{j=1}^{B}jP(Q=j) %= \sum\limits_{j=1}^{B}j\pi_0\sum\limits_{l=1}^k\rho_l^{j}
%=\pi_0\sum\limits_{l=1}^k\sum\limits_{j=1}^{B}j\rho_l^{j}\\
%&=\pi_0\sum\limits_{l=1}^k\frac{\rho_l\left(B\rho_l^{B+1}-(B+1)\rho_l^B+1\right)}{\left(1-\rho_l\right)^2}\\
%&= \frac{\sum\limits_{l=1}^k\frac{\rho_l\left(B\rho_l^{B+1}-(B+1)\rho_l^B+1\right)}{\left(1-\rho_l\right)^2}}{1+\sum\limits_{l=1}^k\sum\limits_{j=1}^{B}\rho_l^{j}}\\
= \frac{\sum\limits_{l=1}^k\frac{\rho_l\left(B\rho_l^{B+1}-(B+1)\rho_l^B+1\right)}{\left(1-\rho_l\right)^2}}{1+\sum\limits_{l=1}^k\frac{\rho_l(1-\rho_l^B)}{1-\rho_l}}.
\end{align*}
%Note that another way to express the switch capacity is by
%\begin{align*}
%C &= q\sum\limits_{l=1}^k \mu_l\sum\limits_{\substack{m=1,\\ m\neq l}}^k\sum\limits_{j=1}^B \pi_m^{(j)}.
%\end{align*}
The rate received by user $l$ (connected to link $l$) is given by
\begin{align}
C_l &= q\left((\gamma-\mu_l)\sum\limits_{j=1}^B \pi_l^{(j)}+\mu_l\sum\limits_{\substack{m=1,\\ m\neq l}}^k\sum\limits_{j=1}^B \pi_m^{(j)}\right),
\label{eq:Clheterog}
\end{align}
where the first term represents the production of entanglements by link $l$ (which get consumed by other links at rate $\gamma-\mu_l$) and the second term represents the consumption by link $l$ of stored entanglements at other links. 
Note then, that if we were to sum all $C_l$, each end-to-end entanglement would be double-counted. Hence, $\sum C_l=2C$. (Note: in the infinite-buffer case, $C_l=q\mu_l$, $l\in\{1,\dots,k\}$; see Appendix \ref{heterogCapApp} for a proof. Then, $\sum C_l = q\gamma = 2C$, another proof of the last equality in Eq. (\ref{eq:heterogCapInfBuf}).) 
The expected number of stored qubits at link $l$, $E[Q_l]$ can be obtained by taking the $l$th component of the sum in the numerator of the expression for $E[Q]$. In other words, when $B=\infty$, 
\begin{align*}
E[Q_l] =\frac{\frac{\rho_l}{(1-\rho_l)^2}}{1+\sum\limits_{l=1}^k\frac{\rho_l}{1-\rho_l}}.
\end{align*}
For a homogeneous system, $E[Q_l]=E[Q]/k$.
%%!TEX PS-program = pdflatexmk
%%!TEX root = quantum_jrnl_short.tex
\subsection{The Homogeneous Case}
\label{sec:analysisHomogCTMC}
Suppose all links (or users) have the same entanglement generation rates, \emph{i.e.} $\mu_l=\mu$, $\forall~ l\in\{1,\dots,k\}$. We can take advantage of this homogeneity as follows: since only one link can be associated with stored qubits at the switch at any given time, and all links have equal rates, it is only necessary to keep track of the \emph{number} of stored entanglements, and not the \emph{identity} of the link (or user). Hence, the state space of the CTMC can be represented by a single variable taking values in $\{0,1,\dots,B\}$
where $B=\infty$ corresponds to
the infinite buffer case, and $B<\infty$ the finite buffer case. We discuss each of these in detail next.
\subsubsection{Infinite Buffer}
\label{sec:ctmcInfBufHomog}
\begin{figure}
\centering
\includegraphics[width=0.35\textwidth]{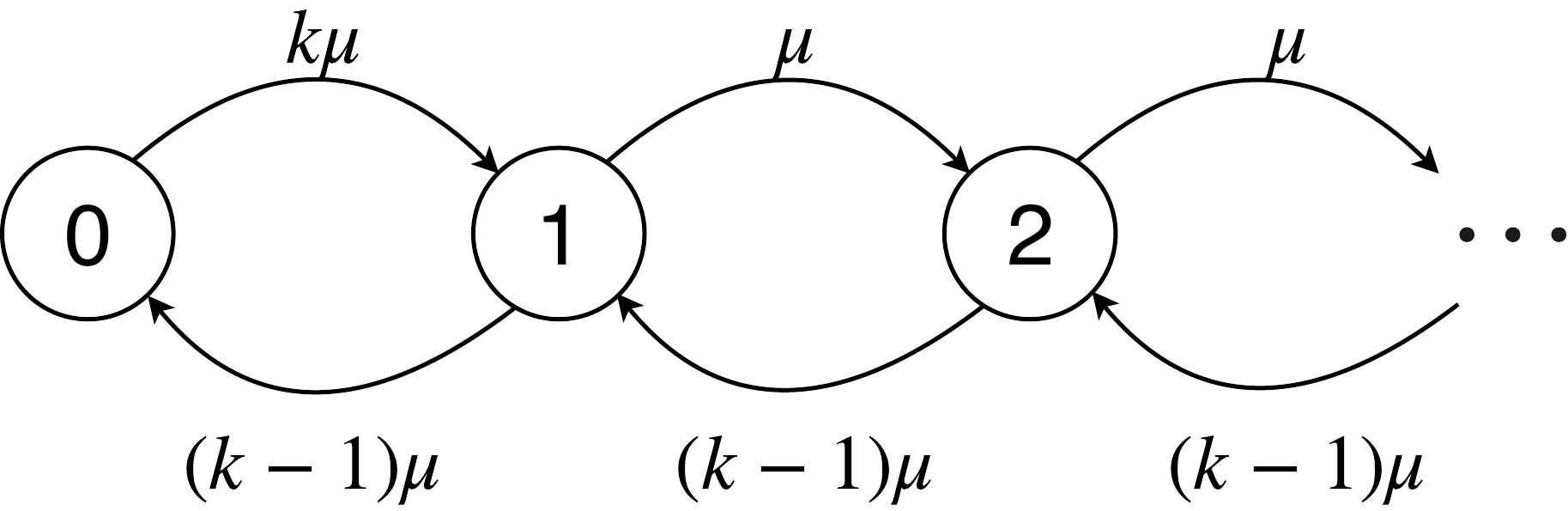}
\caption{A CTMC model with $k$ users, infinite buffer, and homogeneous links. $\mu$ is the entanglement generation rate.}
\label{fig:ctmcInfBufHomog}
\end{figure}

Figure \ref{fig:ctmcInfBufHomog} depicts the CTMC for $k$ homogeneous links and $B=\infty$. 
When no entanglements are stored (system is in state $0$), 
any of the $k$ links can generate a new entanglement, so the transition to state $1$ occurs with rate $k\mu$. Let $S$ represent the link associated with one or more stored entanglements. From states $1$ and above, transitioning ``forward'' (or gaining another entanglement in storage) occurs whenever link $S$ generates a new entanglement. This event occurs with rate $\mu$. Finally, moving ``backward'' through the chain (corresponding to using a stored entanglement, when the switch performs a BSM) occurs whenever any of the $k-1$ links other than $S$ successfully generate an entanglement; this event occurs with rate $(k-1)\mu$.
It is easy to show that when there are two links, the system is not stable (and a stationary distribution does not exist). Henceforth, we only consider $k\geq3$.

%The balance equations for the chain in Figure \ref{fig:ctmcInfBufHomog} are
Note that the CTMC in Figure  \ref{fig:ctmcInfBufHomog} is a birth-death process 
%with balance equations
whose stationary distribution can be obtained using standard techniques found in literature (\emph{e.g.} \cite{kleinrock1975queueing}).
%\begin{align}
%\pi_0 k\mu &= \pi_1(k-1)\mu\label{eq:ctmc1}\\
%\pi_{i-1}\mu &= \pi_{i} (k-1)\mu, ~i=2,3,\dots,\label{eq:ctmc2}\\
%\sum\limits_{i=0}^{\infty}\pi_i &=1\label{eq:ctmc3}
%\end{align}
%and 
The steady-state probability of being in state $0$ is
$\pi_0=(k-2)/(2(k-1))$ and of being in state $j$ is $\pi_j=k(k-2)/(2(k-1)^{j+1})$.
The capacity is
\begin{align*}
C &= q\sum\limits_{i=1}^{\infty}\pi_i (k-1)\mu = q(k-1)\mu(1-\pi_0) 
%= q(k-1)\frac{\mu k }{2(k-1)} 
= \frac{q\mu k}{2}.
\end{align*}
The expected number of stored entangled pairs is given by
\begin{align*}
E[Q] &= \sum\limits_{i=0}^{\infty}i\pi_i = k\pi_0\sum\limits_{i=1}^{\infty}i\left(\frac{1}{k-1}\right)^i
%=\frac{k\pi_0}{k-1}\frac{1}{\left(1-\frac{1}{k-1}\right)^2}\\
%&=\frac{k\pi_0}{k-1}\frac{(k-1)^2}{\left(k-2\right)^2}
%=\frac{k-2}{2(k-1)}\frac{k(k-1)}{\left(k-2\right)^2}
=\frac{k}{2(k-2)}.
\end{align*}
%This corresponds to the amount of memory being used at the switch to store qubits that are associated with link entanglements.
\subsubsection{Finite Buffer}
\label{sec:ctmcFinBufHomog}
\begin{figure}
\centering
\includegraphics[width=0.45\textwidth]{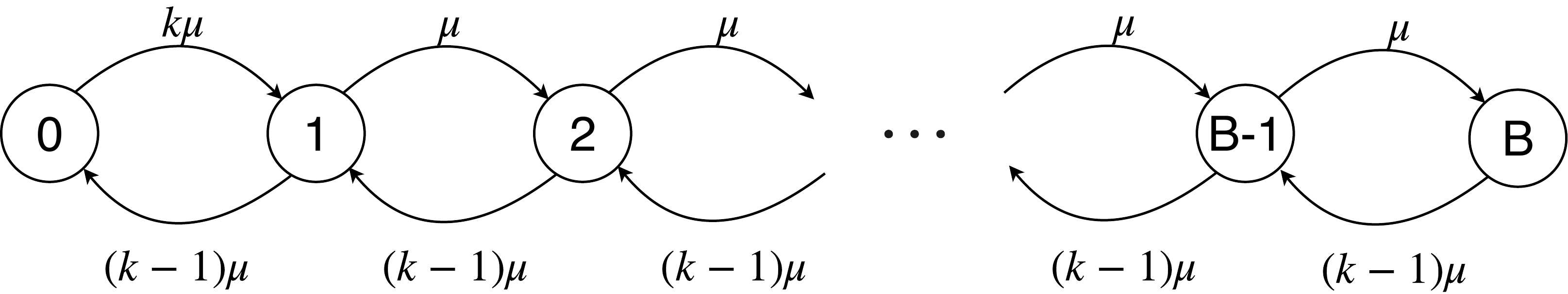}
\caption{A CTMC model with $k$ users, finite buffer of size $B$, and homogeneous links. $\mu$ is the entanglement generation rate.}
\label{fig:ctmcFinBufHomog}
\end{figure}

Figure \ref{fig:ctmcFinBufHomog} illustrates the CTMC for a system with $k$ homogeneous links being served by a switch with finite buffer space $B$. 
%This case is very similar to the infinite buffer case, with the only modification being that we cannot advance beyond state $B$: 
When there are $B$ stored entanglements and a new one is generated on link $S$, we assume that the switch drops the oldest stored entanglement, adhering to the OLEF policy.
This CTMC is also a standard birth-death process whose solution can be found in literature (\emph{e.g.} \cite{kleinrock1975queueing}) and has
%All balance equations are the same as before, except that we stop at $i=B$, \emph{i.e.},
%\begin{align*}
%\pi_0 k\mu &= \pi_1(k-1)\mu,\\
%\pi_{i-1}\mu &= \pi_{i} (k-1)\mu, ~i=2,3,\dots,B,\\
%\sum\limits_{i=0}^{B}\pi_i &=1.
%\end{align*}
%Standard calculations for the stationary distribution yield (see \ref{app:homogInfBufCTMC})
\begin{align*}
\pi_0 &=\frac{(k-2)(k-1)^B}{2(k-1)^{B+1}-k}.
\end{align*}
Using the fact that $\sum_{i=1}^B\pi_i=1-\pi_0$,
the capacity is %of this system is
\begin{align*}
C &= q\sum\limits_{i=1}^B \mu(k-1)\pi_i %= q\mu(k-1)(1-\pi_0)%\\
%& = \mu(k-1)\left(1-\frac{k-2}{2(k-1)-k\left(\frac{1}{k-1}\right)^B}\right)\\
%& = \mu(k-1)\left(\frac{k-k\left(\frac{1}{k-1}\right)^B}{2(k-1)-k\left(\frac{1}{k-1}\right)^B}\right)\\
%& = \mu k(k-1) \left(\frac{1-\left(\frac{1}{k-1}\right)^B}{2(k-1)-k\left(\frac{1}{k-1}\right)^B}\right)\\
 = \frac{q\mu k \left(1-\left(\frac{1}{k-1}\right)^{B}\right)}{2-k\left(\frac{1}{k-1}\right)^{B+1}}.
\end{align*}
Note that as $B\to \infty$, $C$ for the finite buffer case approaches $C$ for the infinite buffer case.
The expected number of stored qubits is
\begin{align*}
E[Q] = \sum\limits_{i=1}^B i\pi_i %= k\pi_0\sum\limits_{i=1}^B i\rho^i = k\pi_0\frac{\rho(B\rho^{B+1}+1-(B+1)\rho^B)}{(1-\rho)^2}\\
%&=k\left(\frac{(k-2)(k-1)^B}{2(k-1)^{B+1}-k}\right)\frac{\left(\frac{1}{k-1}\right)\left(B\left(\frac{1}{k-1}\right)^{B+1}+1-(B+1)\left(\frac{1}{k-1}\right)^B\right)}{\left(1-\frac{1}{k-1}\right)^2}\\
%&=\frac{k(k-2)}{2(k-1)^{B+1}-k}\frac{\left(\frac{1}{k-1}\right)\left(B\left(\frac{1}{k-1}\right)+(k-1)^B-(B+1)\right)(k-1)^2}{(k-2)^2}\\
=\frac{k\left(B+(k-1)^{B+1}-(B+1)(k-1)\right)}{(2(k-1)^{B+1}-k)(k-2)}.
\end{align*}
\subsection{Decoherence}
%%!TEX PS-program = pdflatexmk
%%!TEX root = quantum_jrnl_short.tex
Assume now that quantum states in our system are subject to decoherence. 
Further, assume that all states decohere at the same rate $\alpha$, even in the case of heterogeneous links, since coherence time is dependent on the quantum memories at the switch and not on the links themselves.
%We model this phenomenon by assuming that they decay exponentially at rate $\alpha$, \emph{i.e.}, a quantum state is coherent for an exponentially distributed time with mean $1/\alpha$.  
Under the assumption that coherence time is exponentially distributed with rate $1/\alpha$,
 incorporating decoherence does not change the structure of the CTMC; it merely increases ``backward'' transition rates. Specifically, in the homogeneous case, the transition from any state $j\geq 1$ to state $j-1$ now has rate $(k-1)\mu+j\alpha$, where $j\alpha$ represents the aggregate decoherence rate of all $j$ stored qubits. In the heterogeneous case, the transitions are modified in a similar manner for any state $j\vect{e}_l$, $l\in\{1,\dots,k\}$, $j\geq1$. The derivations of stationary distributions, capacities, and expected number of qubits stored are very similar to those for models without decoherence; we present the final relevant expressions here and leave details to Appendix \ref{app:decoh}.
All expressions below can be computed numerically. 
\begin{description}[style=unboxed,leftmargin=0.5cm]
\item[Heterogeneous Links:] For finite buffer $B$,
 \begin{align*}
 \pi_0 &=\left(1+\sum\limits_{l=1}^k\sum\limits_{j=1}^{B}\prod\limits_{i=1}^j\frac{\mu_l}{\gamma-\mu_l+i\alpha}\right)^{-1},\\
 C &=q\pi_0\sum\limits_{l=1}^k\sum\limits_{j=1}^{B}(\gamma-\mu_l)\prod\limits_{i=1}^j \frac{\mu_l}{\gamma-\mu_l+i\alpha},\\
 E[Q] &=\pi_0\sum\limits_{j=1}^{B}j\sum\limits_{l=1}^k\prod\limits_{i=1}^j \frac{\mu_l}{\gamma-\mu_l+i\alpha}.
 \end{align*}
 For infinite buffers, let $B\to\infty$ in all expressions above.
 \item[Homogeneous Links:] For finite buffer $B$,
 \begin{align*}
\pi_0 &= \left(1+k\sum\limits_{i=1}^{B}\prod\limits_{j=1}^i\frac{\mu}{ ((k-1)\mu+j\alpha)}\right)^{-1},\\
 C &= q(k-1)\mu(1-\pi_0),\\
 E[Q] &= \pi_0k\sum\limits_{i=1}^{B}i\prod\limits_{j=1}^i\frac{\mu}{ ((k-1)\mu+j\alpha)}.
 \end{align*}
 For infinite buffers, let $B\to\infty$ in all expressions above.
\end{description}
\section{DTMC for Bipartite Entanglements}
%%!TEX PS-program = pdflatexmk
%%!TEX root = quantum-sigconf.tex
\label{sec:DTMCBipartite}
In this section, we describe the DTMC model and analyze the simplest switch variant: a switch that serves only bipartite entanglements, has infinite buffer space, and whose quantum states are not subject to decoherence. By studying this basic system, we learn that the DTMC model exhibits limitations such that introducing additional constraints to this model, such as finite buffers or quantum state decoherence, makes the resulting model exceedingly difficult to analyze, and therefore not a viable option for modeling more complex entanglement switching mechanisms. Nevertheless, the results of our study of this basic DTMC scheme serve as a valuable comparison basis against the CTMC model and justify the latter's use as a reasonable approximation model.
\subsection{Model Description}
%%!TEX PS-program = pdflatexmk
%%!TEX root = quantum_jrnl.tex
We model a switch serving $k$ users, each of whom has a separate, dedicated link to the switch, as a slotted system where each slot is of length $\tau$ seconds. Each user (or link) attempts a two-qubit entanglement in each time slot.
 %of length $\tau$ seconds. 
 We assume links are homogeneous, so that the success probabilities of all entanglements are equal and do not depend on the links. Let $p$ denote the probability that an entangled pair is successfully established on a link, and define $\pbar \equiv 1-p$. Then the expected time to successfully create a link entanglement is given by $\tau/p$. We assume that the switch can store an infinite number of qubits. Moreover, we assume that any successful entanglement has fidelity one, and that states do not decohere.
 % are not subject to decoherence.

As before, we assume that any pair of users wishes to ``communicate'' (\emph{i.e.}, share an entangled state) as long as entanglements are available. The switch serves users based on the hybrid OLEF and random policy as described in Section \ref{sec:modelandObj}.
%OLEF policy, but with no additional scheduling or policy constraints: for instance, if there are $n\leq k$ distinct users with one entanglement each, then the switch pairs them at random. 
We also assume that any time the switch performs a BSM, it succeeds with probability $q$.

\begin{figure}[h]
\centering
\includegraphics[width=0.52\textwidth,trim={0 11cm 0 3.5cm},clip]{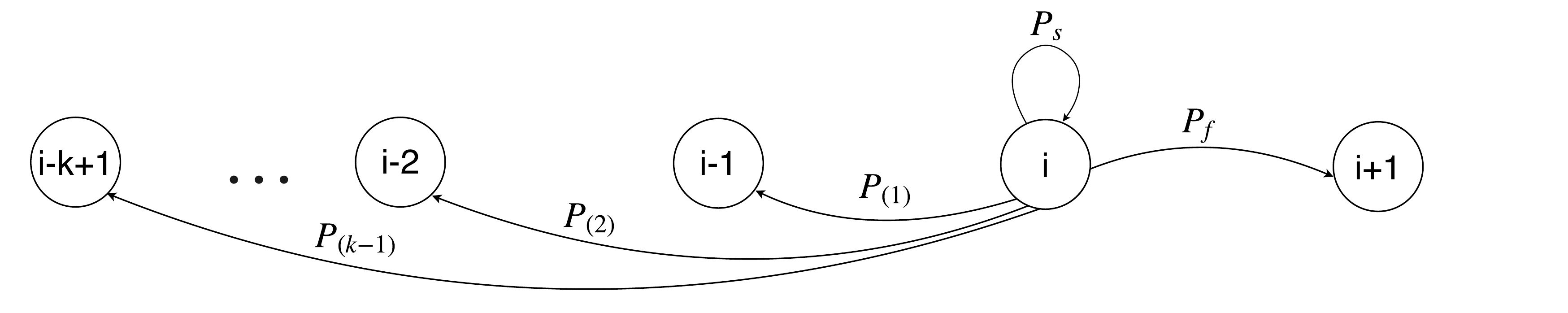}
\caption{A DTMC model with $k$ users, infinite buffer, and homogeneous links. Here, $i\geq k+1$, $P_f$ is the probability of advancing forward in the Markov chain, $P_s$ is the probability of remaining in the current state, and $P_{(j)}$ is the probability of going back $j$ states.}
\label{fig:DTMC}
\end{figure}

Note that only one link will have stored entanglements, since whenever a distinct pair of users have entanglements, they are immediately paired up for a BSM. As a consequence of this, as well as the link homogeneity assumption, it is not necessary to keep track of which link has stored entanglements: one need only keep track of \emph{how many} are stored. Hence, the state space is given by $\Omega=\{0,1,2,\dots\}$. Let $S$ denote the link that has at least one stored entanglement.
Figure \ref{fig:DTMC} illustrates the possible transitions from a state $i\geq k+1$ (as we will see later, transitions for states $i\in\{0,1,\dots,k\}$ require special consideration). Table \ref{tab:dtmc} provides a notation reference that is used in the analysis.
\begin{table}
  \caption{Notation for the DTMC model.}
  \label{tab:dtmc}
  \begin{tabular}{cl}
    \toprule
    Notation & Description\\
    \midrule
    $p$ & probability of a successful link entanglement\\
    $S$ & link with stored entanglements\\
    $P_f$ & probability of gaining an entanglement in memory\\
    $P_s$ & probability of remaining in current state\\
    $P_{(j)}$ & probability of using $j$ of the stored entanglements\\
    $P_{j,0}$ & probability of going from state $j\in\{0,\dots,k-1\}$ to $0$\\
    $P_{j,1}$ & probability of going from state $j\in\{0,\dots,k\}$ to $1$\\
  \bottomrule
\end{tabular}
\end{table}
\subsection{Analysis}
\label{sec:dtmcAnalysis}
First, we fully define the transition probabilities for this chain. We expect the stationary distribution to have a geometric form and show this to be true. However, a closed-form solution is not obtainable for large $k$, as it requires solving a polynomial of degree $k-1$ for an unknown factor, $\beta$. On the other hand, not having a closed-form solution for the stationary probability vector does not preclude us from deriving a simple expression for the capacity of the switch -- it is $qkp/2$. Finally, we also obtain a simple expression for the expected number of qubits in memory at the switch, but are constrained to compute it numerically due to its dependence on $\beta$.
\subsubsection{Transition Probabilities}
%%!TEX PS-program = pdflatexmk
%%!TEX root = quantum_jrnl.tex
\begin{figure*}
\centering
$\setcounter{MaxMatrixCols}{20}
%P =
\begin{bmatrix}
P_{0,0} & P_{0,1} & 0                      & \cdots\\
P_{1,0} & P_{1,1} & P_f  & 0        & \cdots\\
P_{2,0} & P_{2,1} & P_s & P_f     & 0         & \cdots\\
\vdots   & \vdots\\
P_{i-1,0} & P_{i-1,1} & P_{(i-3)} & \cdots &P_{(3)} & P_{(2)} &P_{(1)} & P_s & P_f & 0 & \cdots\\
P_{i,0} & P_{i,1} & P_{(i-2)} & P_{(i-3)} & \cdots &P_{(3)} & P_{(2)} & P_{(1)} & P_s & P_f & 0 & \cdots\\
P_{i+1,0} & P_{i+1,1} & P_{(i-1)} & P_{(i-2)} & P_{(i-3)}& \cdots & P_{(3)} & P_{(2)} & P_{(1)} & P_s & P_f & 0 & \cdots\\
\vdots      & \vdots &  &\ddots  &   & \ddots &   \ddots& \ddots & \ddots & \ddots\\
P_{k-1,0} & P_{k-1,1} & P_{(k-3)}& \cdots & & &  &  & \cdots & P_{(1)} & P_{s} & P_f & 0 & \cdots\\
0 & P_{k,1} & P_{(k-2)}& P_{(k-3)}& \cdots &  &  &  & \cdots & P_{(2)} & P_{(1)} & P_{s} & P_f & 0 & \cdots\\
0 & 0 & P_{(k-1)}& P_{(k-2)} & P_{(k-3)}&\cdots &   &  & \cdots & P_{(3)} & P_{(2)} & P_{(1)} & P_{s} & P_f & 0 & \cdots\\
0 & 0 & 0&  P_{(k-1)} & P_{(k-2)} & P_{(k-3)} & \cdots &   & \cdots & P_{(4)} & P_{(3)} & P_{(2)} & P_{(1)} & P_{s} & P_f & 0 & \cdots\\
\vdots&\vdots&\vdots &\ddots&\ddots&&  &  & \ddots & \ddots & \ddots & \ddots & \ddots & \ddots & \ddots
\end{bmatrix}$
\caption{Transition probability matrix $P$ for the DTMC model.}
\label{fig:dtmcP}
\end{figure*}
Figure \ref{fig:dtmcP} presents the transition probability matrix $P$ for this DTMC. Note that repetition begins after the $k$th row of the matrix. We derive expressions for all non-zero transition probabilities. In the discussion that follows, we say that a link ``succeeds'' or ``fails'' for brevity, when referring to a link that successfully generates an entanglement or fails to do so, respectively.  First, consider any state $i>1$. The transitions for this state are described as follows:
\begin{description}[style=unboxed,leftmargin=0.5cm]
\item[$i\to i+1$:] the only way to advance forward in the chain is if $S$ successfully generates a new entanglement, but all other links fail to do so. This probability is given by $$P_f = p\pbar^{k-1}.$$
\item[$i\to i$:] there are two ways to remain in the current state: (a) all links fail or (b) $S$ succeeds and only one of the $k-1$ other links succeeds. This occurs with probability $$P_s=\pbar^k+(k-1)p^2\pbar^{k-2}.$$
\item[$i\to i-j$], for $j\in\{1,\dots,M\},$ where $M=k-1$ if $i\geq k+1$ and $M=i-2$ otherwise. Here, $M$ signifies the maximum number of stored entanglements that can be used up when starting from state $i$. Note that even in the case all $k$ links succeed and $i\geq k$, only $k-1$ of the stored entanglements get used: the entanglement that was generated by $S$ cannot be paired with another entanglement from $S$. As stated above, we compute transition probabilities to states $0$ or $1$ separately, since they require special consideration. This is why $M=i-2$ for states $i< k+1$. Keeping these constraints in mind, the transition from $i$ to $i-j$ occurs in two types of events:
	\begin{description}
		\item{(a)} $S$ fails and exactly $j$ of the $k-1$ other links succeed,
		\item{(b)} $S$ succeeds and exactly $j+1$ of the $k-1$ other links succeed.
	\end{description}
These events occur with probability
	\begin{align*}
	P_{(j)} &= \pbar{k-1 \choose j} p^j\pbar^{k-1-j}+p{k-1 \choose j+1}p^{j+1}\pbar^{k-1-(j+1)}\\
	&={k-1 \choose j} p^j\pbar^{k-j}+{k-1 \choose j+1}p^{j+2}\pbar^{k-j-2}.
	\end{align*}
\end{description}
Next, we discuss transitions to states $0$ and $1$, which, unlike the probabilities above, depend on the value $i$ of the state from which the transitions occur. To help with this task, we will first need to be able to compute two types of probabilities: the first is the probability that out of $k$ events, $j\geq i$ succeed, where $j$ is either zero or an even number, and we call this probability $P_e(i,k)$; and the second is the probability that out of $k$ events, $j\geq i$ succeed, where $j$ is an odd number, and we call this $P_o(i,k)$. To compute these, we will make use of two indicator functions:
\begin{align*}
\indic\{j~\text{is 0 or even}\} &\coloneqq \frac{1+(-1)^j}{2},~
\indic\{j~\text{is odd}\} \coloneqq \frac{1-(-1)^j}{2}.
\end{align*}
\begin{flalign*}
\text{Then,}\quad
P_e(i,k) &= \sum\limits_{j=i}^k \left(\frac{1+(-1)^j}{2}\right){k \choose j}p^j\pbar^{k-j},\\
P_o(i,k) &= \sum\limits_{j=i}^k \left(\frac{1-(-1)^j}{2}\right){k \choose j}p^j\pbar^{k-j}.
\end{flalign*}
Now, for any state $i$, $1\leq i \leq k$, the transition to state $1$ occurs under the following conditions:
\begin{description}[style=unboxed,leftmargin=0.5cm]
\item{If $i$ is even:} 
	\begin{description}
		\item{1.} $S$ fails and $j\geq i-1$ others succeed, $j$ odd.
		\item{2.} $S$ succeeds and $j\geq i$ others succeed, $j$ even.
	\end{description}
	\begin{align*}
	P_{i,1} &= \pbar P_o(i-1,k-1)+pP_e(i,k-1).
	\end{align*}
\item{If $i$ is odd:}
	\begin{description}
		\item{1.} $S$ fails and $j\geq i-1$ others succeed, $j$ even.
		\item{2.} $S$ succeeds and $j\geq i$ others succeed, $j$ odd.
	\end{description}
	\begin{align*}
	P_{i,1} &= \pbar P_e(i-1,k-1)+pP_o(i,k-1).
	\end{align*}
\end{description}
Similarly, for any state $i\in\{1,\dots,k-1\}$, transitioning to state $0$ occurs under the following conditions:
\begin{description}[style=unboxed,leftmargin=0.5cm]
\item{If $i$ is even:} 
	\begin{description}
		\item{1.} $S$ fails and $j\geq i$ others succeed, $j$ even.
		\item{2.} $S$ succeeds and $j \geq i+1$ others succeed, $j$ odd.
	\end{description}
	\begin{align*}
	P_{i,0} &= \pbar P_e(i,k-1)+pP_o(i+1,k-1).
	\end{align*}
\item{If $i$ is odd:}
	\begin{description}
		\item{1.} $S$ fails and $j\geq i$ others succeed, $j$ odd.
		\item{2.} $S$ succeeds and $j\geq i+1$ others succeed, $j$ even.
	\end{description}
	\begin{align*}
	P_{i,0} &= \pbar P_o(i,k-1)+p P_e(i+1,k-1).
	\end{align*}
\end{description}
In the special case of $0\to0$, either all must fail or there must be an even number of entanglements. Hence, $P_{0,0}=P_e(0,k)$.
Finally, in the special case of $0\to1$, there must be an odd number of entanglements, given by $P_{0,1} = P_o(1,k)$.
\subsubsection{Stationary Distribution}
%%!TEX PS-program = pdflatexmk
%%!TEX root = quantum_jrnl.tex
The balance equations for the DTMC are as follows:
\begin{align}
\sum\limits_{i=0}^{k-1}\pi_iP_{i,0} &= \pi_0,\label{eq:a}\\
\sum\limits_{i=0}^k\pi_iP_{i,1} &= \pi_1\label{eq:b}.
\end{align}
For any state $i\geq 2$, the balance equations have the form:
\begin{align}
\label{eq:ibal}
\pi_{i-1}P_f+\pi_iP_s+\pi_{i+1}P_{(1)}+\cdots+\pi_{i+k-1}P_{(k-1)} = \pi_i.
\end{align}
and finally, the normalizing condition is
\begin{align}
\sum\limits_{i=0}^{\infty}\pi_i = 1.
\label{eq:f}
\end{align}
We postulate that $\pi_i=\beta^{i-1}\pi_1$ for $i\geq 2$, with $\beta\in(0,1)$. Introducing this value of $\pi_1$ in Eq. (\ref{eq:ibal}) yields
(see \ref{app:dtmcPolyproof} for a proof)
\begin{align}
f(\beta) \coloneqq (\beta p+\pbar)^{k-1}(p+\beta\pbar)-\beta.%, \quad \forall i\geq 2.
\label{eq:dtmcPoly}
\end{align}
To show that $\pi_i=\beta^{i-1}\pi_1$ for $i\geq 2$ is indeed the solution to this system, we must prove that:
\begin{description}[style=unboxed,leftmargin=0cm]
\item[1.] There exists  $\beta \in(0,1)$ satisfying Eq. (\ref{eq:dtmcPoly}), and that this $\beta$ is unique,
\item[2.] Given the solution above, note that both Eqs (\ref{eq:a}) and (\ref{eq:b}) can be written in terms of only $\pi_1$ and $\pi_0$. Hence, for the proposed solution to be valid, one of these equations must be redundant, \emph{i.e.}, we must show that Eq. (\ref{eq:a}) is equivalent to Eq. (\ref{eq:b}).
\end{description}
We prove the first statement above in \ref{app:dtmcBetaProof} and the second in \ref{app:dtmcEquiv}, and conclude that the proposed form for $\pi_i$, $i\geq2$ is valid. Moreover, we can derive expressions for $\pi_0$ and $\pi_1$ in terms of $\beta$. From the normalizing condition (\ref{eq:f}), we have
\begin{align}
\pi_0 &= 1-\frac{\pi_1}{1-\beta}.
\label{eq:pi0pi1}
\end{align}
In \ref{app:dtmcEquiv}, we rearranged (\ref{eq:a}) to look as follows:
\begin{align}
\sum\limits_{i=1}^{k-1}\beta^{i}P_{i,0} &= \frac{\beta\pi_0}{\pi_1}P_{0,1}
\label{eq:derivpi1}
\end{align}
and also showed that the left side of Eq. (\ref{eq:derivpi1}) equals
\begin{align*}
%\sum\limits_{i=1}^{k-1}\beta^{i}P_{i,0}=&
&\frac{1}{2}\left[\frac{\beta}{1-\beta}-\frac{2\beta}{1-\beta^2}(p\beta+\pbar)^{k-1}(p+\pbar\beta)-(\pbar-p)^{k}\frac{\beta}{1+\beta}\right]\\
&=\frac{1}{2}\left[\frac{\beta}{1-\beta}-\frac{2\beta^2}{1-\beta^2}-(\pbar-p)^{k}\frac{\beta}{1+\beta}\right]\text{~by Eq. (\ref{eq:dtmcPoly})},\\
%=&\frac{1}{2}\left[\frac{\beta(1+\beta)-2\beta^2}{1-\beta^2}-(\pbar-p)^{k}\frac{\beta}{1+\beta}\right]\\\\
%=&\frac{1}{2}\left[\frac{\beta-\beta^2}{1-\beta^2}-(\pbar-p)^{k}\frac{\beta}{1+\beta}\right]
&=\frac{1}{2}\left[\frac{\beta}{1+\beta}-(\pbar-p)^{k}\frac{\beta}{1+\beta}\right],
\end{align*}
and therefore, Eq. (\ref{eq:derivpi1}) becomes
\begin{align}
\frac{\beta\pi_0}{\pi_1}P_{0,1} &= \frac{1}{2}\left[\frac{\beta}{1+\beta}-(\pbar-p)^{k}\frac{\beta}{1+\beta}\right],\text{~or~}\nonumber\\
\frac{\pi_0}{\pi_1}P_{0,1} &= \frac{1-(\pbar-p)^{k}}{2(1+\beta)}.\label{eq:derivpi1New}
\end{align}
Next, we compute
\begin{align*}
P_{0,1} &= P_o(1,k) = \sum\limits_{i=1}^k \frac{1-(-1)^i}{2}{k \choose i} p^i\pbar^{k-i}\\
&=\sum\limits_{i=0}^k \frac{1-(-1)^i}{2}{k \choose i} p^i\pbar^{k-i}
=\frac{1}{2}-\frac{1}{2}(\pbar-p)^k.
\end{align*}
Substituting this into Eq. (\ref{eq:derivpi1New}),
\begin{align}
\frac{\pi_0}{\pi_1}\left(\frac{1-(\pbar-p)^k}{2}\right) &= \frac{1-(\pbar-p)^{k}}{2(1+\beta)},\nonumber\\
\frac{\pi_0}{\pi_1}&= \frac{1}{1+\beta},\nonumber\\
1-\frac{\pi_1}{1-\beta}&= \frac{\pi_1}{1+\beta}\text{~by Eq. (\ref{eq:pi0pi1})},\nonumber\\
%1&= \frac{\pi_1}{1+\beta}+\frac{\pi_1}{1-\beta},\\\\
%1&= \pi_1\frac{2}{1-\beta^2},\\\\
\pi_1 &=\frac{1-\beta^2}{2}.
\label{eq:dtmcPi1}
\end{align}
Now, we can compute $\pi_0$ in terms of only $\beta$:
\begin{align*}
\pi_0 &= 1-\frac{\pi_1}{1-\beta} = 1-\left(\frac{1}{1-\beta}\right)\frac{1-\beta^2}{2}
%= 1-\frac{1+\beta}{2}
= \frac{1-\beta}{2}.
\end{align*}
\subsubsection{Capacity and Qubits in Memory}
%%!TEX PS-program = pdflatexmk
%%!TEX root = quantum_jrnl.tex
\label{subsec:dtmcCapEQ}
As with the CTMC models let $Q$ represent the number of stored qubits at the switch. Let $N$ denote the number of entanglement pairs generated in one time step of the DTMC.
Then the capacity is defined as follows:
\begin{align*}
C &= q\sum\limits_{i=0}^{\infty}\pi_i E\lbrack N\lvert Q=i\rbrack.
\end{align*}
To compute this expression, we consider two separate cases: case 1 is when $i\geq k-1$ and case 2 is when $i<k-1$. In case 1, there can be at most $k-1$ entanglements; the expected number is given by
\begin{align*}
%E\lbrack  N\lvert Q=i,  i\geq k-1\rbrack = \sum\limits_{j=0}^{k-1}j{k-1 \choose j}p^j\pbar^{k-1-j} = (k-1)p.
E\lbrack  N\lvert Q=i\geq k-1\rbrack = \sum\limits_{j=0}^{k-1}j{k-1 \choose j}p^j\pbar^{k-1-j} = (k-1)p.
\end{align*}
For case 2, we can have up to $i+m$ entanglements, where $m=\lfloor\left(\frac{k-i}{2}\right)^+\rfloor$. The expected number is then given by
\begin{align*}
%&E\lbrack  N\lvert Q=i, ~ i< k-1\rbrack = \sum\limits_{j=0}^{i+m} jP(N = j \lvert Q=i,~ i\leq k-2)=\\
&E\lbrack  N\lvert Q=i< k-1\rbrack = \sum\limits_{j=0}^{i+m} jP(N = j \lvert Q=i\leq k-2).%=\\
%&\sum\limits_{j=0}^i jP(N = j\lvert Q=i\leq k-2)+\sum\limits_{j=i+1}^{i+m} jP(N = j\lvert Q=i\leq k-2).
%&E\lbrack  N\lvert Q=i, ~ i< k-1\rbrack = \sum\limits_{j=0}^{i+m} jP(j \text{~ents}\lvert Q=i,~ i\leq k-2)=\\
%&\sum\limits_{j=0}^i jP(j \text{~ents}\lvert Q=i,~ i\leq k-2)+\sum\limits_{j=i+1}^{i+m} jP(j \text{~ents}\lvert Q=i, ~i\leq k-2).
\end{align*}
For the sum above, consider first $j\in\{0,\dots,i\}$.
%For the first sum above, 
Here,
we are looking for the probability that there are fewer new entanglements than the number stored, so the probability that we generate $j$ pairs is given by
\begin{align*}
P(N = j \lvert Q=i,~ i\leq k-2) &= {k-1 \choose j} p^j\pbar^{k-1-j}.
%P(j \text{~ents}\lvert Q=i,~ i\leq k-2) &= {k-1 \choose j} p^j\pbar^{k-1-j}.
\end{align*}
However, note that the case $j=i$ is a special one: another way we can generate $i$ entanglements is if there are a total of $i+1$ successes from the $k-1$ links that have nothing stored, while $S$ fails. Then, the extra entanglement has no pair, and the total number of pairs generated is still $i$. This is given by
\begin{align*}
i{k-1 \choose i+1}p^{i+1}\pbar^{k-i-1}.
\end{align*}
%Next, we focus on the second sum. 
Next, we focus on the case where $j\in\{i+1,\dots,i+m\}$.
After the first $i$ successes, there need to be anywhere from 2 to at most $k-i$ ``extra'' successes to generate new pairs. Denote the number of these extra successes by the variable $l \in\{2,\dots,k-i\}$, and the number of new pairs (or BSMs) generated from them is $\left\lfloor\frac{l}{2}\right\rfloor$. Then we can write the second sum as follows:
\begin{align*}
\sum\limits_{l=2}^{k-i}\left(\left\lfloor\frac{l}{2}\right\rfloor+i\right){k \choose i+l}p^{i+l}\pbar^{k-i-l}.
\end{align*}
Combining everything we have learned, we obtain
\begin{align}
C %&= \sum\limits_{i=0}^{k-2}\pi_i E\lbrack N\lvert Q=i\rbrack +\sum\limits_{i=k-1}^{\infty}\pi_iE\lbrack N\lvert Q=i\rbrack\\
&=q\sum\limits_{i=0}^{k-2}\pi_i \left(\sum\limits_{j=0}^i j{k-1 \choose j} p^j\pbar^{k-1-j}+i{k-1 \choose i+1}p^{i+1}\pbar^{k-i-1}\right.\nonumber\\
&+\left.\sum\limits_{l=2}^{k-i}\left(\left\lfloor\frac{l}{2}\right\rfloor+i\right){k \choose i+l}p^{i+l}\pbar^{k-i-l}\right)
 +q(k-1)p\sum\limits_{i=k-1}^{\infty}\pi_i.
 \label{eq:dtmcCap}
\end{align}
In \ref{app:dtmcCapEQ}, we show that the above evaluates to 
\begin{align}
C=\frac{qkp}{2}.
\end{align}
Next, we derive the expected number of qubits stored at the switch, $E[Q]$. This is given by 
\begin{align}
E[Q] &= \sum\limits_{i=1}^{\infty}i\pi_i = \pi_1\sum\limits_{i=1}^{\infty}i\beta^{i-1} =
\frac{\pi_1}{\beta}\sum\limits_{i=1}^{\infty}i\beta^{i}\nonumber\\
& =\frac{\pi_1}{\beta}\frac{\beta}{(1-\beta)^2}
%=\frac{\pi_1}{(1-\beta)^2}
=\frac{1-\beta^2}{2(1-\beta)^2}=\frac{1+\beta}{2(1-\beta)}.\label{eq:EQdtmc}
\end{align}
\subsection{Comparison of DTMC models with CTMC models}
%%!TEX PS-program = pdflatexmk
%%!TEX root = quantum_jrnl.tex
\label{sec:ctmcVsdtmc}
Recall that in the discrete model, the amount of time it takes to successfully generate a link entanglement is given by $\tau/p$. In the continuous model, the rate of successful entanglement generation is $\mu$, so the time to generate an entanglement is $1/\mu$. Hence, $\tau/p=1/\mu$ or equivalently, $\mu=p/\tau$. Then, note that the DTMC capacity that we derived in Section \ref{subsec:dtmcCapEQ} is the capacity per time slot of length $\tau$ seconds. Therefore, in order to make a comparison against the CTMC capacity, we must perform a unit conversion: divide the discrete capacity by $\tau$ in order to obtain the number of entanglement pairs per \emph{second}, as opposed to per \emph{time slot}. This yields
\begin{align*}
C_{\text{DTMC}} &= \frac{qkp}{2\tau} = \frac{qk\mu}{2} = C_{\text{CTMC}}.
\end{align*}
We conclude that the capacities produced by the DTMC and CTMC models match exactly.

\begin{figure}
    \centering
    \subfloat{
        \includegraphics[width=0.24\textwidth]{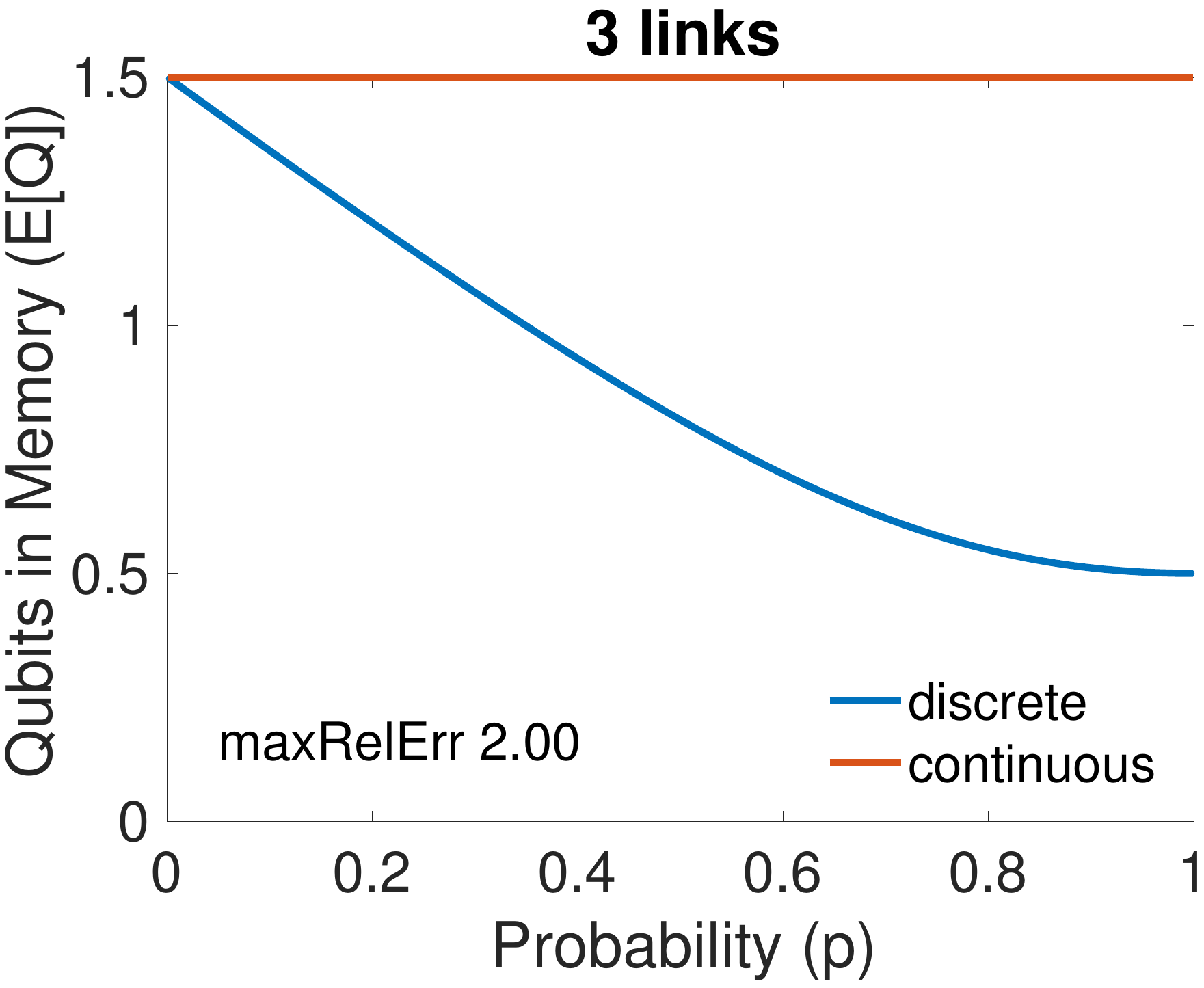}}
    \subfloat{
        \includegraphics[width=0.24\textwidth]{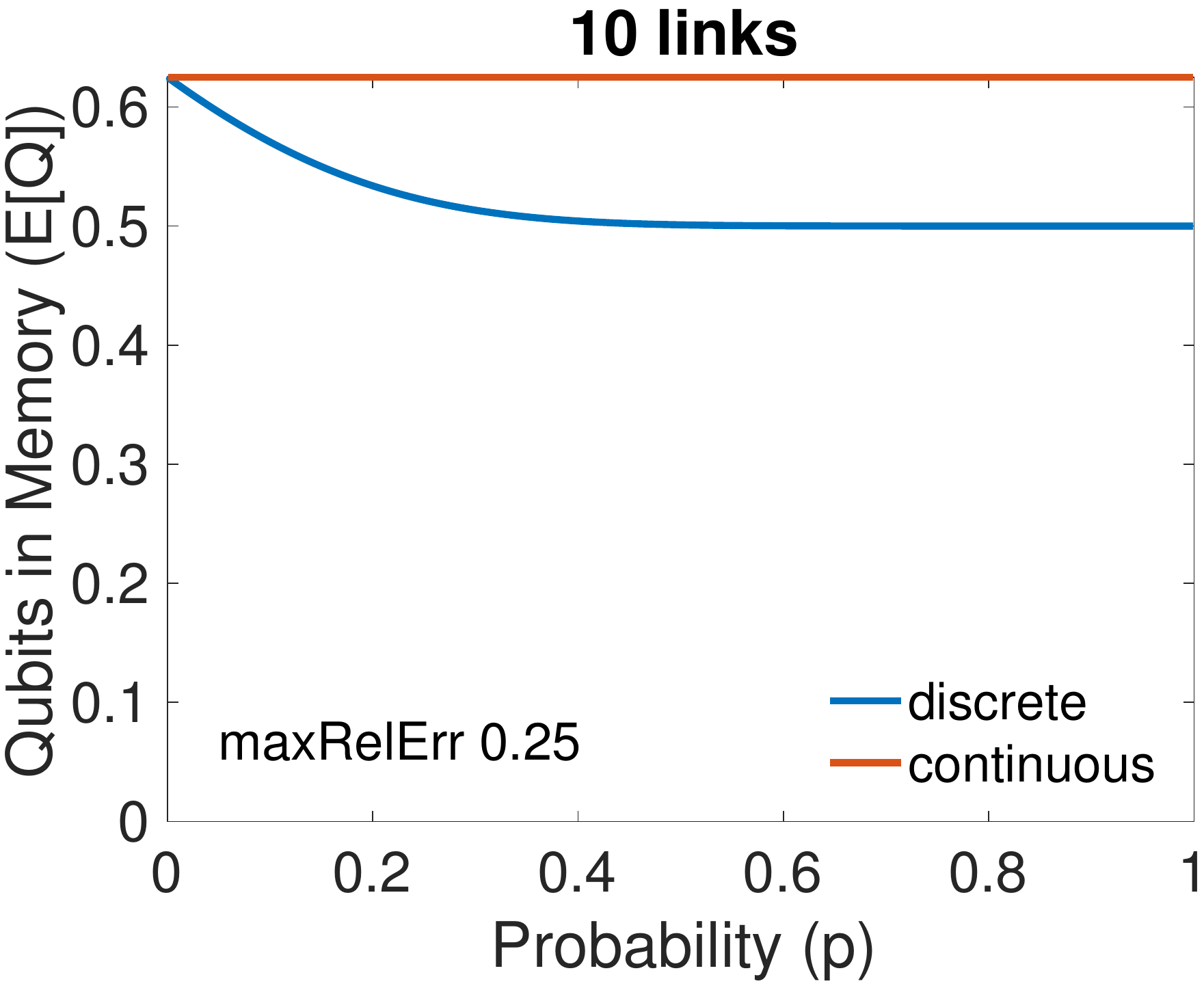}}
        \qquad
        \subfloat{
        \includegraphics[width=0.24\textwidth]{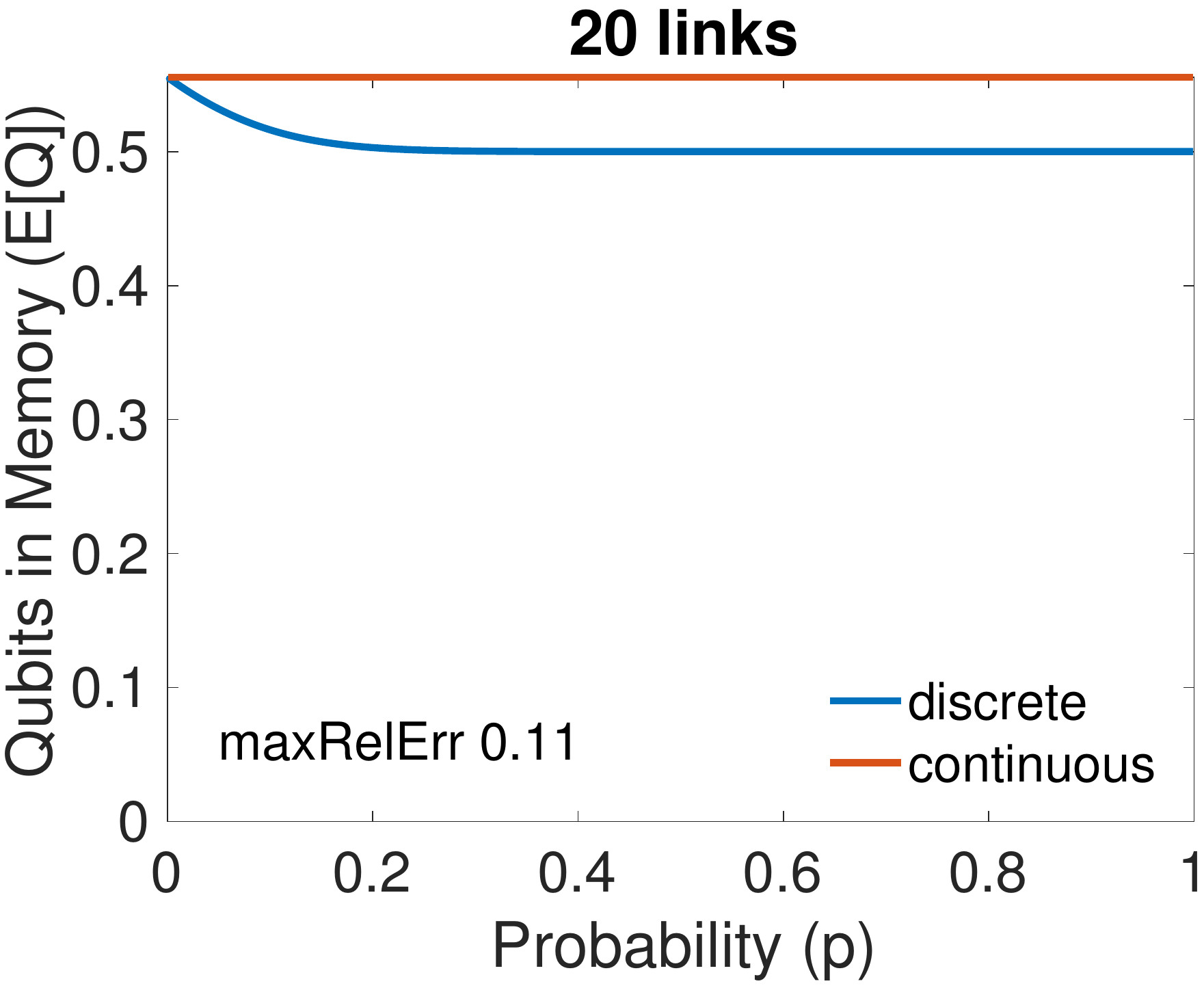}}
        \subfloat{
        \includegraphics[width=0.24\textwidth]{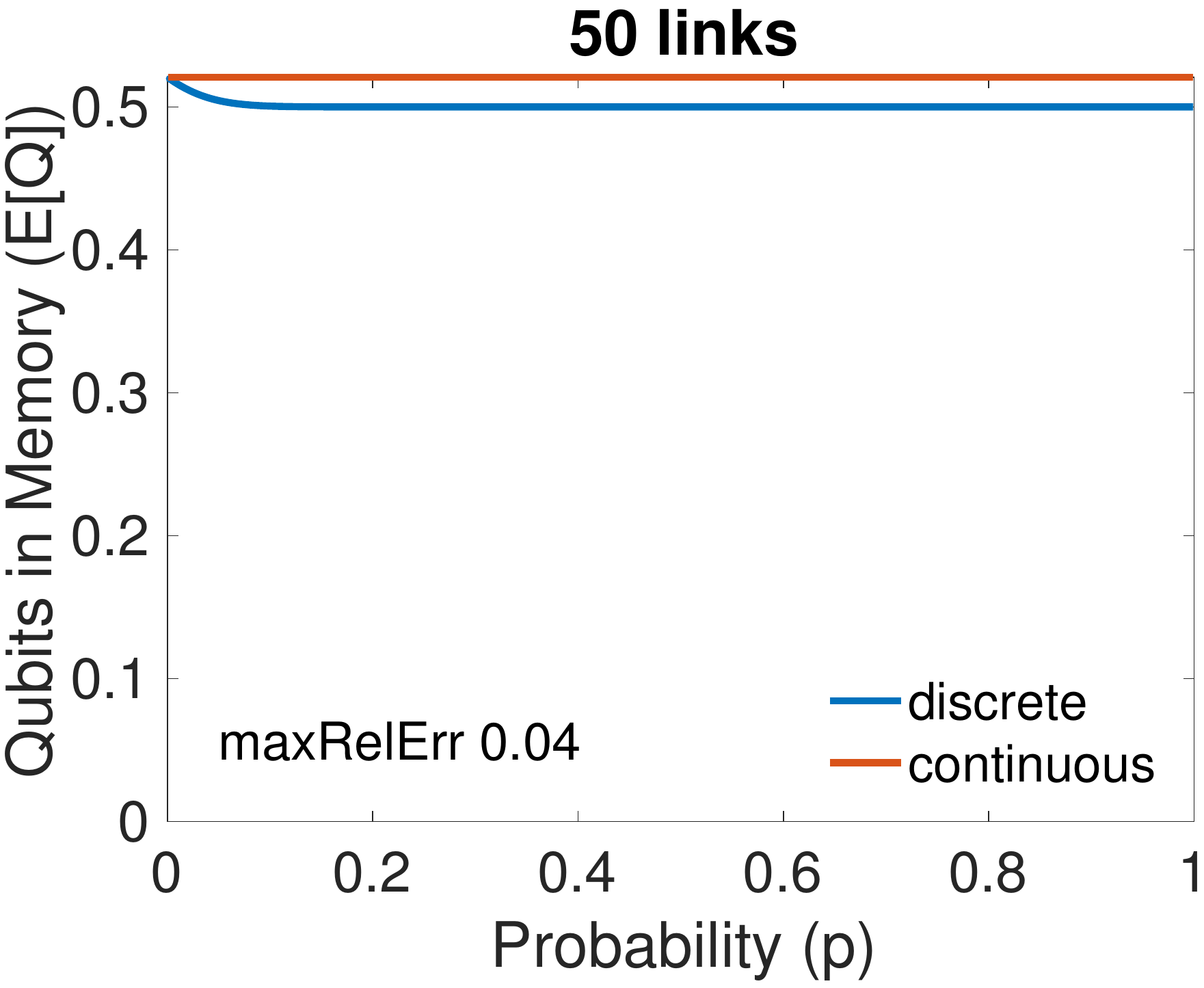}}
    \caption{Comparison of the expected number of qubits in memory $E[Q]$ for the DTMC and CTMC models, as the number of links is varied $\in\{3,10,20,50\}$ and for entanglement generation probabilities $p\in(0,1)$. maxRelErr is the maximum relative error between the discrete and continuous expressions for $E[Q]$.}
    \label{fig:dtmcVSctmc}
\end{figure}

Next, we compare the expected number of qubits in memory at the switch, $E[Q]$ as predicted by the DTMC and the CTMC models.
Figure \ref{fig:dtmcVSctmc} compares numerically the discrete and continuous $E[Q]$'s as the number of users $k$ and probability $p$ vary.
For each value of $p$ and $k$, we use Eq. (\ref{eq:dtmcPoly}) to numerically solve for $\beta$. For each value of $k$, we report the maximum relative error, defined as
\begin{align*}
maxRelErr(k) &= \max\limits_{p\in(0,1)}\frac{|E[Q]_{\text{DTMC}}(k,p)-E[Q]_{\text{CTMC}}(k,p)|}{E[Q]_{\text{DTMC}}(k,p)},
\end{align*}
where $E[Q]_{\text{DTMC}}$ and $E[Q]_{\text{CTMC}}$ are the discrete and continuous functions for $E[Q]$, respectively. We observe that the error is largest when $p$ is close to 1. Note that
\begin{align*}
\lim_{p\to1}f(\beta) = \lim_{p\to1}(\beta p+\pbar)^{k-1}(p+\beta\pbar)-\beta = \beta^{k-1}-\beta.
\end{align*}
Since $f(\beta)=0$, we conclude that as $p\to1$ and $k\to\infty$, $\beta\to0$ (note: $\beta=1$ is always a root of $f(\beta)$, but we always discard this root because it is not in $(0,1)$). As $\beta\to0$, $E[Q]\to 1/2$ according to Eq. (\ref{eq:EQdtmc}), which is consistent with the numerical observations. Meanwhile, as $k\to\infty$, the continuous $E[Q]$ also approaches 1/2. We conclude that as $k\to\infty$, $maxRelErr\to0$, which can be observed in Figure~\ref{fig:dtmcVSctmc}. Also, the largest $maxRelErr$ occurs for the lowest value of $k=3$, when $p\to1$. But even in this (worst case), although the error is $maxRelErr(3)=2$, it corresponds to discrete and continuous versions of $E[Q]$ differing by a prediction of only a single qubit. From these analytic and numerical observations, we conclude that the CTMC model is 
sufficiently accurate so as to be used to explore issues such as decoherence, link heterogeneity, and switch buffer constraints.
%a reasonable approximation of the DTMC model.

Recall that in Section \ref{sec:CTMCBipartite}, we introduced CTMCs for systems where the switch has buffer constraints, links are heterogeneous, and coherence time is finite. The construction and analyses of these models is relatively simple compared to the DTMC model of this section. Even if one were to introduce a finite buffer into this model, several changes would be required to state transitions and balance equations, resulting in even more complex expressions for the stationary distribution (recall that even in the infinite-buffer case, we must solve for it numerically). Attempting to model decoherence in discrete time would require one to consider all possible combinatoric settings of stored qubit decoherence, further complicating the transition probabilities, but also increasing the \emph{number} of possible transitions from each state. Consider, for instance, state $i$ in Figure \ref{fig:DTMC}: each of the existing ``backward'' transitions $P_{(j)}$, $j\in\{1,\dots,k-1\}$ would have to be modified based on the number of ways that $l$ qubits can decohere and $m$ new entanglements can be generated such that $l+m=j$, and in addition, extra transitions must be added from state $i$ to states $\{0,1,\dots,i-k+2\}$ because any number of the stored qubits can decohere. This process is highly cumbersome and prone to mistakes, quickly outweighing the advantages of using DTMCs. On the other hand, by using CTMCs we gain much in modeling power and lose little in accuracy.
%\section{CTMC for Tripartite Measurements}
%\input{CTMCTripartite}
\section{CTMC for $n$-Partite Entanglements}
%%!TEX PS-program = pdflatexmk
%%!TEX root = quantum_jrnl.tex
\label{sec:nPartite}
Consider a switch that exclusively serves $n$-partite entanglements to its users, \emph{i.e.}, whenever there are enough available link entanglements, the switch performs $n$-way GHZ measurements according to the OLEF policy. We assume that at any time, any group of $n$ users wishes to ``communicate'', or share an entangled state. In this work, we consider the simplest variant of this system, in which all links are homogeneous, the switch has infinite buffer, and there is no decoherence. Note that in such a system, at most $n-1$ links can have stored entanglements at any given time. Let the vector $\vect{i}=(i_1~i_2~\dots~i_{n-1})$ represent the number of stored entanglements at the links, where some entries may equal 0 (in cases where fewer than $n-1$ links have stored entanglements). We construct a CTMC with state space $S\coloneqq \{(i_1~i_2~\dots~i_{n-1}): i_1\geq 0, \dots,i_{n-1}\geq0\}$, which can be partitioned as follows:
\begin{align*}
S = S_0\cup S_1\cup \cdots\cup S_{n-1},
\end{align*}
where $S_j$ contains the set of states $\vect{i}$ wherein $j$ entries are 0 and $n-1-j$ entries are $\geq 1$. We will borrow some of the notation from Section \ref{sec:CTMCBipartite} to describe the non-zero transitions for this CTMC. First, from any state $\vect{i}\in S_0$, in which $n-1$ distinct links have at least one entanglement each, an $n$-partite entanglement will be generated when the $n$th link generates one; this occurs with rate $(k-(n-1))\mu$. Now, consider a state $\vect{i}\in S_j$, for $j\in\{1,\dots,n-1\}$. Note that due to the $j$ zero entries, some of these states are ``equivalent'' in the sense that their steady-state probabilities are equal. For instance, for $j=n-2$, states $\vect{e}_l$, $l=1,\dots,n-1$, are all equally likely in steady state. To account for this symmetry, whenever a link with no stored entanglements generates one, we divide the rate of transition by $j$. Hence, for example, the chain transitions out of state $\vect{0}$ with rate $k\mu$ into one of states $\vect{e}_l$, $l\in\{1,\dots,n-1\}$, but since there are $n-1$ such states, each individual transition has rate $k\mu/(n-1)$. In general, for a state with $j$ zero entries in $\vect{i}$, note that there are a total of $k-(n-1-j)$ links with no stored entanglements. Hence, when one of these links generates one, the rate is given by $(k-(n-1-j))\mu/j$.
Finally, for any state in $S\setminus S_{n-1}$, a link with an already-stored entanglement may generate another one; this occurs with rate $\mu$.

Now, let $Q_t^l$ be the number of stored qubits at the switch for link $l$ at time $t$, and consider the DTMC $X \coloneqq \{\vect{Q}_t\coloneqq (Q_t^1~Q_t^2~\cdots~Q_t^{n-1}),t=1,2,\dots\}$ that results from the uniformization of this CTMC. The non-zero transition probabilities for the DTMC are as follows:
\begin{align*}
&\text{for } \vect{i}\in S_0\to \begin{cases}
    \vect{i}-\vect{1},& \hspace{-0.6em}\text{with prob. } \frac{k-(n-1)}{k},\\
    \vect{i}+\vect{e}_l,& \hspace{-0.6em}\text{with prob. } \frac{1}{k},~l=1,\dots,n-1,
\end{cases}\\
&\text{for } \vect{i}\in S_j\to\begin{cases}
	\vect{i}+\vect{e}_l, & \hspace{-0.6em}\text{with prob. }\frac{k-(n-1-j)}{kj},\text{for }l\text{ s.t. } i_l=0,\\
	\vect{i}+\vect{e}_l, & \hspace{-0.6em}\text{with prob. }\frac{1}{k},\text{for }l \text{ s.t. } i_l \neq 0,
\end{cases}\\
&\text{for }j=1,\dots,n-2,\text{ and}\\
&\text{for }\vect{0}\to \vect{e}_l \text{ with prob. } \frac{1}{n-1}, ~l=1,\dots,n-1.
\end{align*}

%Define $\vect{Q}\coloneqq \lim_{t\to\infty}\vect{Q}_t$, so that $|\vect{Q}|$ is the number of qubits stored at the switch in steady state.
Note that  $X$ is irreducible and aperiodic. 
When in addition $X$ is positive recurrent (\emph{i.e.} stable), we denote by $\vect{Q}$ the stationary version of the vector $\vect{Q}_t$.
\begin{prop}
If $X$ is stable and if in steady state ${E[|\vect{Q}|]<\infty}$ then
\begin{align}
C = \frac{q\mu k}{n}.
\label{eq:Cnpart}
\end{align}
Further, if $X$ is stable, $k>n$, and if in steady state ${E[|\vect{Q}|^2]<\infty}$ then
\begin{align}
E[|\vect{Q}|] = \frac{(n-1)k}{2(k-n)}.
\label{eq:EQnpart}
\end{align}
%A necessary condition for stability is $k>n-1$.
\end{prop}
See proof of Proposition 1 in Appendix \ref{app:npartite}. We conjecture that $X$ is stable for $k>n$. We prove this conjecture for $n=3$ in Appendix \ref{app:tripartiteStab}.

The result in Eq. (\ref{eq:Cnpart}) can be interpreted as follows: in a system with $k$ links, link-level entanglements are generated at rate $k\mu$. For each $n$-partite entanglement, $n$ link-level entanglements are consumed, hence, we divide $k\mu$ by $n$, and since measurements succeed with probability $q$, multiply by $q$ to obtain the number of successfully-generated $n$-partite entanglements per time unit. The quantity $q\mu k/n$ is an upper bound on the capacity of the switch, and we are able to achieve it as a consequence of the OLEF policy and the assumption that any $n$ users wish to share an entangled state at any time, which allows the switch to line up link-level entanglements in any configuration. For this reason, it follows that OLEF is an optimal policy. The same argument can be used in systems with heterogeneous links and systems with decoherence to prove that the resulting capacity is a tight upper bound.
\section{Numerical Observations}
%%!TEX PS-program = pdflatexmk
%%!TEX root = quantum_jrnl.tex
\label{sec:numerObs}
In this section, we investigate the capacity and buffer requirements of a bipartite entanglement switch
based on our model. 
In particular, we are interested in how the buffer capacity $B$ and number of users $k$ affect
% convergence rate of %a finite-buffer system's 
capacity and $E[Q]$.
% as a function of $B$ to their infinite-buffer versions, as well as how these quantities behave as a function of $k$.
%observe asymptotic results as both $B$ and $k$ are varied. 
We then examine the effect of decoherence on homogeneous and heterogeneous switches with infinite buffer capacity.

Throughout this section, we denote the distance of user $l$ from the switch as $L_l$ (measured in km). We assume that each user is connected to the switch with single mode optical fiber of loss coefficient $\beta = 0.2$ dB/km.
We also assume that the switch is equipped with a photonic entanglement source with a raw (local) entanglement generation rate of 1 Giga-ebits\footnote{An ebit is one unit of bipartite entanglement corresponding to the state of two maximally entangled qubits, the so-called Bell or EPR state.} per second. So, in every (1 ns long) time slot, one photon of a Bell state is loaded into a memory local to the switch, and the other photon is transmitted (over a lossy optical fiber) to a user, who loads the received photon into a memory (held by the user), which has a trigger which lets the user know the time slots in which their memory successfully loads a photon. 
Let us denote $\tau = 1$ ns as the time duration of one qubit of each entangled pair, and the entanglement generation rate between the switch and the user $l$, $\mu_l = c \eta_l / \tau$ ebits per second. Here, we take ${c = 0.1}$ to account for various losses other than the transmission loss in fiber, for example inefficiencies in loading the entangled photon pair in the two memories (at the switch and at the user), and any inefficiency in a detector in the memory at the user used for heralding the arrival of a photon (\emph{e.g.}, by doing a Bell measurement over the received photon pulse and one photon of a locally-generated two-photon entangled state produced by the user). Here, $\eta_l$, the transmissivity of the optical fiber connecting user $l$ and the switch is given by $\eta_l = 10^{-0.1\beta L_l}$. Channel loss to user $l$, measured in dB, is $10\log_{10}(1/\eta_l)$.
%the rate of entanglement generation on link $l$ is given by $\mu_l=c\eta_l/\tau$, where $\eta_l$ is the transmissivity of fiber for link $l$, $\tau$ is the amount of time it takes to attempt an entanglement, and $c$ is a constant that accounts of inefficiencies on the link. We let $c=0.1$ and $\tau=10^{-9}$ seconds. $\eta_l \coloneqq e^{-\beta L_l}$, where $L_l$ is the length of link $l$ and $\beta$ its fiber loss coefficient. For all links, we assume that {$\beta=0.2$ dB/km}. 
Unless otherwise stated, all $\mu_l$ discussed in this section have units of Mega-ebits/sec.
% Let us assume $L_1 = L_2 = 22.8$, $L_3 = L_4 = 41.2$ and $L_5 = 76.1$.   With the above assumptions, we get the entanglement generation rates for the $5$ users as: $\mu_1 = \mu_2 = 35$ Mebits/sec, $\mu_3 = \mu_4 = 15$ Mebits/sec and $\mu_5 = 3$ Mebits/sec. We use these rates for the plots shown in
\subsection{Effect of Buffer Size: Homogeneous Links}
\begin{figure}
\centering
\subfloat{\includegraphics[width=0.235\textwidth]{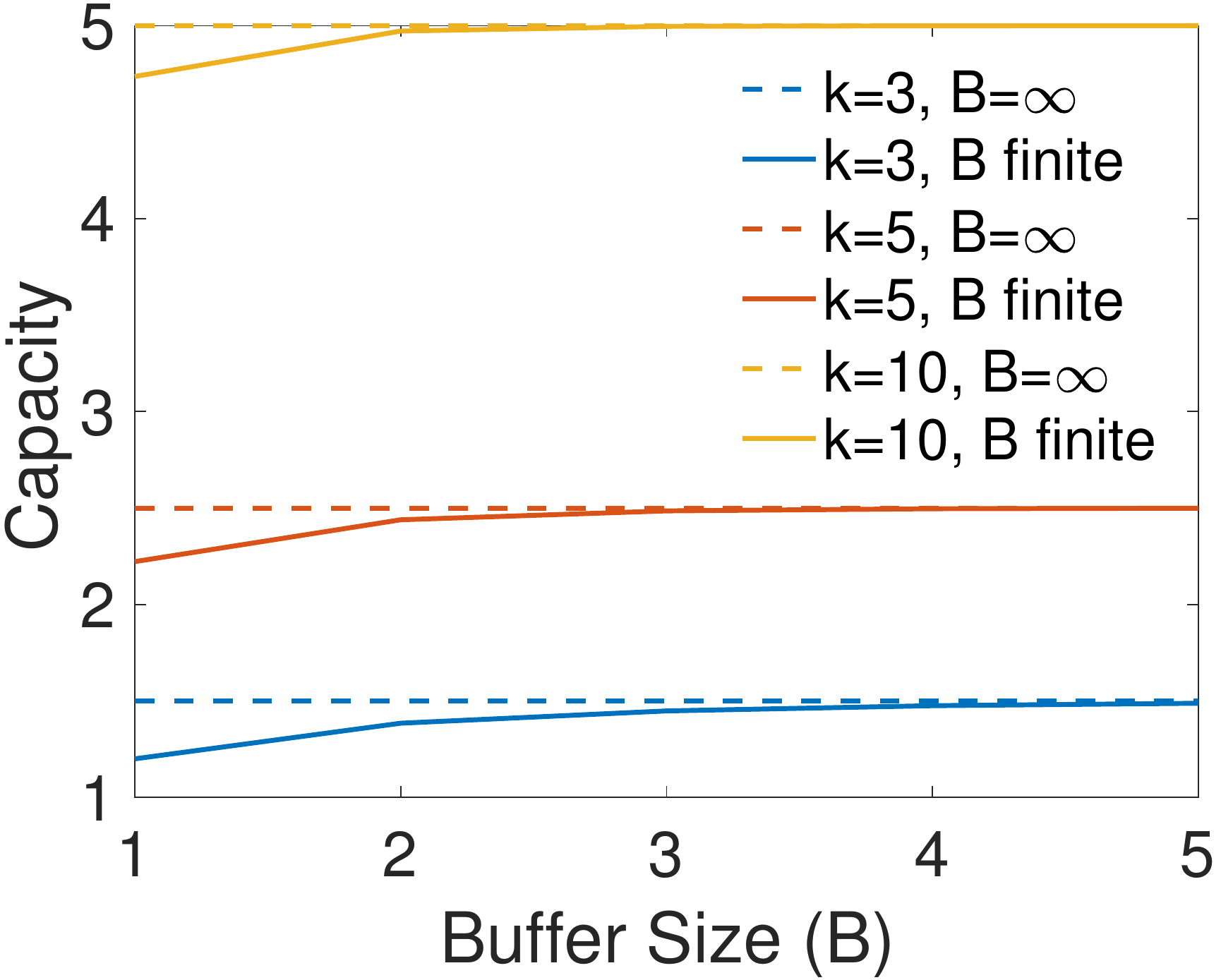}\label{fig:CplotHomog}}
\subfloat{\includegraphics[width=0.245\textwidth]{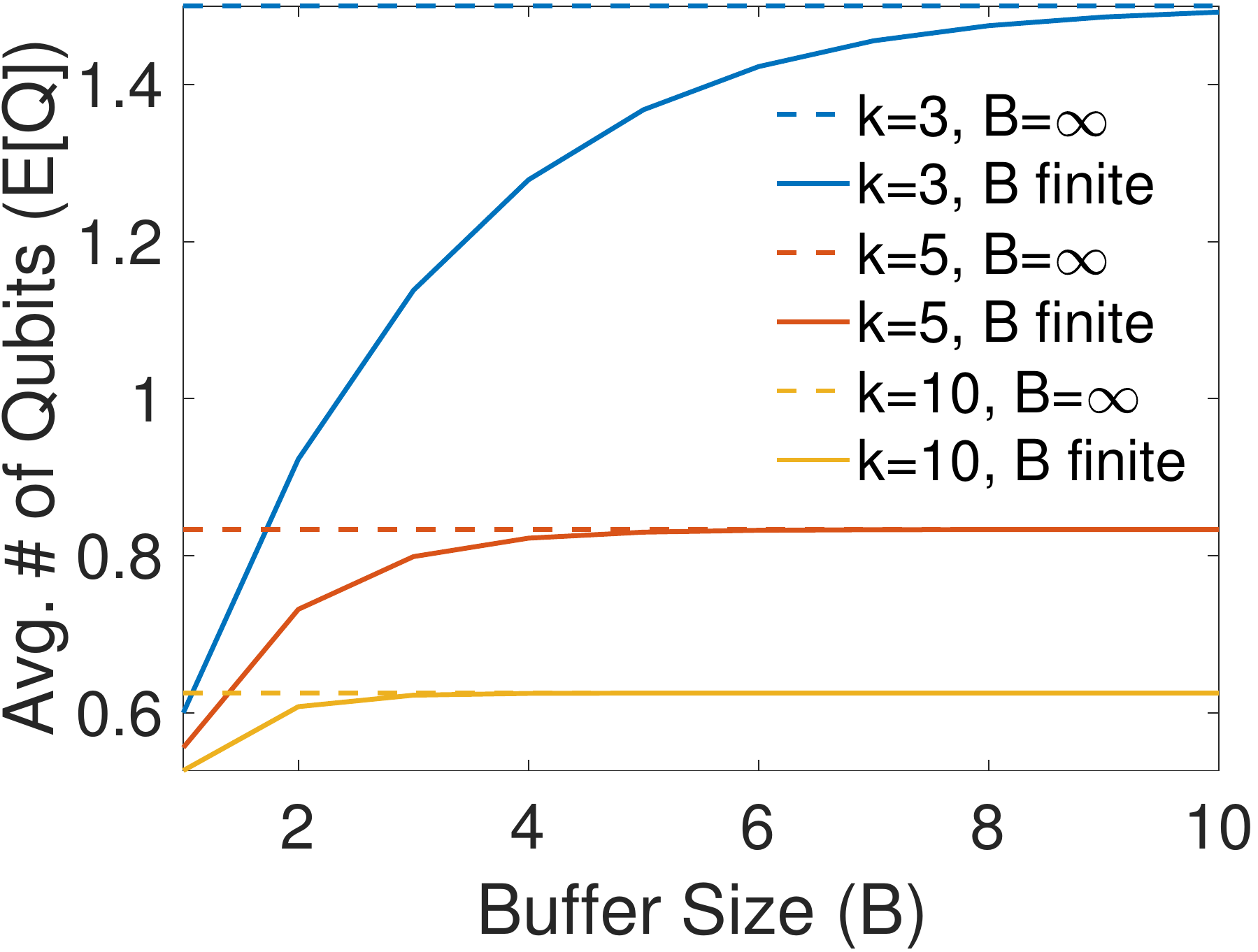}\label{fig:EQplotHomog}}
\caption{The effect of buffer size on capacity (left) and on the expected number of stored entanglements (right) in systems with homogeneous links. Capacity is in Mega-ebits/sec.}
\label{fig:plotsHomog}
\end{figure}

In homogeneous-link systems, all users are equidistant from the switch (\emph{i.e.} $L_l=L_m$, $\forall l,m\in\{1,\dots,k\}$).
In Figure~\ref{fig:plotsHomog}, we compare models with infinite and finite buffer sizes as the number of links $k$ is varied.
Note that when links are homogeneous, $q\mu$ is simply a multiplicative factor in the expressions for $C$, and does not factor into  formulas for $E[Q]$. Hence, we set $q\mu=1$ for Figure \ref{fig:plotsHomog} (left), and with $\mu=1$, the links are 100 km long.
For the finite buffer models, $B$ is varied from one to five. Recall from Section \ref{sec:ctmcFinBufHomog} that as $B\to\infty$, the capacity of the finite-buffer model approaches that of the infinite-buffer model, as expected, and note that the same is true when $k\to\infty$. Interestingly, this convergence seems to occur quite rapidly, even for the smallest value of $k$ (3), and the maximum relative difference between the two capacities is $0.25$ (even as $\mu$ increases). From this, we conclude that buffer does not play a major role in capacity for homogeneous systems under the OLEF policy and only a small quantum memory is required.
%\begin{figure}[h]
%\centering
%\includegraphics[width=0.4\textwidth]{EQplotHomog.pdf}
%\caption{The effect of buffer size on the expected number of stored entanglements in a system with homogeneous links.}
%\label{fig:EQplotHomog}
%\end{figure}

Figure \ref{fig:plotsHomog} (right) shows the behavior of $E[Q]$ for infinite and finite buffer sizes and different values of $k$. As with capacity, the effect of buffer capacity on $E[Q]$ diminishes as $k$ grows, and largest relative difference occurs for $k=3$ and $B=1$, and equals $1.5$ -- less than two qubits. Note from the expressions for $E[Q]$ in Sections \ref{sec:ctmcInfBufHomog} and \ref{sec:ctmcFinBufHomog} that as $k\to\infty$, $E[Q]\to1/2$. Numerically, we observe that convergence to this value occurs quickly: even for $k=25$, $E[Q]$ is already $0.54$ for both the infinite and finite models.
\subsection{Effect of Buffer Size: Heterogeneous Links}\label{sec:bufSizeHeterog}
\begin{figure}
\centering
\subfloat{\includegraphics[width=0.23\textwidth]{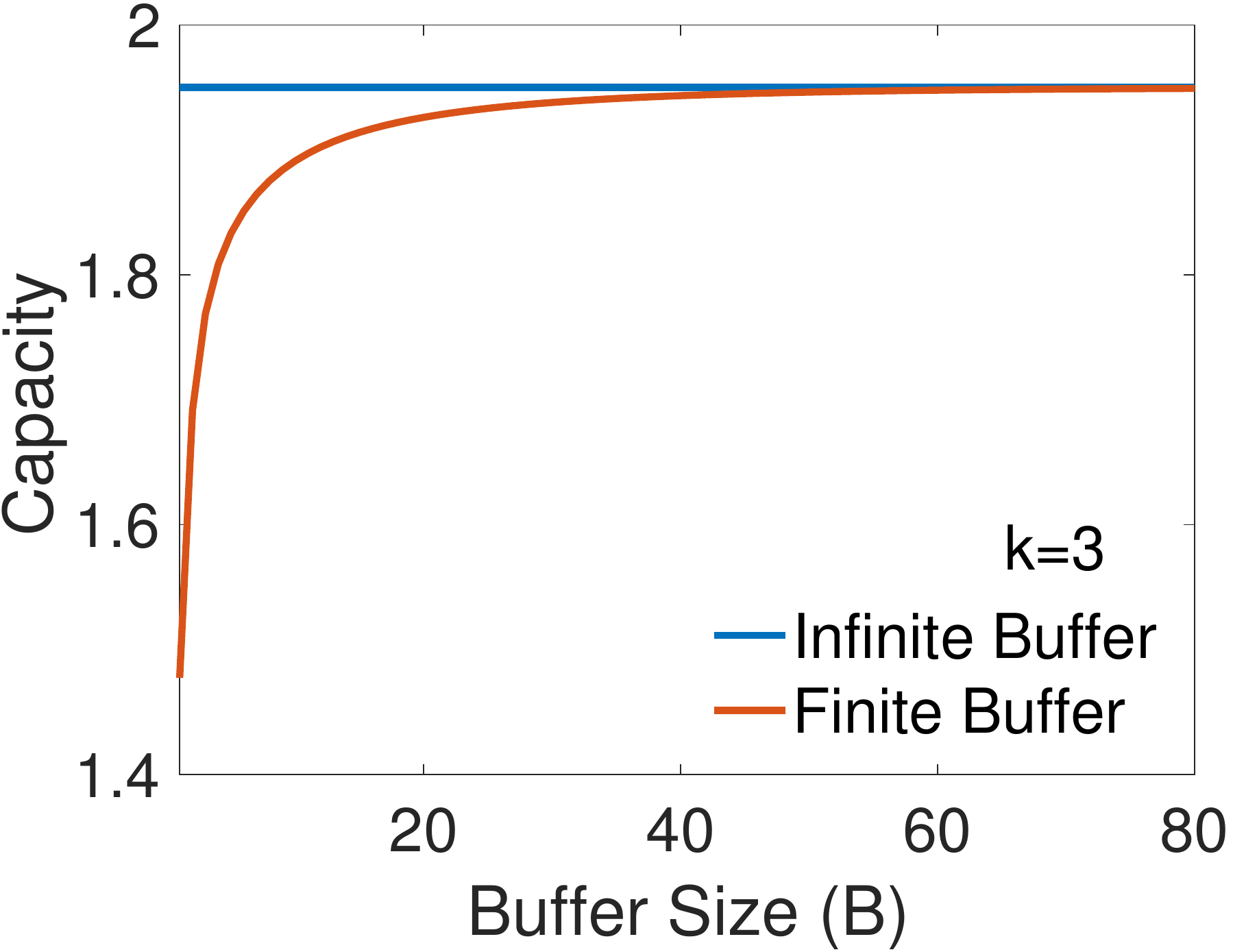}}
\subfloat{\includegraphics[width=0.23\textwidth]{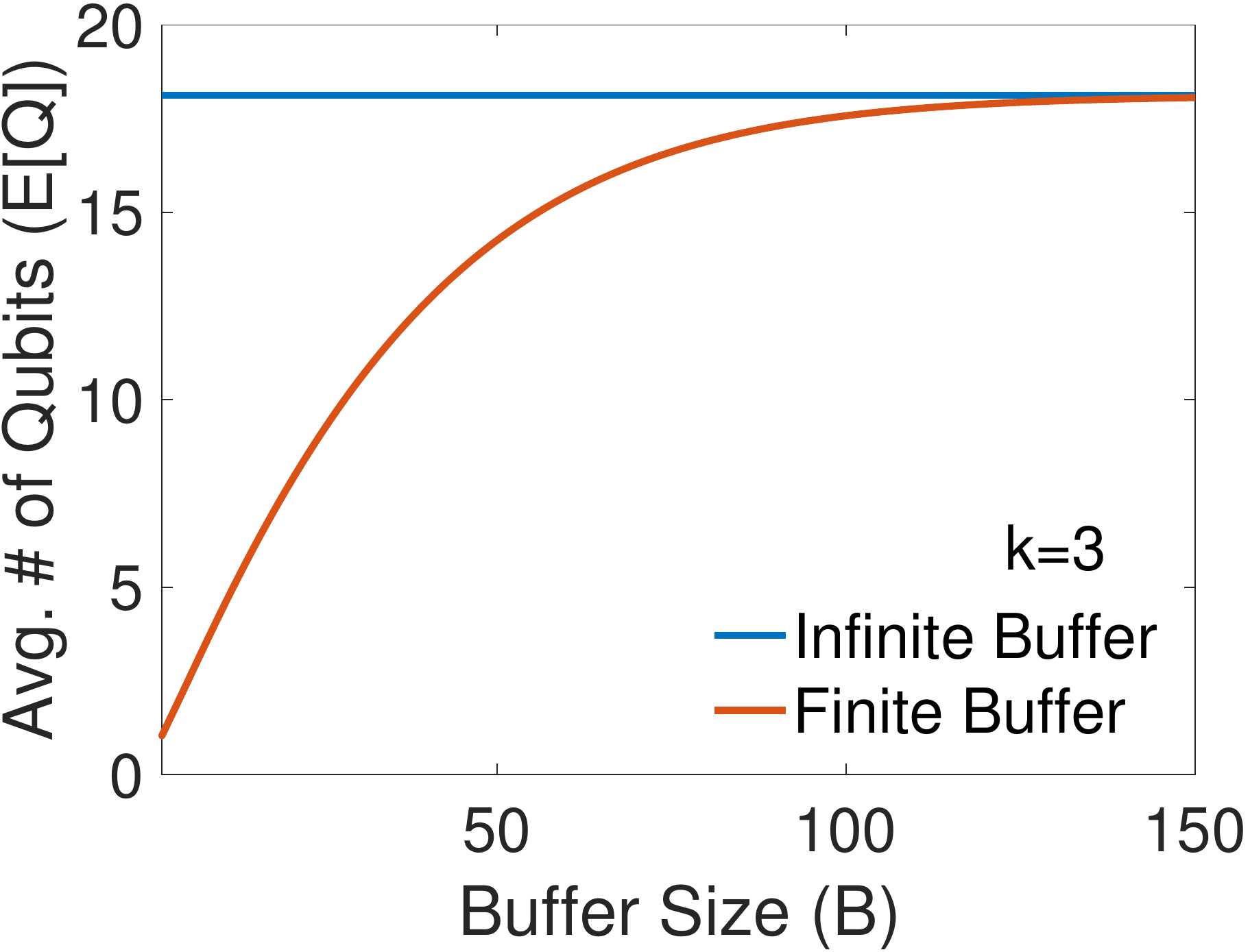}}
\qquad
\subfloat{\includegraphics[width=0.23\textwidth]{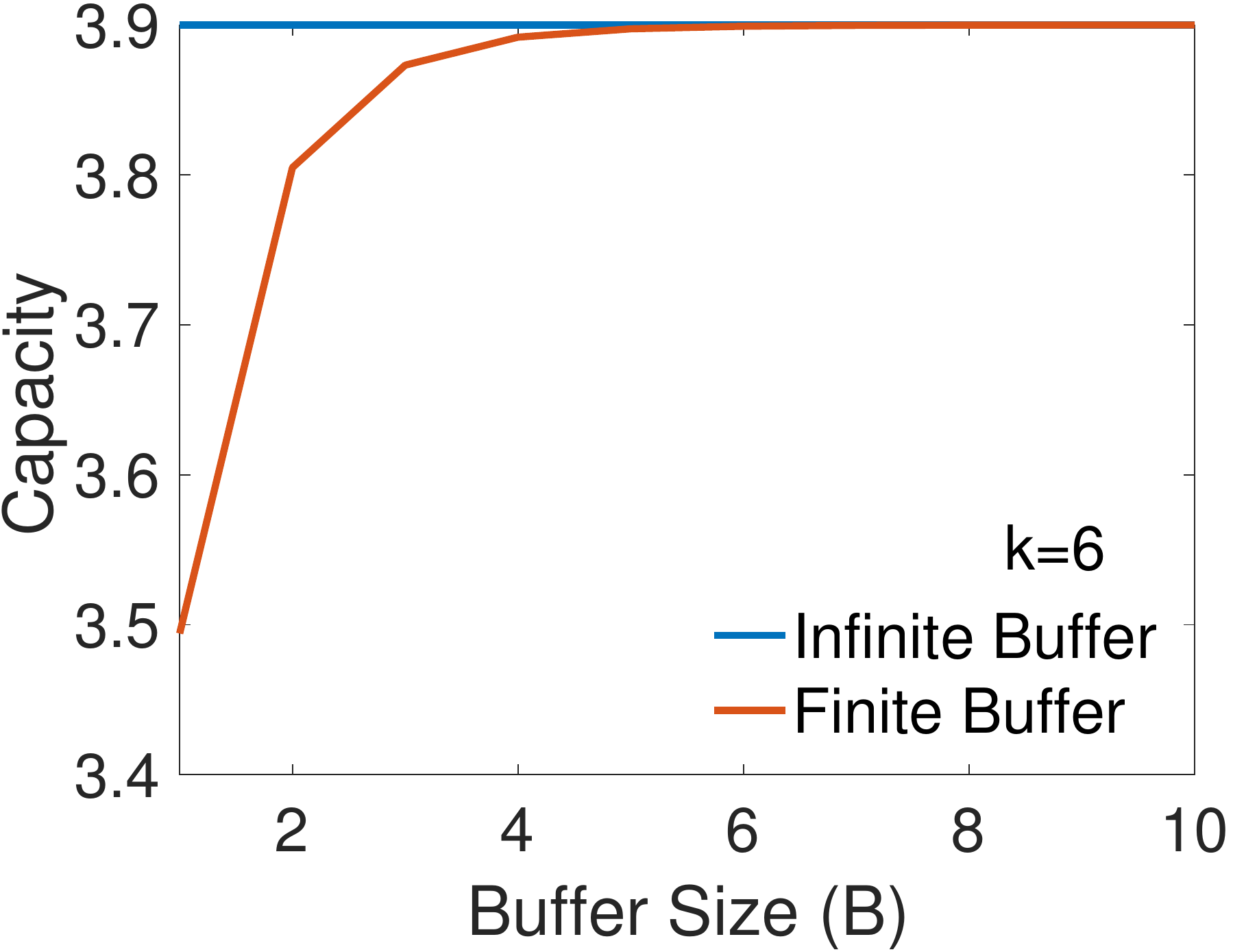}}
\subfloat{\includegraphics[width=0.23\textwidth]{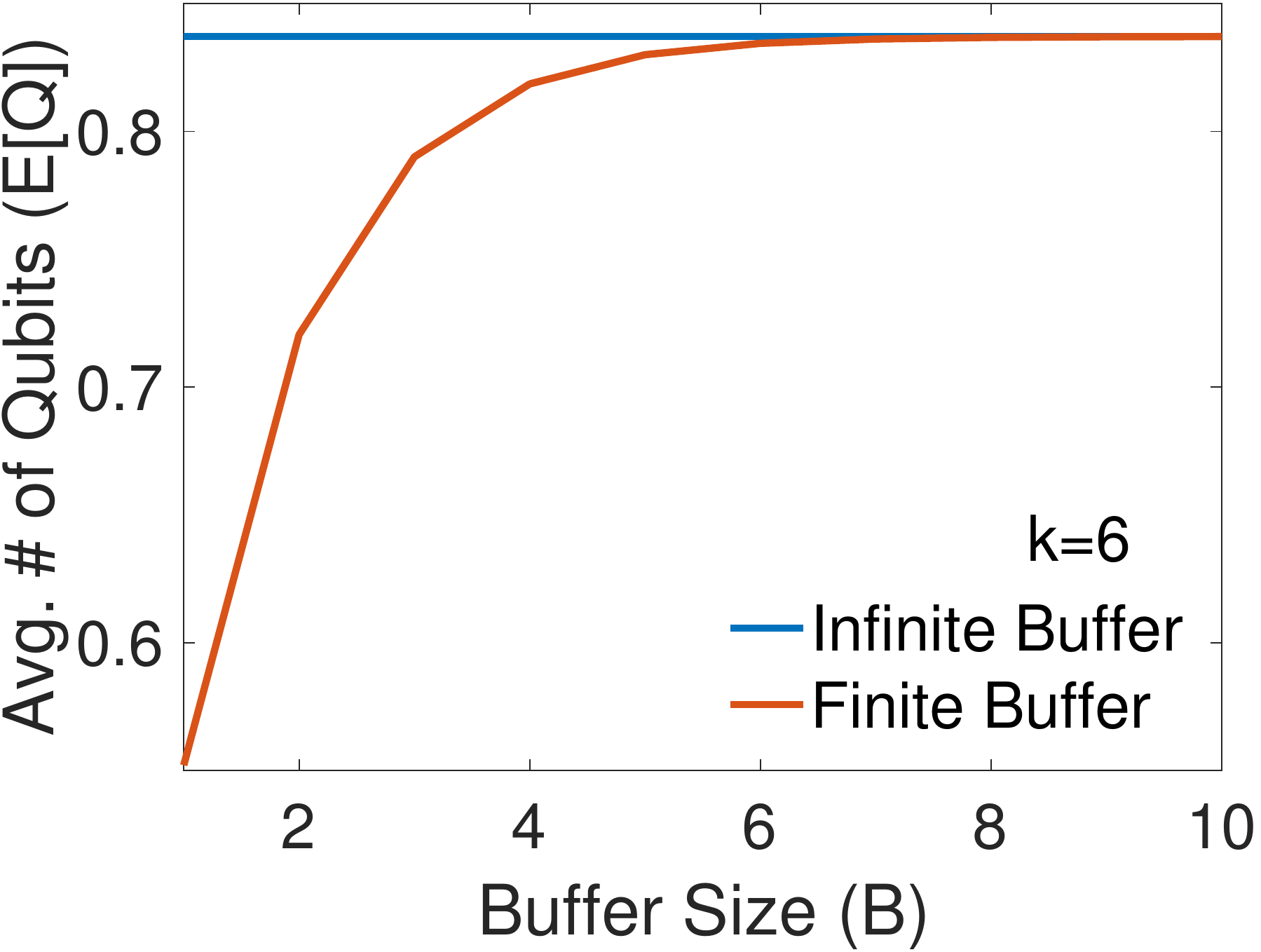}}
\qquad
\subfloat{\includegraphics[width=0.23\textwidth]{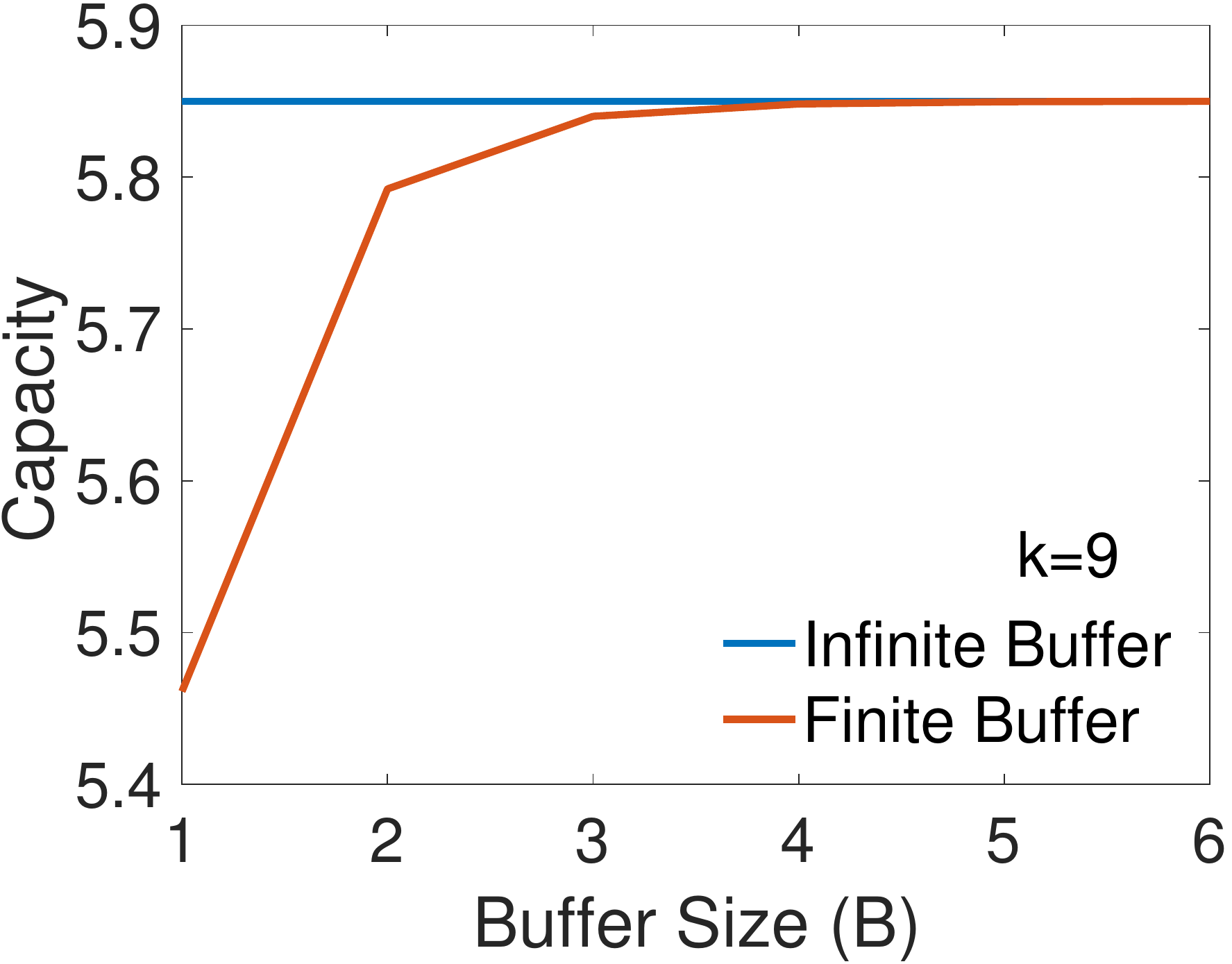}}
\subfloat{\includegraphics[width=0.23\textwidth]{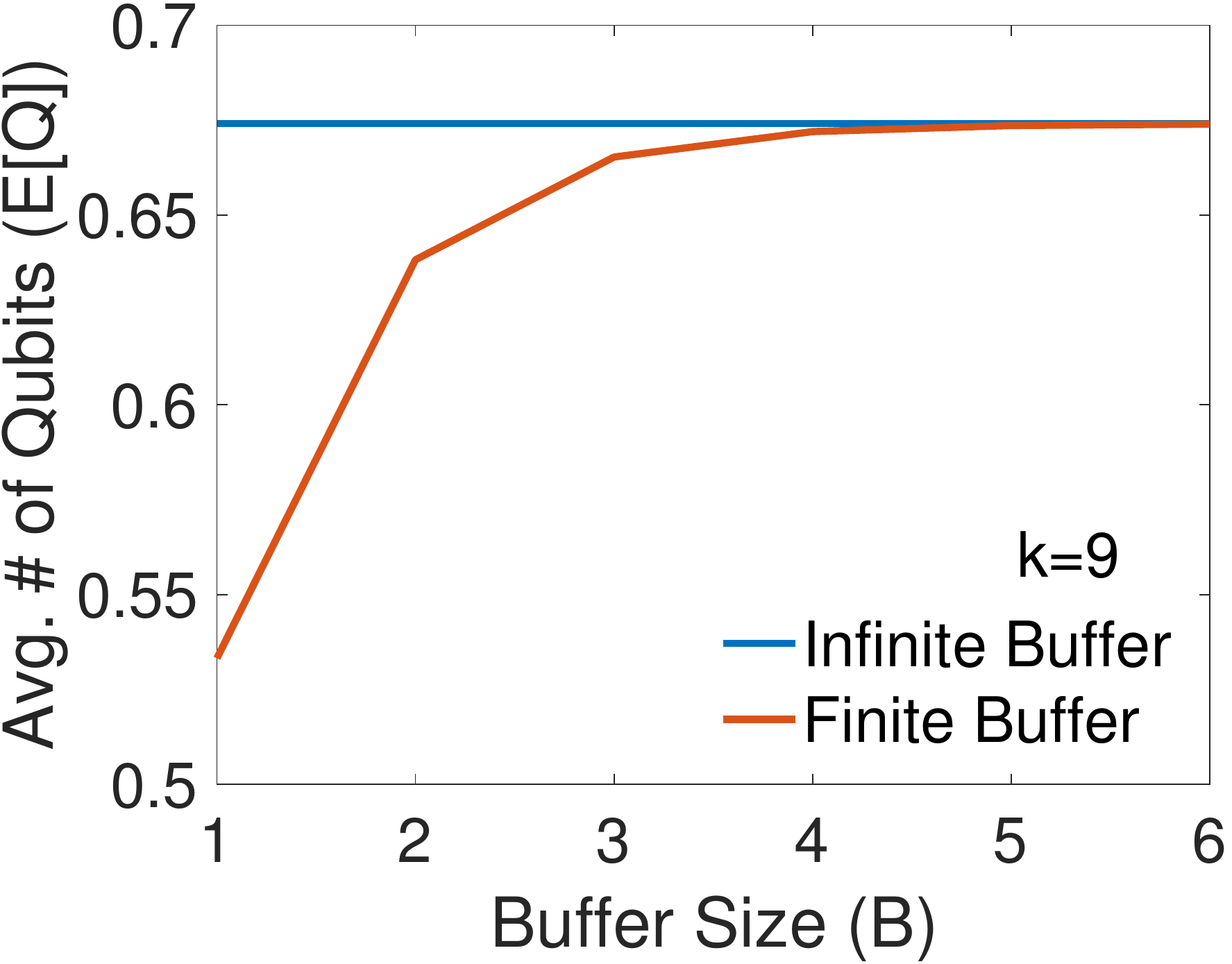}}
\caption{Capacity (Mega-ebits/sec) and expected number of qubits in memory $E[Q]$ for heterogeneous systems with varied number of links and buffer sizes. Links are divided into two classes: one class generates entanglements approximately twice as quickly as the other class.}
\label{fig:heterogCandEQ}
\end{figure}

Figure \ref{fig:heterogCandEQ} illustrates how buffer size and number of users affect $C$ and $E[Q]$ for a set of heterogeneous systems. We vary the number of links from three to nine. For each value of $k$, the links are split into two classes: links in the first class successfully generate entanglements at rate $\mu_1$ and those in the second class at rate $\mu_2$. We set $\mu_1=1.9\mu_2$ and $\mu_2=1$. This setting corresponds to links in class one having lengths $86$ km and links in class two having lengths $100$ km.
Values of $\mu_1$ and $\mu_2$
%a faster rate than the second class (approximately twice as fast). The values of $\mu_l$ 
are chosen in a manner that satisfies the stability condition for heterogeneous systems: recall from Section \ref{sec:infBufHeterog} that for all $l\in\{1,\dots,k\}$, $\mu_l$ must be strictly less than half the aggregate entanglement generation rate. For all experiments, $q=1$ since it only scales capacity.
%For $k=3$, the entanglement generation rate for one of the links is $\mu_f=0.95$ and $\mu_s=0.5$ for the other two links. For $k=5$, two of the links have rates $\mu_f$ and the others $\mu_s$, and for $k=10$, half of the links have rates $\mu_f$ and the other half have rates $\mu_s$. 

For each value of $k$, the ratio of class 1 to class 2 links is $1{:}2$ (so $k=3,6,9$ have one, two, and three class 1 links, respectively).
As with the homogeneous-link systems, we observe that the slowest convergence is for smaller values of $k$ and the largest relative difference is for smaller values of $B$.  However, the rate of convergence speeds up quickly as $k$ increases from $3$ to $6$: with the latter, convergence is already observed for $B<10$. Meanwhile, when $k=9$, there is little benefit in having storage for more than two qubits. Another interesting observation is that quantum memory usage is large when $k=3$ but not for larger values of $k$. This is due to the system operating closer to the stability constraints for $k=3$ than larger values of $k$. In the next section, we will see another example of a system that operates close to its stability constraints. In such cases, $C$ and $E[Q]$ can be affected significantly as $B$ is varied.
%This is likely due to the relatively large entanglement rate of the class one link, whose generated entanglements need to be stored while waiting for the other two links to generate entanglements to consume them. }
\subsection{Effect of Decoherence}
\begin{figure}
\centering
\includegraphics[width=0.25\textwidth]{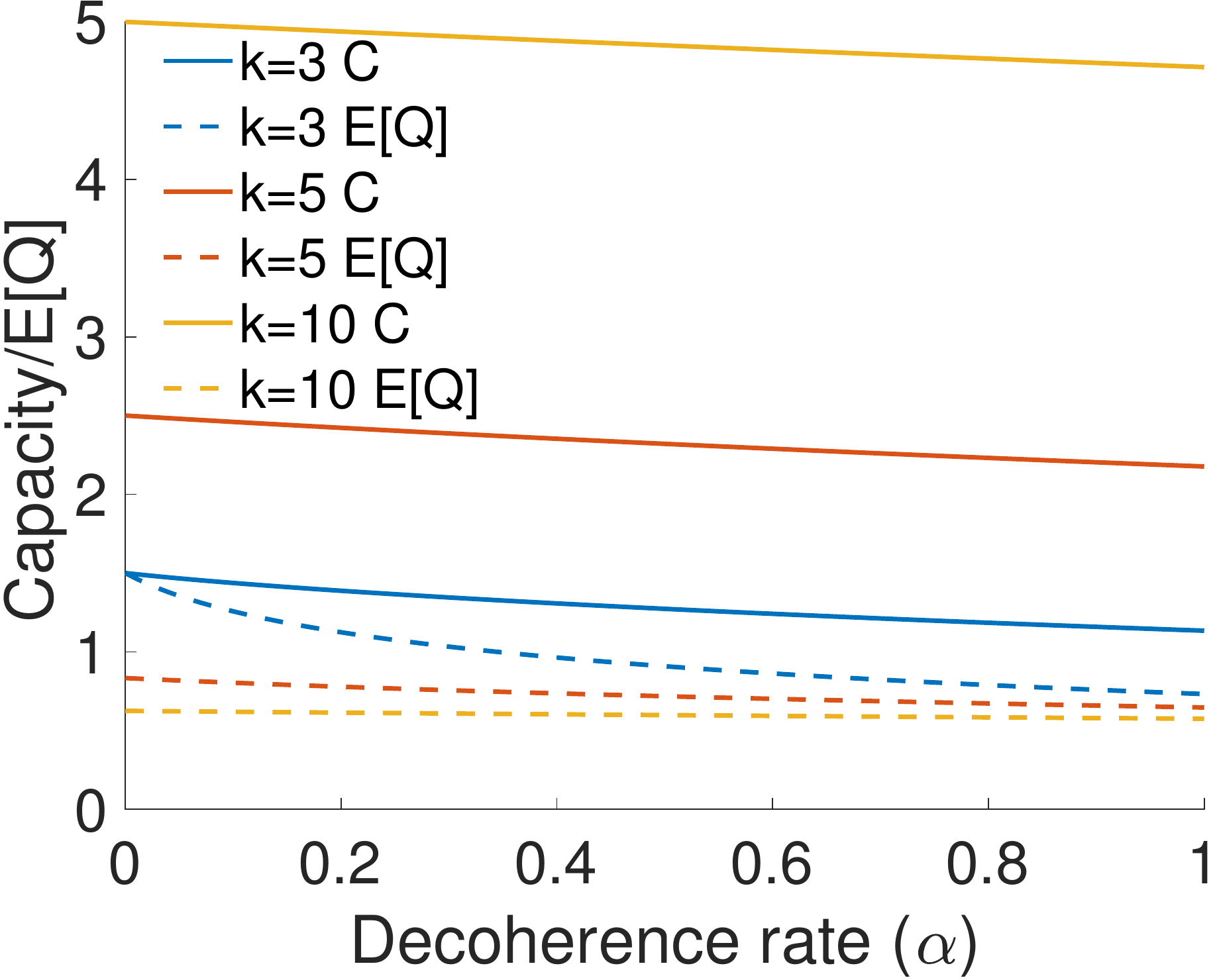}
\caption{Effect of decoherence on capacity (Mega-ebits/sec) and expected number of stored qubits $E[Q]$, for varying number of users $k$. For all experiments, the entanglement generation rate is $\mu=1$ for all links.}
\label{fig:decohHomog}
\end{figure}

In this section, we study the effect of decoherence on capacity and expected number of stored qubits $E[Q]$. 
%In previous sections, we discussed the effect of buffer size on $C$ and $E[Q]$ and determined that in general, buffer capacity does not significantly affect the performance metrics.
%infinite-buffer models serve as reasonable approximations to finite-buffer systems.
%Hence, in this section we focus only on models with infinite buffer sizes. 
We set $q=1$ for all experiments since it only scales capacity. Figure~\ref{fig:decohHomog} presents $C$ and $E[Q]$ for homogeneous systems with $\mu=1$ (corresponding to 100 km long links), $B=\infty$ and different values of $k$, as  decoherence rate $\alpha$ varies from 0 (the equivalent of previous models that did not incorporate decoherence) to $\mu=1$. Note that in practice, $\alpha$ is expected to be much smaller than $\mu$. We observe that even as $\alpha$ approaches $\mu$ decoherence does not cause major degradation in capacity for homogeneous systems, and likewise does not introduce drastic variations in $E[Q]$.

\begin{figure}
\centering
\subfloat{\includegraphics[width=0.235\textwidth]{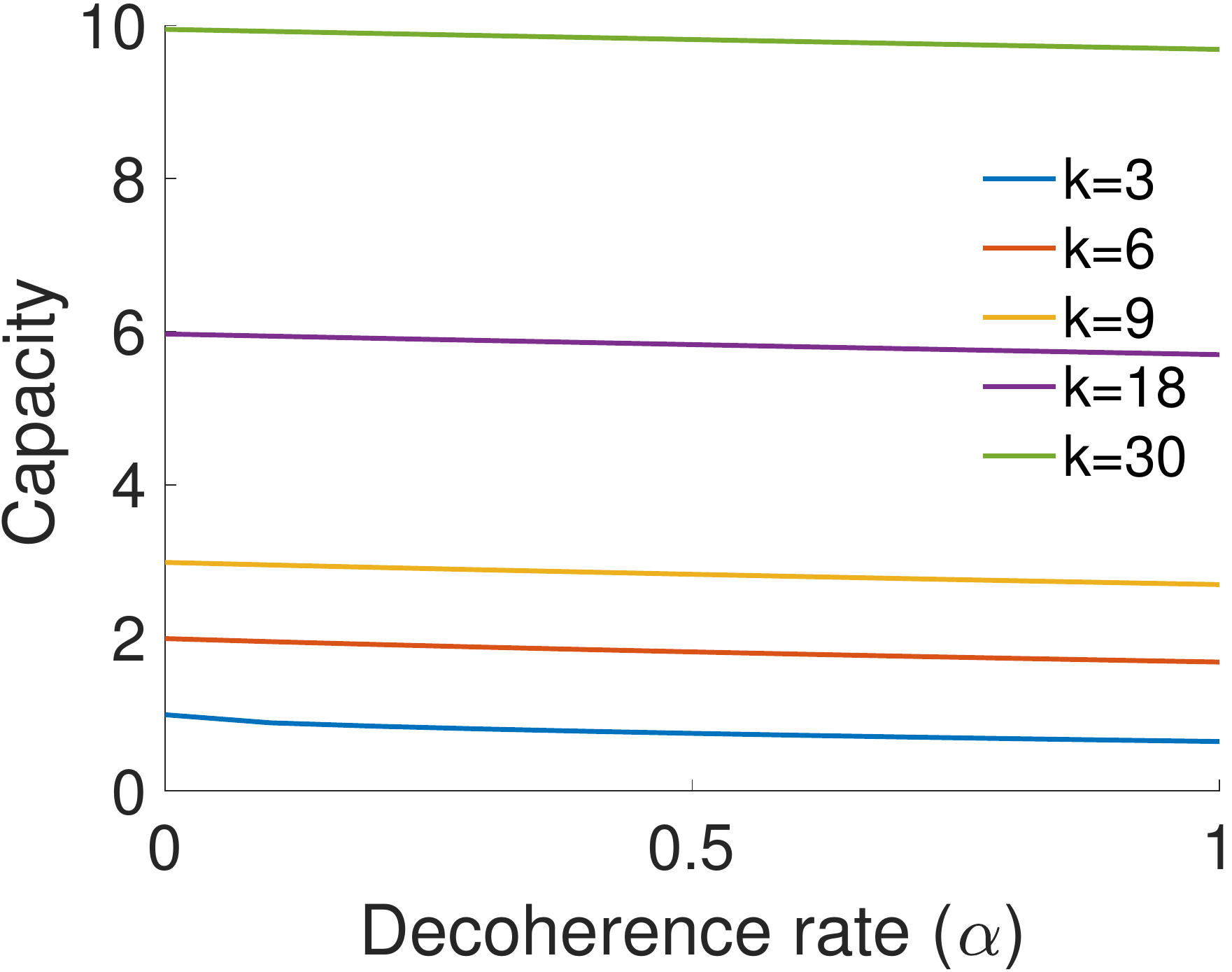}}
\subfloat{\includegraphics[width=0.235\textwidth]{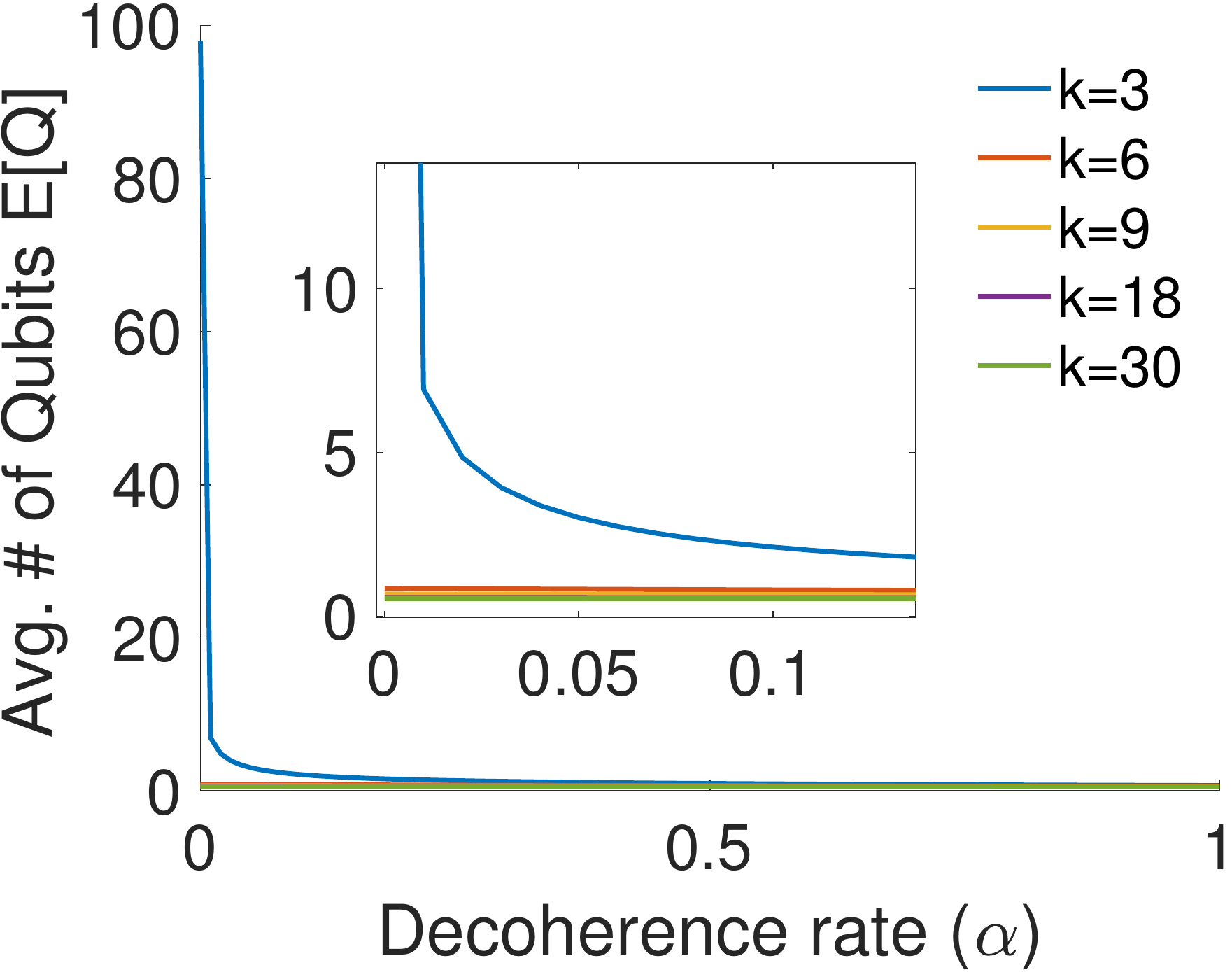}}
\caption{Effect of decoherence on capacity (Mega-ebits/sec) and expected number of stored qubits $E[Q]$, for varying number of users $k$. In all experiments, the links are heterogeneous.}
\label{fig:decohHeterog}
\end{figure}

Figure \ref{fig:decohHeterog} presents the effects of $\alpha$ in a set of heterogeneous systems with infinite buffer. In these experiments, entanglement generation rates are set in a similar manner to that of Section \ref{sec:bufSizeHeterog}, with two classes of links configured so that the first class generates entanglements almost twice as fast as the second class (here, $\mu_1=0.99$ and $\mu_2=0.5$, corresponding to 100.2 km and 115 km long links for class one and two, respectively), and the number of links in class one to those in class two is $1{:}2$. In these experiments, for each value of $k$, capacity behaves much as it would in a homogeneous system with $\mu$ set as the average of the $\mu_l$ from the heterogeneous system. Note that for $k=3$, $E[Q]$ is very large when $\alpha=0$; similar to the experiment in Figure \ref{fig:heterogCandEQ} this is because the system is operating close to the stability constraints. In all other cases, $E[Q]$ is close to 0.

\begin{figure}
\centering
\subfloat{\includegraphics[width=0.235\textwidth]{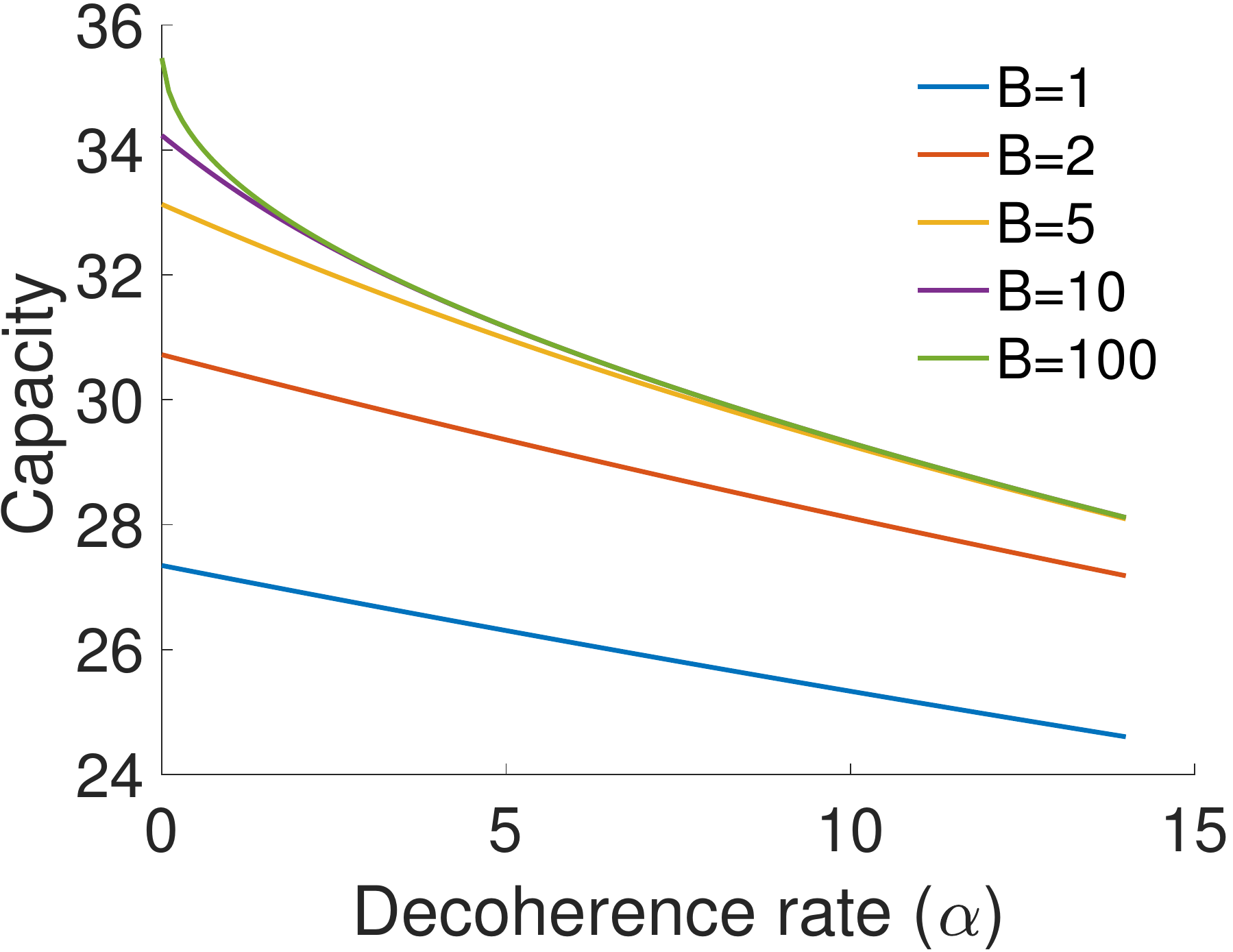}}
\subfloat{\includegraphics[width=0.235\textwidth]{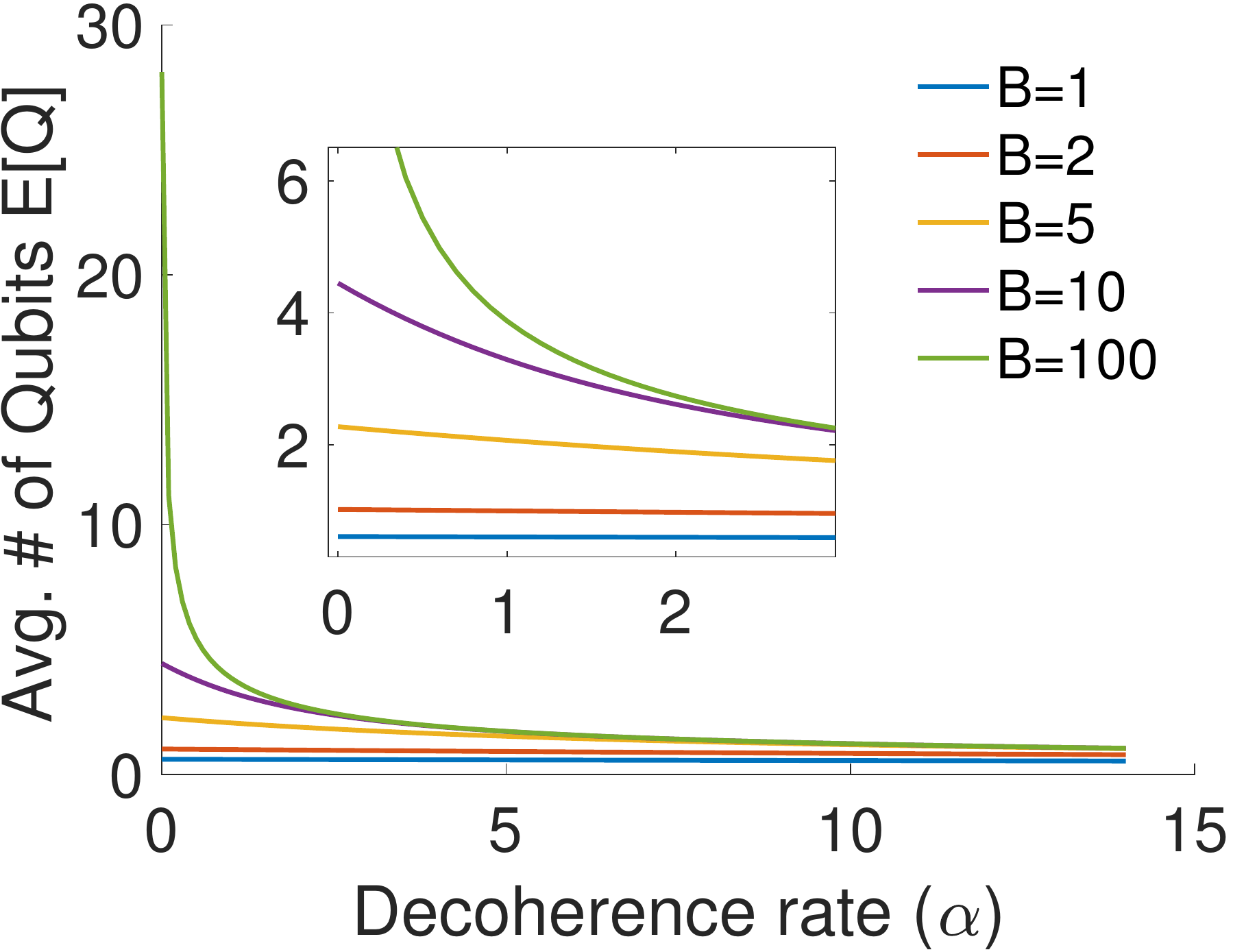}}
\qquad
{\footnotesize (a) Heterogeneous-link system}
\subfloat{\includegraphics[width=0.235\textwidth]{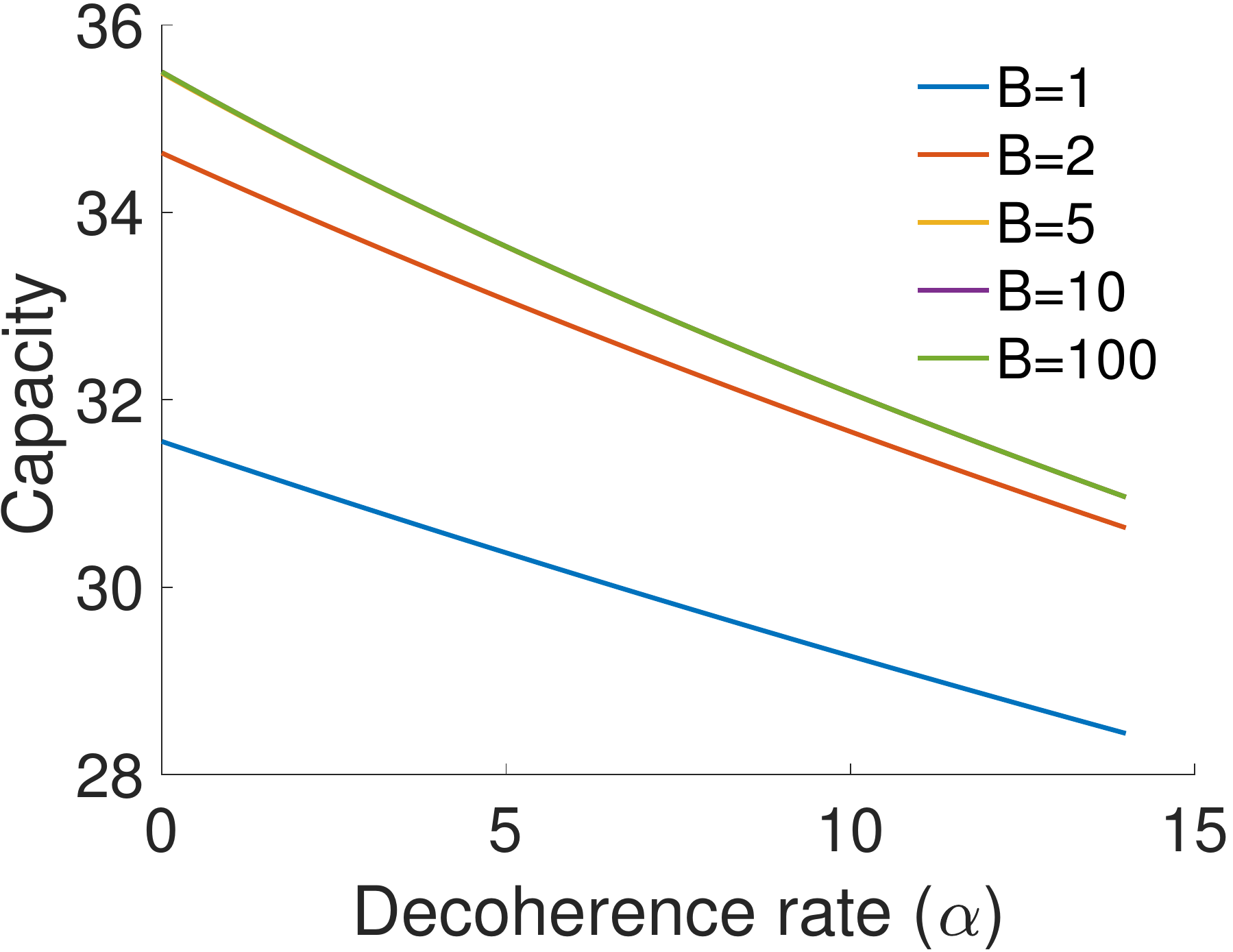}}
\subfloat{\includegraphics[width=0.235\textwidth]{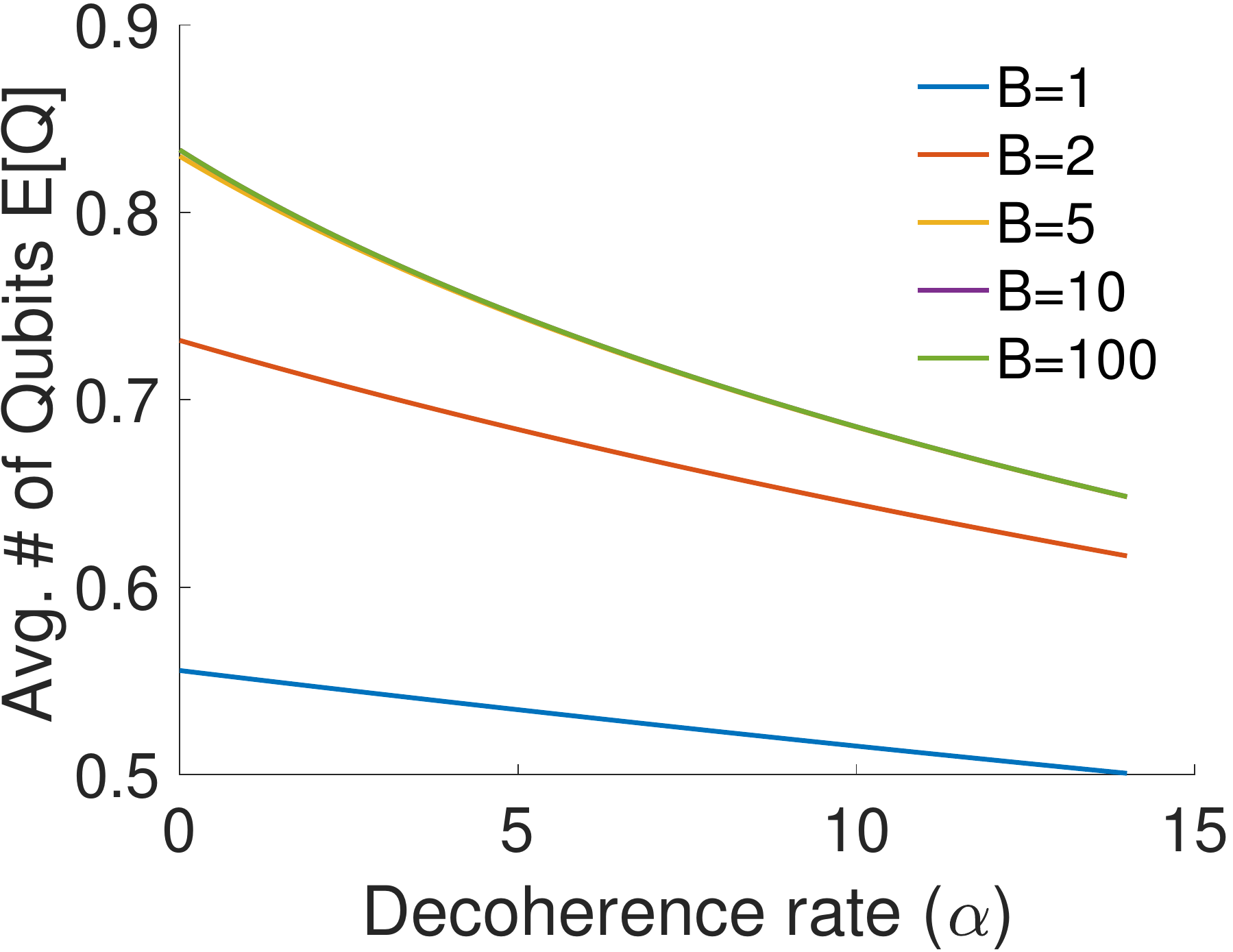}}
\qquad
{\footnotesize (b) Homogeneous-link system}
\caption{Effect of decoherence on capacity (Mega-ebits/sec) and expected number of stored qubits $E[Q]$ for $k=5$ links and varying buffer sizes $B$. In (a), $\mu_l$ are $(35~ 15~ 15~ 3~ 3)$, and in (b), $\mu$ is the average of $\mu_l$, $l=1,\dots,5$, \emph{i.e.} 14.2. For all plots above, $B=100$ curves behave equivalently to $B=\infty$.}
\label{fig:decohHeterogCloseToStab}
\end{figure}

In Figure \ref{fig:decohHeterogCloseToStab}(a), we focus on a heterogeneous system that operates close to the stability constraints and observe the effects of both decoherence and buffer size on $C$ and $E[Q]$. There are five links, with entanglement generation rates $(35~ 15~ 15~ 3~ 3)$ Mega-ebits/sec, corresponding to link lengths of $22.8$, $41.2$, $41.2$, $76$, and $76$ km, respectively. For this system, $\gamma/2=35.5$, so the fastest link is just below the constraint when $\alpha=0$. The average of the $\mu_l$ is 14.2, so $\alpha$ is varied from 0 to this value. $B$ is varied from 1 to 100, with the latter being close enough to mimic infinite buffer behavior for $C$ and $E[Q]$. Figure \ref{fig:decohHeterogCloseToStab}(b) presents a homogeneous system with $k=5$ and $\mu=14.2$ for a comparison. We observe that the homogeneous system achieves better capacity for all values of $B$, even though the average entanglement generation rate is the same for both systems. Further, the homogeneous system is more robust to changes in buffer size than the heterogeneous system: for the former, $B=5,10$ are equivalent to $B=100$. Further, note that for $B=100$ and $\alpha=0$ the heterogeneous system performs almost as well as the homogeneous system in terms of capacity, but the memory usage is much higher for the former.
Finally, for this buffer size, as $\alpha$ increases, the homogeneous system is more robust to the effects of decoherence: capacity degrades by 7.35 Mega-ebits/sec for the heterogeneous system between $\alpha=0$ and $\alpha=14$, while it degrades by $4.54$ Mega-ebits/sec for the homogeneous system.

%These experiments demonstrate that the effect of decoherence in a heterogeneous system can be quite pronounced, especially for larger values of $k$. For instance, in the system with 10 users, capacity degrades from almost 44 BSMs per time unit (when $\alpha=0$) to only 7 BSMs per time unit (when $\alpha=1$). Moreover, the decline in capacity when $k=5,10$ for this experiment occurs rapidly for $\alpha\in(0,0.1)$. Similar behavior is seen for $E[Q]$ when $k=3$. On the other hand, $\alpha$ makes little difference to $C$ when $k=3$ or to $E[Q]$ when $k=5,10$.
%\section{A Conjecture on General Infinite Buffer Homogeneous Link Systems}
%\input{Conjecture}
\section{Conclusion}
%%!TEX PS-program = pdflatexmk
%%!TEX root = quantum_jrnl_short.tex
\label{sec:concl}
In this work, we examine variants of a system with $k$ users who are being served by a quantum entanglement switch in a star topology. Each user is connected to the switch via a dedicated link; we consider both the case of homogeneous and heterogeneous links. We also analyze cases in which the switch has finite or infinite buffer space for storing entangled qubits. For all variants of the problem, we focus on a specific scheduling policy, where the oldest entanglements are given priority over more recently-generated ones. We obtain simple and intuitive expressions for switch capacity, as well as for the expected number of qubits in memory. 

In the case of bipartite entanglement switching, we make numerical comparisons of these two metrics while varying the number of users $k$ and buffer sizes $B$. We observe that in most cases, little memory is required to achieve the performance of an infinite-memory system.
%the finite-buffer systems converge to their infinite-buffer counterparts quite quickly as $k$ and $B$ are increased. 
We also make numerical observations for models that incorporate decoherence and conclude that the degradation of quantum states in homogeneous systems has little effect on performance metrics, while it can have more significant consequences in heterogeneous systems that operate close to their stability constraints.
We also obtain expressions for capacity and expected number of stored qubits for a homogeneous system with infinite buffer and no decoherence. The results hold when the system is stable; we conjecture that this occurs whenever $k>n$ and prove the conjecture in the case of $n=3$. Finally, we argue that the entanglement switching policy presented in this work achieves a tight upper bound on the switch capacity.

% or


\appendices
\section{DTMC Derivations}
\subsection{Stationary Distribution}
%%!TEX PS-program = pdflatexmk
%%!TEX root = quantum_jrnl.tex
\subsubsection{Proof of Eq. (\ref{eq:dtmcPoly})}\label{app:dtmcPolyproof}
Introducing the value of $\pi_i=\beta^{k-1}\pi_1$ into Eq. (\ref{eq:ibal}) yields
\begin{align*}
\beta^{i-1}\pi_1= \beta^{i-2}\pi_1 P_f  + \beta^{i-1}\pi_1 P_s + \beta^{i-1} \pi_1 \sum_{j=1}^{k-1} \beta^j P_{(j)}
\end{align*}
or equivalently
\begin{align}
\beta &= P_f +\beta P_s + \beta\sum_{j=1}^{k-1} \beta^j P_{(j)}\nonumber\\
&=  \beta(\pbar^k +(k-1)p^2 \pbar^{k-2})
+ \beta\sum_{j=1}^{k-1}{k-1 \choose j} (\beta p)^j \pbar^{k-j}\nonumber\\
&+ \frac{1}{\beta}\sum_{j=1}^{k-2}{k-1\choose j+1}(\beta p)^{j+2}\pbar^{k-j-2}+p\pbar^{k-1}. \label{eq50}
\end{align}
\begin{flalign*}
%\sum_{j=1}^{k-1} {k-1 \choose j} (\beta p)^j \pbar^{k-j} = \pbar \sum_{j=1}^{k-1}{k-1 \choose j} (\beta p)^j \pbar^{k-1-j}= \pbar\left((\beta p+\pbar)^{k-1}- \pbar^{k-1}\right)
\text{With~}\sum_{j=1}^{k-1} {k-1 \choose j} (\beta p)^j \pbar^{k-j} =  \pbar\left((\beta p+\pbar)^{k-1}- \pbar^{k-1}\right)\text{~and}&&
\end{flalign*}
\vspace{-1em}
\begin{flalign*}
&\sum_{j=1}^{k-2}{k-1\choose j+1}(\beta p)^{j+2}\pbar^{k-j-2}
= \sum_{i=2}^{k-1} {k-1\choose i}(\beta p)^{i+1} \pbar^{k-1-i}&&\\
%&= \beta p  \sum_{i=2}^{k-1} {k-1\choose i}(\beta p)^{i} \pbar^{k-1-i}
&= \beta p\left( (\beta p+\pbar)^{k-1} - (k-1) \beta p \pbar^{k-2} - \pbar^{k-1}\right),
\end{flalign*}
Eq. (\ref{eq50}) becomes
\begin{align*}
\beta &=  \beta(\pbar^k +(k-1)p^2 \pbar^{k-2})  + \beta\pbar\left((\beta p+\pbar)^{k-1}- \pbar^{k-1}\right)\\
&+  p\left( (\beta p+\pbar)^{k-1} - (k-1) \beta p \pbar^{k-2} - \pbar^{k-1}\right)+p\pbar^{k-1}\\
&=  p\pbar^{k-1} + \beta\pbar^k  + \beta (k-1)p^2 \pbar^{k-2} +\beta\pbar (\beta p+\pbar)^{k-1} \\
&- \beta \pbar^{k}+ p (\beta p+\pbar)^{k-1} - \beta (k-1)p^2 \pbar^{k-2} -p \pbar^{k-1}\\
&=(\beta p+\pbar)^{k-1}(p+\beta\pbar).
\end{align*}
Hence, $\beta$ satisfies the equation $f(\beta)=0$ with 
\[
\pushQED{\qed}
f(\beta)\coloneqq (\beta p+\pbar)^{k-1}(p+\beta\pbar)-\beta.
\qedhere
\popQED
\]
\subsubsection{Proof that Eq. (\ref{eq:dtmcPoly}) has a unique solution in $(0,1)$}\label{app:dtmcBetaProof}
We have
\[
f'(\beta)=(k-1)p(\beta p+\pbar)^{k-2}(p+\beta\pbar) +\pbar (\beta p+\pbar)^{k-1}-1
\]
and $f''(\beta)$ is given by
\begin{align*}
%f''(\beta)&=
&(k-1)(k-2)p^2 (\beta p+\pbar)^{k-3}(p+\beta\pbar) + 2 (k-1)p\pbar (\beta p+\pbar)^{k-2}\\
&=(k-1)p (\beta p+\pbar)^{k-3}\left[ (k-2)p + 2(k-1)\pbar(\beta p+\pbar)\right]>0.
\end{align*}
This shows that the mapping $\beta \to f '(\beta)$ is strictly increasing in $[0,1]$. On the other hand,
\[
f'(0)=(k-1)p^2 \pbar ^{k-2} + \pbar^k-1
\]
and $f'(1)=  (k-1)p + \pbar -1=(k-2)p>0$. Let us show that $f'(0)<0$.
Define $g(p)=(k-1)p^2 \pbar ^{k-2} + \pbar^k-1 = f'(0)$. We find
\[
g'(p)= -\pbar^{k-3} (p^2k^2 +2p(1-2k)+k).
\]
Define $h(p)=p^2k^2 +2p(1-2k)+k$ so that $g'(p)=-\pbar^{k-3} h(p)$.
We have $h'(p)=2(pk^2+1-2k)$, which vanishes for $p=p_0\coloneqq(2k-1)/k^2$. Also, $h''(p)=2k^2>0$. We deduce from this  that $h(p)$ decreases in $[0,p_0)$ and increases in $(p_0,1]$.
Therefore, $h(p)$ is minimized in $[0,1]$ for $p =p_0$.
We have $h(p_0)=(-(2k-1)^2 +k^3)/k^2$ which is easily seen to be strictly positive for all $k\geq 3$. This shows that  $h(p)>0$ for $p\in [0,1]$, which implies that
$g'(p)<0$ for $p\in [0,1]$, so that $g(p)<g(0)=0$ for $p\in (0,1]$ and, finally, $f'(0)<0$. 

From  $f'(0)<0$, $f'(1)>0$ and the fact that the continuous mapping $\beta \to f'(\beta)$ is strictly increasing in $[0,1]$, we deduce that there exists $\beta_0\in (0,1)$ such that $f'(\beta)<0$ for $\beta\in[0,\beta_0)$, $f'(\beta_0)=0$  and $f'(\beta)>0$ for $\beta\in (\beta_0,1]$. This in turn shows that $\beta \to f(\beta)$ is strictly decreasing in $[0,\beta_0)$ and strictly increasing in $(\beta_0,1]$.  But since
$f(0)>0$ and $f(1)=0$, this implies that $f$ has a unique zero in $(0,1)$. This zero is actually located in $(0,\beta_0)$.\hfill $\qed$
\subsubsection{Equivalence of Eqs (\ref{eq:a}) and (\ref{eq:b})}
\label{app:dtmcEquiv}
%%!TEX PS-program = pdflatexmk
%%!TEX root = quantum_jrnl.tex
We start by rearranging (\ref{eq:a}):
\begin{align*}
\sum\limits_{i=0}^{k-1}\pi_iP_{i,0} &= \pi_0,\\
\sum\limits_{i=1}^{k-1}\pi_iP_{i,0} &= \pi_0(1-P_{0,0}),\\
\pi_1\sum\limits_{i=1}^{k-1}\beta^{i-1}P_{i,0} &= \pi_0P_{0,1},\\
\sum\limits_{i=1}^{k-1}\beta^{i}P_{i,0} &= \frac{\beta\pi_0}{\pi_1}P_{0,1}.
\end{align*}
Then, we rearrange (\ref{eq:b}) in a similar fashion:
\begin{align*}
\sum\limits_{i=0}^k\pi_iP_{i,1} &= \pi_1,\\
\pi_0P_{0,1}+\pi_1\sum\limits_{i=1}^k\beta^{i-1}P_{i,1} &= \pi_1,\\
\pi_0P_{0,1} &= \pi_1\left(1-\frac{1}{\beta}\sum\limits_{i=1}^k\beta^{i}P_{i,1}\right),\\
\frac{\beta\pi_0}{\pi_1}P_{0,1} &= \beta-\sum\limits_{i=1}^k\beta^{i}P_{i,1}.
\end{align*}
Hence, to show that one of (\ref{eq:a}) and (\ref{eq:b}) is redundant, it suffices to show that
\begin{align}
\sum\limits_{i=1}^{k-1}\beta^{i}P_{i,0}  &= \beta-\sum\limits_{i=1}^k\beta^{i}P_{i,1},
\label{eq:dtmcWTS}
\end{align}
or equivalently,
\begin{align}
\sum\limits_{i=1}^{k-1}\beta^{i}(P_{i,0}+P_{i,1})+\beta^kP_{k,1}  &= \beta.
\label{eq:redproof}
\end{align}
Before we continue, we derive a few useful expressions. The first is as follows:
\begin{align*}
P_e(i,k-1)+P_o(i,k-1) 
%&= \sum\limits_{j=i}^{k-1}\left(\frac{1+(-1)^j}{2}\right){k-1 \choose j} p^j\pbar^{k-1-j}+\sum\limits_{j=i}^{k-1}\left(\frac{1-(-1)^j}{2}\right){k-1 \choose j} p^j\pbar^{k-1-j}\\
&=\sum\limits_{j=i}^{k-1}{k-1 \choose j} p^j\pbar^{k-1-j}.
\end{align*}
Next, we have
\begin{align*}
P_e(i,k-1)-P_o(i,k-1) 
%&= \sum\limits_{j=i}^{k-1}\left(\frac{1+(-1)^j}{2}\right){k-1 \choose j} p^j\pbar^{k-1-j}-\sum\limits_{j=i}^{k-1}\left(\frac{1-(-1)^j}{2}\right){k-1 \choose j} p^j\pbar^{k-1-j}\\
&=\sum\limits_{j=i}^{k-1}{k-1 \choose j} p^j\pbar^{k-1-j}(-1)^j.
\end{align*}
Finally,
\begin{align*}
P_o(i,k-1)-P_e(i,k-1) 
%&= \sum\limits_{j=i}^{k-1}\left(\frac{1-(-1)^j}{2}\right){k-1 \choose j} p^j\pbar^{k-1-j}-\sum\limits_{j=i}^{k-1}\left(\frac{1+(-1)^j}{2}\right){k-1 \choose j} p^j\pbar^{k-1-j}\\
&=-\sum\limits_{j=i}^{k-1}{k-1 \choose j} p^j\pbar^{k-1-j}(-1)^j.
\end{align*}
Now, consider the left side of Eq. (\ref{eq:dtmcWTS}): $\sum\limits_{i=1}^{k-1}\beta^iP_{i,0}$ is equal to
%{\footnotesize
\begin{align*}
%&\sum\limits_{i=1}^{k-1}\beta^iP_{i,0} = 
&\sum\limits_{i=1}^{k-1}\beta^i\left[\left(\frac{1+(-1)^i}{2}\right)(\pbar P_e(i,k-1)+pP_o(i+1,k-1))\right.\\
&+\left. \left(\frac{1-(-1)^i}{2}\right)(\pbar P_o(i,k-1)+pP_e(i+1,k-1))\right]\\
%=& \sum\limits_{i=1}^{k-1}\beta^i\Big[\frac{\pbar}{2}(P_e(i,k-1)+P_o(i,k-1)\\
%&+ (-1)^i(P_e(i,k-1)-P_o(i,k-1)))\Big]+\\
%&\sum\limits_{i=1}^{k-2}\beta^i\left[\frac{p}{2}(P_e(i+1,k-1)+P_o(i+1,k-1)+(-1)^i(P_o(i+1,k-1)-P_e(i+1,k-1)))\right]\\\\
%&= \frac{\pbar}{2}\sum\limits_{i=1}^{k-1}\beta^i\left[\sum\limits_{j=i}^{k-1}{k-1 \choose j} p^j\pbar^{k-1-j}+(-1)^i\sum\limits_{j=i}^{k-1}{k-1 \choose j} (-p)^j\pbar^{k-1-j}\right]\\
%&+\frac{p}{2}\sum\limits_{i=1}^{k-2}\beta^i\left[\sum\limits_{j=i+1}^{k-1}{k-1 \choose j} p^j\pbar^{k-1-j}-(-1)^i\sum\limits_{j=i+1}^{k-1}{k-1 \choose j} (-p)^j\pbar^{k-1-j}\right]\\\\
&= \frac{\pbar}{2}\sum\limits_{i=1}^{k-1}\beta^i\left[\sum\limits_{j=i}^{k-1}{k-1 \choose j} p^j\pbar^{k-1-j}(1+(-1)^i(-1)^j)\right]\\
&+\frac{p}{2}\sum\limits_{i=1}^{k-2}\beta^i\left[\sum\limits_{j=i+1}^{k-1}{k-1 \choose j} p^j\pbar^{k-1-j}(1-(-1)^i(-1)^j)\right]\\
%=& \frac{\pbar}{2}\sum\limits_{i=1}^{k-1}\beta^i\left[\sum\limits_{j=i}^{k-1}{k-1 \choose j} p^j\pbar^{k-1-j}(1+(-1)^i(-1)^j)\right]+
%\frac{p}{2}\sum\limits_{i=1}^{k-2}\beta^i\left[\sum\limits_{j=i+1}^{k-1}{k-1 \choose j} p^j\pbar^{k-1-j}(1-(-1)^i(-1)^j)\right]\\\\
&= \frac{\pbar}{2}\sum\limits_{j=1}^{k-1}{k-1 \choose j} p^j\pbar^{k-1-j}\sum\limits_{i=1}^{j}\beta^i(1+(-1)^i(-1)^j)\\
&+\frac{p}{2}\sum\limits_{j=2}^{k-1}{k-1 \choose j} p^j\pbar^{k-1-j}\sum\limits_{i=1}^{j-1}\beta^i(1-(-1)^i(-1)^j)=\\%%%%%
&\frac{1}{2}
\left[\frac{\beta}{1-\beta}
-\frac{2\beta}{1-\beta^2}(p\beta+\pbar)^{k-1}(p+\pbar\beta)
-(\pbar-p)^{k}\frac{\beta}{1+\beta}\right].
\end{align*}
%}
Next, we look at $\sum\limits_{i=1}^{k-1}\beta^i P_{i,1}$, which is equal to 
%{\footnotesize
\begin{align*}
%&\sum\limits_{i=1}^{k-1}\beta^i P_{i,1} = 
&\sum\limits_{i=1}^{k-1}\beta^i \left[\left(\frac{1+(-1)^i}{2}\right)(\pbar P_o(i-1,k-1)+pP_e(i,k-1))\right.\\
&+\left.\left(\frac{1-(-1)^i}{2}\right)(\pbar P_e(i-1,k-1)+pP_o(i,k-1))\right]\\
%&=\frac{\pbar}{2}\sum\limits_{i=1}^{k-1}\beta^i\left[P_o(i-1,k-1)+P_e(i-1,k-1)+(-1)^i(P_o(i-1,k-1)-P_e(i-1,k-1))\right]\\
%&+\frac{p}{2}\sum\limits_{i=1}^{k-1}\beta^i\left[P_o(i,k-1)+P_e(i,k-1)+(-1)^i(P_e(i,k-1)-P_o(i,k-1))\right]\\\\
%&=\frac{\pbar}{2}\sum\limits_{i=1}^{k-1}\beta^i\left[\sum\limits_{j=i-1}^{k-1}{k-1 \choose j} p^j\pbar^{k-1-j}-(-1)^i\sum\limits_{j=i-1}^{k-1}{k-1 \choose j} (-p)^j\pbar^{k-1-j}\right]\\
%&+\frac{p}{2}\sum\limits_{i=1}^{k-1}\beta^i\left[\sum\limits_{j=i}^{k-1}{k-1 \choose j} p^j\pbar^{k-1-j}+(-1)^i\sum\limits_{j=i}^{k-1}{k-1 \choose j} (-p)^j\pbar^{k-1-j}\right]\\\\
&=\frac{\pbar}{2}\sum\limits_{i=1}^{k-1}\beta^i\left[\sum\limits_{j=i-1}^{k-1}{k-1 \choose j} p^j\pbar^{k-1-j}
\left(1-(-1)^i(-1)^j\right)\right]\\
&+\frac{p}{2}\sum\limits_{i=1}^{k-1}\beta^i\left[\sum\limits_{j=i}^{k-1}{k-1 \choose j} p^j\pbar^{k-1-j}
\left(1+(-1)^i(-1)^j\right)\right]\\
&=\beta\pbar((\beta p+\pbar)^{k-1}-(\beta p)^{k-1})\\
&+\frac{1}{2}\left(\frac{\beta}{1-\beta}-\frac{2\beta}{1-\beta^2}(p\beta+\pbar)^{k}+\frac{\beta}{1+\beta}(\pbar-p)^{k}\right).
\end{align*}%}
Summing these two expressions, we obtain $\sum\limits_{i=1}^{k-1}\beta^i(P_{i,0}+P_{i,1})$,
%{\footnotesize
\begin{align*}
%\sum\limits_{i=1}^{k-1}\beta^i(P_{i,0}+P_{i,1}) &=
%\frac{1}{2}\left[\frac{\beta}{1-\beta}-\frac{2\beta}{1-\beta^2}(p\beta+\pbar)^{k-1}(p+\pbar\beta)-(\pbar-p)^{k}\frac{\beta}{1+\beta}\right]\\
%&+\frac{1}{2}\left(\frac{\beta}{1-\beta}-\frac{2\beta}{1-\beta^2}(p\beta+\pbar)^{k}+\frac{\beta}{1+\beta}(\pbar-p)^{k}\right)+\beta\pbar((\beta p+\pbar)^{k-1}-(\beta p)^{k-1})\\\\
%&=\frac{\beta}{1-\beta}-\frac{\beta}{1-\beta^2}(p\beta+\pbar)^{k-1}(p+\beta\pbar+p\beta+\pbar)+\beta\pbar((\beta p+\pbar)^{k-1}-(\beta p)^{k-1})\\\\
%&=\frac{\beta}{1-\beta}-\frac{\beta}{1-\beta^2}(p\beta+\pbar)^{k-1}(1+\beta)+\beta\pbar((\beta p+\pbar)^{k-1}-(\beta p)^{k-1})\\\\
%&=\frac{\beta}{1-\beta}-\frac{\beta}{1-\beta}(p\beta+\pbar)^{k-1}+\beta\pbar((\beta p+\pbar)^{k-1}-(\beta p)^{k-1})\\\\
%&=\frac{\beta}{1-\beta}+(p\beta+\pbar)^{k-1}\left(\beta\pbar-\frac{\beta}{1-\beta}\right)-\beta\pbar(\beta p)^{k-1}\\\\
%&=
\frac{\beta}{1-\beta}-\frac{\beta}{1-\beta}(p\beta+\pbar)^{k-1}(p+\pbar\beta)-\beta\pbar(\beta p)^{k-1}.
\end{align*}%}
Next, we compute
\begin{align*}
P_{k,1} &=
\left(\frac{1+(-1)^k}{2}\right)\pbar P_o(k-1,k-1)\\
&+\left(\frac{1-(-1)^k}{2}\right)\pbar P_e(k-1,k-1)\\
&= \pbar\left(\left(\frac{1+(-1)^k}{2}\right) \left(\frac{1-(-1)^{k-1}}{2}\right)p^{k-1}\right.\\
&+\left.\left(\frac{1-(-1)^k}{2}\right) \left(\frac{1+(-1)^{k-1}}{2}\right)p^{k-1}\right) = \pbar p^{k-1}.
\end{align*}
Finally, the left side of Eq. (\ref{eq:redproof}) becomes
%we add $\beta^kP_{k,1}$ to the above and obtain
%{\footnotesize
\begin{align*}
&\sum\limits_{i=1}^{k-1}\beta^i(P_{i,0}+P_{i,1})+\beta^kP_{k,1} =\\
%&\frac{\beta}{1-\beta}-\frac{\beta}{1-\beta}(p\beta+\pbar)^{k-1}(p+\pbar\beta)-\beta\pbar(\beta p)^{k-1}\\
%&+\beta^k\left(
%\left(\frac{1+(-1)^k}{2}\right)\pbar P_o(k-1,k-1)+\left(\frac{1-(-1)^k}{2}\right)\pbar P_e(k-1,k-1)\right)\\\\
%&=\frac{\beta}{1-\beta}-\frac{\beta}{1-\beta}(p\beta+\pbar)^{k-1}(p+\pbar\beta)-\beta\pbar(\beta p)^{k-1}\\
%&+\beta^k\pbar\left(\left(\frac{1+(-1)^k}{2}\right) \left(\frac{1-(-1)^{k-1}}{2}\right)p^{k-1}
%+\left(\frac{1-(-1)^k}{2}\right) \left(\frac{1+(-1)^{k-1}}{2}\right)p^{k-1}\right)\\\\
&\frac{\beta}{1-\beta}-\frac{\beta}{1-\beta}(p\beta+\pbar)^{k-1}(p+\pbar\beta)-\pbar\beta^k p^{k-1}+\beta^k\pbar p^{k-1}\\
&=\frac{\beta}{1-\beta}-\frac{\beta}{1-\beta}(p\beta+\pbar)^{k-1}(p+\pbar\beta).
\end{align*}%}
Recall from (\ref{eq:redproof}) that the expression above must equal to $\beta$. Using Eq. (\ref{eq:dtmcPoly}), we know that
\begin{align*}
(p\beta+\pbar)^{k-1}(p+\pbar\beta)=\beta,
\end{align*}
and therefore,
\begin{align*}
\pushQED{\qed}
\sum\limits_{i=1}^{k-1}\beta^i(P_{i,0}+P_{i,1})+\beta^kP_{k,1} = \frac{\beta}{1-\beta}-\frac{\beta^2}{1-\beta}
%=\frac{\beta-\beta^2}{1-\beta} 
%= \beta\frac{1-\beta}{1-\beta}
=\beta.
\end{align*}
\qed
%\subsection{Capacity and Number of Qubits Stored}
\subsection{Proof of DTMC Capacity}
%%!TEX PS-program = pdflatexmk
%%!TEX root = quantum_jrnl.tex
%\subsubsection{Proof of DTMC Capacity}
\label{app:dtmcCapEQ}
For simplicity, let us first derive $C$ with the assumption that $q=1$. Since $q$ simply scales the capacity, we will multiply the resulting expression by $q$ at the end. 
Consider the first term of Eq. (\ref{eq:dtmcCap}):
\begin{align*}
&\sum\limits_{i=0}^{k-2}\pi_i \sum\limits_{j=0}^i j{k-1 \choose j} p^j\pbar^{k-1-j}\\
&=\sum\limits_{j=1}^{k-2}j{k-1 \choose j}p^j\pbar^{k-1-j}\sum\limits_{i=j}^{k-2}\pi_i\\
&=\sum\limits_{j=1}^{k-2}j{k-1 \choose j}p^j\pbar^{k-1-j}\pi_1\sum\limits_{i=j}^{k-2}\beta^{i-1}\\
&=\frac{\pi_1}{\beta}\sum\limits_{j=1}^{k-2}j{k-1 \choose j}p^j\pbar^{k-1-j}\left(\sum\limits_{i=0}^{k-2}\beta^{i}-\sum\limits_{i=0}^{j-1}\beta^i\right)\\
&=\frac{\pi_1}{\beta}\sum\limits_{j=1}^{k-2}j{k-1 \choose j}p^j\pbar^{k-1-j}\left(\frac{1-\beta^{k-1}}{1-\beta}-\frac{1-\beta^j}{1-\beta}\right)\\
&=\frac{\pi_1}{\beta}\sum\limits_{j=1}^{k-2}j{k-1 \choose j}p^j\pbar^{k-1-j}\left(\frac{\beta^j-\beta^{k-1}}{1-\beta}\right)\\
&=\frac{\pi_1}{\beta}\sum\limits_{j=1}^{k-1}j{k-1 \choose j}p^j\pbar^{k-1-j}\left(\frac{\beta^j-\beta^{k-1}}{1-\beta}\right)\\
&=\frac{\pi_1}{\beta(1-\beta)}\left((k-1)(\beta p+\pbar)^{k-2}\beta p-\beta^{k-1}(k-1)p\right)\\
&=\frac{\pi_1(k-1)p}{\beta(1-\beta)}\left((\beta p+\pbar)^{k-2}\beta -\beta^{k-1}\right).
\end{align*}
Next, keeping in mind that $k-1\geq 2$, the last term of Eq. (\ref{eq:dtmcCap}) is 
\begin{align*}
&(k-1)p\sum\limits_{i=k-1}^{\infty}\pi_i
=\frac{(k-1)p\pi_1}{\beta}\sum\limits_{i=k-1}^{\infty}\beta^{i}\\
&=\frac{(k-1)p\pi_1}{\beta}\left(\sum\limits_{i=0}^{\infty}\beta^{i}-\sum\limits_{i=0}^{k-2}\beta^i\right)\\
&=\frac{(k-1)p\pi_1}{\beta}\left(\frac{1}{1-\beta}-\frac{1-\beta^{k-1}}{1-\beta}\right)
=\frac{(k-1)p\pi_1}{\beta}\frac{\beta^{k-1}}{1-\beta}.
\end{align*}
Hence, so far, 
\begin{align}
C&=\sum\limits_{i=0}^{k-2}\pi_i \left(i{k-1 \choose i+1}p^{i+1}\pbar^{k-i-1}\right.\nonumber\\
&\left.+\sum\limits_{l=2}^{k-i}\left(\left\lfloor\frac{l}{2}\right\rfloor+i\right){k \choose i+l}p^{i+l}\pbar^{k-i-l}\right)\nonumber\\
 &+\frac{\pi_1(k-1)p}{\beta(1-\beta)}\left((\beta p+\pbar)^{k-2}\beta -\beta^{k-1}\right)
 +\frac{(k-1)p\pi_1}{\beta}\frac{\beta^{k-1}}{1-\beta}\nonumber\\
 &=\frac{\pi_1(k-1)p}{(1-\beta)}(\beta p+\pbar)^{k-2}
 +\sum\limits_{i=0}^{k-2}\pi_i \left(i{k-1 \choose i+1}p^{i+1}\pbar^{k-i-1}\right.\nonumber\\
&\left.+\sum\limits_{l=2}^{k-i}\left(\left\lfloor\frac{l}{2}\right\rfloor+i\right){k \choose i+l}p^{i+l}\pbar^{k-i-l}\right).%\nonumber\\
\label{eq:dtmcCnew}
\end{align}
Next in Eq. (\ref{eq:dtmcCnew}) we have the term
\begin{align*}
&\sum\limits_{i=0}^{k-2}\pi_i i{k-1 \choose i+1}p^{i+1}\pbar^{k-i-1} = 
\pi_1\sum\limits_{i=1}^{k-2}\beta^{i-1} i{k-1 \choose i+1}p^{i+1}\pbar^{k-i-1}\\
&=\pi_1\sum\limits_{j=2}^{k-1}\beta^{j-2} (j-1){k-1 \choose j}p^{j}\pbar^{k-j}\\
%&=\frac{\pbar\pi_1}{\beta^2}\left(\sum\limits_{j=2}^{k-1}j {k-1 \choose j}(\beta p)^{j}\pbar^{k-j-1}-\sum\limits_{j=2}^{k-1}{k-1 \choose j}(\beta p)^{j}\pbar^{k-j-1}\right)\\
%&=\frac{\pbar\pi_1}{\beta^2}\Bigg(\sum\limits_{j=1}^{k-1}j {k-1 \choose j}(\beta p)^{j}\pbar^{k-j-1}-\sum\limits_{j=0}^{k-1}{k-1 \choose j}(\beta p)^{j}\pbar^{k-j-1}\\
%&-(k-1)\beta p\pbar^{k-2}+\pbar^{k-1}+(k-1)\beta p\pbar^{k-2}\Bigg)\\
&=\frac{\pbar\pi_1}{\beta^2}\left((k-1)(\beta p+\pbar)^{k-2}\beta p-(\beta p +\pbar)^{k-1}+\pbar^{k-1}\right).
\end{align*}
Substituting this into Eq. (\ref{eq:dtmcCnew}), we have
\begin{align}
&C=\sum\limits_{i=0}^{k-2}\pi_i \sum\limits_{l=2}^{k-i}\left(\left\lfloor\frac{l}{2}\right\rfloor+i\right){k \choose i+l}p^{i+l}\pbar^{k-i-l}\nonumber\\
&+\frac{\pi_1(k-1)p}{(1-\beta)}(\beta p+\pbar)^{k-2}\nonumber\\
 &+\frac{\pbar\pi_1}{\beta^2}\left((k-1)(\beta p+\pbar)^{k-2}\beta p-(\beta p +\pbar)^{k-1}+\pbar^{k-1}\right)\nonumber\\
%&=\sum\limits_{i=0}^{k-2}\pi_i \sum\limits_{l=2}^{k-i}\left(\left\lfloor\frac{l}{2}\right\rfloor+i\right){k \choose i+l}p^{i+l}\pbar^{k-i-l}\nonumber\\
% &+\frac{\pbar\pi_1}{\beta^2}\left(\pbar^{k-1}-(\beta p +\pbar)^{k-1}\right)
%+\pi_1(k-1)p(\beta p+\pbar)^{k-2}\left(\frac{1}{1-\beta}+\frac{\pbar}{\beta}\right)\nonumber\\\nonumber\\
&=\sum\limits_{i=0}^{k-2}\pi_i \sum\limits_{l=2}^{k-i}\left(\left\lfloor\frac{l}{2}\right\rfloor+i\right){k \choose i+l}p^{i+l}\pbar^{k-i-l}\nonumber\\
&+\frac{\pbar\pi_1}{\beta^2}\left(\pbar^{k-1}-(\beta p +\pbar)^{k-1}\right)
+\pi_1(k-1)p(\beta p+\pbar)^{k-2}\frac{(\beta p+\pbar)}{\beta(1-\beta)}\nonumber\\
&=\sum\limits_{i=0}^{k-2}\pi_i \sum\limits_{l=2}^{k-i}\left(\left\lfloor\frac{l}{2}\right\rfloor+i\right){k \choose i+l}p^{i+l}\pbar^{k-i-l}\nonumber\\
&+\frac{\pbar\pi_1}{\beta^2}\left(\pbar^{k-1}-(\beta p +\pbar)^{k-1}\right)+\pi_1(k-1)p\frac{(\beta p+\pbar)^{k-1}}{\beta(1-\beta)}.
\label{eq:dtmcCnew2}
\end{align}
Consider the remaining sum above.
Let $m=i+l$. Then
\begin{align}
&\sum\limits_{i=0}^{k-2}\pi_i\sum\limits_{l=2}^{k-i}\left(\left\lfloor\frac{l}{2}\right\rfloor+i\right){k \choose i+l}p^{i+l}\pbar^{k-i-l}\nonumber\\
%&=\sum\limits_{i=0}^{k-2}\pi_i\sum\limits_{m=i+2}^{k}\left\lfloor\frac{m-i}{2}\right\rfloor{k \choose m}p^{m}\pbar^{k-m}
%+\sum\limits_{i=0}^{k-2}\pi_i\sum\limits_{m=i+2}^{k}i{k \choose m}p^{m}\pbar^{k-m}\\
&=\sum\limits_{m=2}^{k}{k \choose m}p^{m}\pbar^{k-m}\left(\sum\limits_{i=0}^{m-2}\pi_i\left(\left\lfloor\frac{m-i}{2}\right\rfloor+i\right)\right)\nonumber\\
&=\sum\limits_{m=2}^{k}{k \choose m}p^{m}\pbar^{k-m}\Bigg(\sum\limits_{i=0}^{m}\pi_i\left(\left\lfloor\frac{m-i}{2}\right\rfloor+i\right)\nonumber\\
&-(m-1)\pi_{m-1}-m\pi_m\Bigg) \coloneqq S.
\label{eq:dtmcLongSum}
\end{align}
The inner sum above can be rewritten as follows:
\begin{align*}
&\sum\limits_{i=0}^{m}\pi_i\left(\left\lfloor\frac{m-i}{2}\right\rfloor+i\right) =
\sum\limits_{i=0}^{m}\pi_i\Bigg(i+\\
&+\left(\frac{m-i}{2}\right)\frac{1+(-1)^{m-i}}{2}+\left(\frac{m-i-1}{2}\right)\frac{1-(-1)^{m-i}}{2}\Bigg)\\
&=\sum\limits_{i=0}^{m}\pi_i\left(i+\frac{m-i}{2}-\frac{1}{2}\frac{(1-(-1)^{m-i})}{2}\right)\\
&=\sum\limits_{i=0}^{m}\pi_i\left(\frac{2m-1}{4}+\frac{i}{2}+\frac{(-1)^{m-i}}{4}\right)\\
&=\frac{\pi_1}{\beta}\sum\limits_{i=1}^{m}\beta^i\left(\frac{2m-1}{4}+\frac{i}{2}+\frac{(-1)^{m-i}}{4}\right)\\
&+\pi_0\left(\frac{2m-1+(-1)^m}{4}\right)\\
%&=\frac{\pi_1}{\beta}\left(\sum\limits_{i=1}^{m}\beta^i\frac{2m-1}{4}+\sum\limits_{i=1}^{m}\beta^i\frac{i}{2}+(-1)^m\sum\limits_{i=1}^{m}\frac{(-\beta)^i}{4}\right)\\
%&+\pi_0\left(\frac{2m-1+(-1)^m}{4}\right)\\
%&=\frac{\pi_1}{\beta}\left(\frac{2m-1}{4}\sum\limits_{i=0}^{m}\beta^i+\sum\limits_{i=1}^{m}\beta^i\frac{i}{2}+\frac{(-1)^m}{4}\sum\limits_{i=0}^{m}(-\beta)^i
%-\frac{2m-1}{4}-\frac{(-1)^m}{4}\right)\\&+\pi_0\left(\frac{2m-1+(-1)^m}{4}\right)\\
&=\frac{\pi_1}{\beta}\left(\frac{2m-1}{4}\left(\frac{1-\beta^{m+1}}{1-\beta}-1\right)+\frac{1}{2}\sum\limits_{i=1}^{m}i\beta^i\right.\\
&+\left.\frac{(-1)^m}{4}\left(\frac{1-(-\beta)^{m+1}}{1+\beta}-1\right)\right)+\pi_0\left(\frac{2m-1+(-1)^m}{4}\right)\\
%&=\frac{\pi_1}{\beta}\left(\frac{2m-1}{4}\left(\frac{\beta-\beta^{m+1}}{1-\beta}\right)
%+\frac{\beta}{2}\frac{(m\beta^{m+1}-(m+1)\beta^m+1)}{(1-\beta)^2}\right.\\
%&-\left.\frac{(-1)^m}{4}\left(\frac{(-\beta)^{m+1}+\beta}{1+\beta}\right)\right)+\pi_0\left(\frac{2m-1+(-1)^m}{4}\right)\\
%&=\pi_1\left(\frac{2m-1}{4}\left(\frac{1-\beta^{m}}{1-\beta}\right)
%+\frac{1}{2}\frac{(m\beta^{m+1}-(m+1)\beta^m+1)}{(1-\beta)^2}
%-\frac{(-1)^m}{4}\left(\frac{1-(-\beta)^{m}}{1+\beta}\right)\right)+\pi_0\left(\frac{2m-1+(-1)^m}{4}\right)\\
&=\pi_1\left(\frac{2m-1}{4}\left(\frac{1-\beta^{m}}{1-\beta}\right)
+\frac{1}{2}\frac{(m\beta^{m+1}-(m+1)\beta^m+1)}{(1-\beta)^2}\right.\\
&+\left.\frac{\beta^{m}-(-1)^m}{4(1+\beta)}\right)+\pi_0\left(\frac{2m-1+(-1)^m}{4}\right).
\end{align*}
Now, we can use the fact that $\pi_0+\pi_1/(1-\beta)=1$ to obtain
\begin{align*}
\pi_1\frac{2m-1}{4}\frac{1}{1-\beta}+\frac{2m-1}{4}\pi_0 %= \frac{2m-1}{4}\left(\pi_0+\frac{\pi_1}{1-\beta}\right) 
= \frac{2m-1}{4}.
\end{align*}
Using the same relation, we have
\begin{align*}
\frac{(-1)^m}{4}\left(\pi_0-\frac{\pi_1}{1+\beta}\right) %&=\frac{(-1)^m}{4}\left(1-\frac{\pi_1}{1-\beta}-\frac{\pi_1}{1+\beta}\right)
=\frac{(-1)^m}{4}\left(1-\frac{2\pi_1}{1-\beta^2}\right).
\end{align*}
Therefore,
\begin{align*}
&\sum\limits_{i=0}^{m}\pi_i\left(\left\lfloor\frac{m-i}{2}\right\rfloor+i\right) = \\
&\pi_1\left(\frac{2m-1}{4}\left(\frac{-\beta^{m}}{1-\beta}\right)
+\frac{m\beta^{m+1}-(m+1)\beta^m+1}{2(1-\beta)^2}\right.\\
&\left.+\frac{\beta^{m}}{4(1+\beta)}\right)
+\frac{2m-1}{4}+\frac{(-1)^m}{4}\left(1-\frac{2\pi_1}{1-\beta^2}\right)\\
&=-\pi_1\beta^m\frac{\beta}{(1-\beta)^2(1+\beta)}-\pi_1\frac{m\beta^m}{1-\beta}
+\frac{m}{2}\\&+\frac{(-1)^m}{4}\left(1-\frac{2\pi_1}{1-\beta^2}\right)+\frac{\pi_1}{2(1-\beta)^2}-\frac{1}{4}.
\end{align*}
From Eq. (\ref{eq:dtmcPi1}), we know that $\pi_1=(1-\beta^2)/2$. Using this,
\begin{align*}
\sum\limits_{i=0}^{m}\pi_i\left(\left\lfloor\frac{m-i}{2}\right\rfloor+i\right)
&=-\pi_1\beta^m\frac{\beta}{(1-\beta)^2(1+\beta)}\\
&-\pi_1\frac{m\beta^m}{1-\beta}
+\frac{m}{2}+\frac{\pi_1}{2(1-\beta)^2}-\frac{1}{4}.
\end{align*}
Next, we can write
\begin{align*}
&\sum\limits_{i=0}^{m}\pi_i\left(\left\lfloor\frac{m-i}{2}\right\rfloor+i\right)-(m-1)\pi_{m-1}-m\pi_m =\\
&\frac{-\pi_1\beta^m\beta}{(1-\beta)^2(1+\beta)}
-\pi_1\frac{m\beta^m}{1-\beta}+\frac{m}{2}+\frac{\pi_1}{2(1-\beta)^2}-\frac{1}{4}\\
&-\frac{\pi_1}{\beta}m\beta^{m}\left(\frac{1}{\beta}+1\right)+\frac{\pi_1}{\beta^2}\beta^{m}=
\frac{m}{2}+\frac{\pi_1}{2(1-\beta)^2}-\frac{1}{4}\\
&+\pi_1\beta^m\left(\frac{1}{\beta^2}-\frac{\beta}{(1-\beta)^2(1+\beta)}\right)
-\frac{\pi_1m\beta^m}{\beta^2(1-\beta)}.
\end{align*}
Hence, Eq. (\ref{eq:dtmcLongSum}) becomes
\begin{align*}
&S=
%&\sum\limits_{m=2}^{k}{k \choose m}p^{m}\pbar^{k-m}\left(\sum\limits_{i=0}^{m}\pi_i\left(\left\lfloor\frac{m-i}{2}\right\rfloor+i\right)-(m-1)\pi_{m-1}-m\pi_m\right)=\\
\sum\limits_{m=2}^{k}{k \choose m}p^{m}\pbar^{k-m}\left(\pi_1\beta^m\left(\frac{1}{\beta^2}-\frac{\beta}{(1-\beta)^2(1+\beta)}\right)\right.\\
&-\left.\frac{\pi_1m\beta^m}{\beta^2(1-\beta)}
+\frac{m}{2}+\frac{\pi_1}{2(1-\beta)^2}-\frac{1}{4}\right)=\\
&\pi_1\left(\frac{1}{\beta^2}-\frac{\beta}{(1-\beta)^2(1+\beta)}\right)(p\beta+\pbar)^k
-\frac{\pi_1kp\beta(p\beta+\pbar)^{k-1}}{\beta^2(1-\beta)}\\
&-\pbar^k\left(\pi_1\left(\frac{1}{\beta^2}-\frac{\beta}{(1-\beta)^2(1+\beta)}\right)+\frac{\pi_1}{2(1-\beta)^2}-\frac{1}{4}\right)\\
&-kp\pbar^{k-1}\left(\pi_1\beta\left(\frac{1}{\beta^2}-\frac{\beta}{(1-\beta)^2(1+\beta)}\right)
-\frac{\pi_1\beta}{\beta^2(1-\beta)}\right.\\
&+\left.\frac{1}{2}+\frac{\pi_1}{2(1-\beta)^2}-\frac{1}{4}\right)+\frac{kp}{2}
+\frac{\pi_1}{2(1-\beta)^2}-\frac{1}{4}=\\
%&\pi_1\left(\frac{1}{\beta^2}-\frac{\beta}{(1-\beta)^2(1+\beta)}\right)(p\beta+\pbar)^k
%-\frac{\pi_1kp}{\beta(1-\beta)}(p\beta+\pbar)^{k-1}+\frac{kp}{2}\\
%&+\frac{\pi_1}{2(1-\beta)^2}-\frac{1}{4}\\
%&-\pbar^k\left(\pi_1\left(\frac{1}{\beta^2}-\frac{\beta}{(1-\beta)^2(1+\beta)}\right)+\frac{\pi_1}{2(1-\beta)^2}-\frac{1}{4}\right)\\
%&-kp\pbar^{k-1}\left(\pi_1\left(\frac{1}{\beta}-\frac{\beta^2}{(1-\beta)^2(1+\beta)}-\frac{1}{\beta(1-\beta)}+\frac{1}{2(1-\beta)^2}\right)+\frac{1}{2}-\frac{1}{4}\right)=\\\\%%%%%
&\pi_1\left(\frac{1}{\beta^2}-\frac{\beta}{(1-\beta)^2(1+\beta)}\right)(p\beta+\pbar)^k
-\frac{\pi_1kp(p\beta+\pbar)^{k-1}}{\beta(1-\beta)}\\
&-\pbar^k\left(\frac{\pi_1}{\beta^2}+\frac{\pi_1}{2(1-\beta^2)}-\frac{1}{4}\right)-kp\pbar^{k-1}\left(\frac{-\pi_1}{2(1-\beta^2)}+\frac{1}{4}\right)\\
&+\frac{kp}{2}+\frac{\pi_1}{2(1-\beta)^2}-\frac{1}{4}.
\end{align*}
Substituting $\pi_1=(1-\beta^2)/2$ above and simplifying yields
\begin{align*}
%&\sum\limits_{m=2}^{k}{k \choose m}p^{m}\pbar^{k-m}\left(\sum\limits_{i=0}^{m}\pi_i\left(\left\lfloor\frac{m-i}{2}\right\rfloor+i\right)-(m-1)\pi_{m-1}-m\pi_m\right)=\\
&S=
%&\pi_1\left(\frac{1}{\beta^2}-\frac{\beta}{(1-\beta)^2(1+\beta)}\right)(p\beta+\pbar)^k
%-\frac{\pi_1kp}{\beta(1-\beta)}(p\beta+\pbar)^{k-1}+\frac{kp}{2}+\frac{\pi_1}{2(1-\beta)^2}-\frac{1}{4}\\
%&-\pbar^k\left(\frac{\pi_1}{\beta^2}+\frac{1-\beta^2}{2}\frac{1}{2(1-\beta^2)}-\frac{1}{4}\right)-kp\pbar^{k-1}\left(\frac{1-\beta^2}{2}\frac{-1}{2(1-\beta^2)}+\frac{1}{2}-\frac{1}{4}\right)\\\\
\pi_1\left(\frac{1}{\beta^2}-\frac{\beta}{(1-\beta)^2(1+\beta)}\right)(p\beta+\pbar)^k\\
&-\frac{\pi_1kp}{\beta(1-\beta)}(p\beta+\pbar)^{k-1}+\frac{kp}{2}+\frac{\pi_1}{2(1-\beta)^2}-\frac{1}{4}-\pbar^k\frac{\pi_1}{\beta^2}.
\end{align*}
Finally, substituting this result into Eq. (\ref{eq:dtmcCnew2}), $C$ becomes
\begin{align*}
&C = \frac{\pbar\pi_1}{\beta^2}\left(\pbar^{k-1}-(\beta p +\pbar)^{k-1}\right)+\pi_1(k-1)p\frac{(\beta p+\pbar)^{k-1}}{\beta(1-\beta)}\\
&+\pi_1\left(\frac{1}{\beta^2}-\frac{\beta}{(1-\beta)^2(1+\beta)}\right)(p\beta+\pbar)^k\\
&-\frac{\pi_1kp}{\beta(1-\beta)}(p\beta+\pbar)^{k-1}+\frac{kp}{2}+\frac{\pi_1}{2(1-\beta)^2}-\frac{1}{4}-\pbar^k\frac{\pi_1}{\beta^2}\\
& =\frac{-\pi_1(\beta p +\pbar)^{k-1}\left(\pbar\beta+p\right)}{(1-\beta)^2(1+\beta)}
+\frac{kp}{2}+\frac{\pi_1}{2(1-\beta)^2}-\frac{1}{4}.
\end{align*}
We know from Eq. (\ref{eq:dtmcPoly}) that
\begin{align*}
(\beta p +\pbar)^{k-1}(p+\beta \pbar)-\beta=0.
\end{align*}
Using this above, we obtain
\begin{align*}
C &= -\frac{\pi_1\beta}{(1-\beta)^2(1+\beta)}+\frac{kp}{2}+\frac{\pi_1}{2(1-\beta)^2}-\frac{1}{4}\\
%&= \frac{\pi_1}{(1-\beta)^2}\left(\frac{1}{2}-\frac{\beta}{1+\beta}\right)-\frac{1}{4}+\frac{kp}{2}\\
%&= \frac{\pi_1}{(1-\beta)^2}\left(\frac{1+\beta-2\beta}{2(1+\beta)}\right)-\frac{1}{4}+\frac{kp}{2}\\
%&= \frac{\pi_1}{(1-\beta)^2}\left(\frac{1-\beta}{2(1+\beta)}\right)-\frac{1}{4}+\frac{kp}{2}\\
%&= \frac{\pi_1}{(1-\beta)}\left(\frac{1}{2(1+\beta)}\right)-\frac{1}{4}+\frac{kp}{2}\\
&= \frac{\pi_1}{2(1-\beta^2)}-\frac{1}{4}+\frac{kp}{2}.
\end{align*}
Recall that $\pi_1=(1-\beta^2)/2$. Hence,
\begin{align*}
C &= \frac{1-\beta^2}{2}\frac{1}{2(1-\beta^2)}-\frac{1}{4}+\frac{kp}{2},\\
C &=\frac{kp}{2}.
\end{align*}
Finally, recall that we earlier assumed $q=1$. Removing this assumption, we obtain
\begin{align*}
\pushQED{\qed}
C &= \frac{qkp}{2}.
\qedhere
\popQED
\end{align*}
\section{CTMC Derivations for Bipartite Entanglements}
\subsection{Capacity for Heterogeneous Systems with $B=\infty$}
%%!TEX PS-program = pdflatexmk
%%!TEX root = quantum_jrnl_short.tex
\label{heterogCapApp}
\subsubsection*{Proof of the last equality in Eq. (\ref{eq:heterogCapInfBuf})} From the first part of this equation, we have
\begin{align*}
&C = q\sum\limits_{l=1}^k\sum\limits_{j=1}^{\infty}\pi_l^{(j)}(\gamma-\mu_l)=
q\sum\limits_{l=1}^k\sum\limits_{j=1}^{\infty}\pi_0\rho_l^j(\gamma-\mu_l)=\\
&q\pi_0\sum\limits_{l=1}^k\frac{(\gamma-\mu_l)\rho_l}{1-\rho_l}=
q\pi_0\sum\limits_{l=1}^k\left(\frac{\gamma}{2}\frac{\rho_l}{1-\rho_l}+\left(\frac{\gamma}{2}-\mu_l\right)\frac{\rho_l}{1-\rho_l}\right)\\
&=q\pi_0\sum\limits_{l=1}^k\left(\frac{\gamma}{2}\frac{\rho_l}{1-\rho_l}+\left(\frac{\gamma-2\mu_l}{2}\right)\frac{\mu_l(\gamma-\mu_l)}{(\gamma-\mu_l)(\gamma-2\mu_l)}\right)=\\
&q\pi_0\sum\limits_{l=1}^k\left(\frac{\gamma}{2}\frac{\rho_l}{1-\rho_l}+\frac{\mu_l}{2}\right)=
q\pi_0\frac{\gamma}{2}\left(\sum\limits_{l=1}^k\frac{\rho_l}{1-\rho_l}+1\right)=\frac{q\gamma}{2}.
\end{align*}
\subsubsection*{Proof that $C_l=q\mu_l$} Letting $B\to\infty$ in Eq. (\ref{eq:Clheterog}),
\begin{align*}
C_l &= q\pi_0\left((\gamma-\mu_l)\frac{\rho_l}{1-\rho_l}+\mu_l\sum\limits_{\substack{m=1,\\ m\neq l}}^k\frac{\rho_m}{1-\rho_m}\right)\\
&=q\pi_0\mu_l\left(\frac{1}{1-\rho_l}+\sum\limits_{\substack{m=1,\\ m\neq l}}^k\frac{\rho_m}{1-\rho_m}+\frac{\rho_l}{1-\rho_l}-\frac{\rho_l}{1-\rho_l}\right)\\
&=q\pi_0\mu_l\left(1+\sum\limits_{m=1}^k\frac{\rho_m}{1-\rho_m}\right)=q\mu_l.
\end{align*}
\subsection{Decoherence}
%%!TEX PS-program = pdflatexmk
%%!TEX root = quantum_jrnl_short.tex
\label{app:decoh}
 \subsubsection*{Homogeneous, Infinite Buffer}
For this system, the balance equations are as follows:
 \begin{align*}
\pi_0 k \mu &= \pi_1(\alpha+(k-1)\mu),\\
\pi_{i-1}\mu &= \pi_i(i\alpha+(k-1)\mu), \quad i=2,3,\dots,\\
\sum\limits_{i=0}^{\infty}\pi_i &= 1.
 \end{align*}
 Solving for the stationary distribution, we have:
 \begin{align*}
 \pi_1 &= \frac{k\mu}{(k-1)\mu+\alpha}\pi_0,\\
 \pi_2 &= \frac{\mu \pi_1}{(k-1)\mu+2\alpha} = \frac{k\mu^2 \pi_0}{((k-1)\mu+2\alpha)((k-1)\mu+\alpha)},
 \end{align*}
 and so on. In general, for $i=1,2,\dots$ we can write
 \begin{align*}
 \pi_i &= \frac{\pi_0k\mu^i}{\prod\limits_{j=1}^i ((k-1)\mu+j\alpha)}=\pi_0k\prod\limits_{j=1}^i\frac{\mu}{ ((k-1)\mu+j\alpha)}.
 \end{align*}
 Using the normalizing condition, we have
 \begin{align*}
& \pi_0+k\pi_0\sum\limits_{i=1}^{\infty}\prod\limits_{j=1}^i\frac{\mu}{ ((k-1)\mu+j\alpha)} = 1,
% \end{align*}
\quad \text{so that}\\
% \begin{align*}
&\pi_0 = \left(1+k\sum\limits_{i=1}^{\infty}\prod\limits_{j=1}^i\frac{\mu}{ ((k-1)\mu+j\alpha)}\right)^{-1}.
 \end{align*}
 The capacity and $E[Q]$ can be computed numerically using the following formulas:
 \begin{align*}
 C &= \sum\limits_{i=1}^{\infty}\pi_i(k-1)\mu = (k-1)\mu(1-\pi_0),\\
 E[Q] &= \sum\limits_{i=1}^{\infty}i\pi_i = \pi_0k\sum\limits_{i=1}^{\infty}i\prod\limits_{j=1}^i\frac{\mu}{ ((k-1)\mu+j\alpha)}.
 \end{align*}
 \subsubsection*{Homogeneous, Finite Buffer}
 The derivations are very similar to the previous case, with the only difference being that the balance equations are truncated at state $i=B$. The resulting expressions are almost identical to those above, with the exception of $i$ being in $\{1,\dots,B\}$ instead of $\{1,2,\dots\}$:
\begin{align*}
\pi_0 &= \left(1+k\sum\limits_{i=1}^{B}\prod\limits_{j=1}^i\frac{\mu}{ ((k-1)\mu+j\alpha)}\right)^{-1},\\
 C &= \sum\limits_{i=1}^{B}\pi_i(k-1)\mu = (k-1)\mu(1-\pi_0),\\
 E[Q] &= \sum\limits_{i=1}^{B}i\pi_i = \pi_0k\sum\limits_{i=1}^{B}i\prod\limits_{j=1}^i\frac{\mu}{ ((k-1)\mu+j\alpha)}.
 \end{align*}
 \subsubsection*{Heterogeneous, Infinite Buffer}
 The balance equations are:
 \begin{align*}
& \pi_0\mu_l =\pi_l^{(1)}(\gamma-\mu_l+\alpha),~ l\in\{1,\dots,k\},\\
& \pi_l^{(j-1)}\mu_l = \pi_l^{(j)}(\gamma-\mu_l+j\alpha),~ l\in\{1,\dots,k\}, ~j\in\{2,3,\dots\},\\
&\pi_0+ \sum\limits_{l=1}^k\sum\limits_{j=1}^{\infty}\pi_l^{(j)} =1.
 \end{align*}
 For $j=1,2,\dots$, we can write
 \begin{align*}
 \pi_l^{(j)} &= \pi_0\prod\limits_{i=1}^j \frac{\mu_l}{\gamma-\mu_l+i\alpha}.
 \end{align*}
 Using the normalizing condition, we obtain
 \begin{align*}
 \pi_0 &=\left(1+\sum\limits_{l=1}^k\sum\limits_{j=1}^{\infty}\prod\limits_{i=1}^j\frac{\mu_l}{\gamma-\mu_l+i\alpha}\right)^{-1}.
 \end{align*}
 The capacity and $E[Q]$ can be computed numerically using
 \begin{align*}
&C \hspace{-0.15em}= \hspace{-0.45em}\sum\limits_{l=1}^k\sum\limits_{j=1}^{\infty}\pi_l^{(j)}(\gamma-\mu_l) = \pi_0\sum\limits_{l=1}^k\sum\limits_{j=1}^{\infty}(\gamma-\mu_l)\prod\limits_{i=1}^j \frac{\mu_l}{\gamma-\mu_l+i\alpha},\\
& E[Q] = \sum\limits_{j=1}^{\infty}jP(Q=j) = \sum\limits_{j=1}^{\infty}j\sum\limits_{l=1}^k\pi_l^{(j)}\\
 &\qquad=\pi_0\sum\limits_{j=1}^{\infty}j\sum\limits_{l=1}^k\prod\limits_{i=1}^j \frac{\mu_l}{\gamma-\mu_l+i\alpha}.
 \end{align*}
 \subsubsection*{Heterogeneous, Finite Buffer}
 The derivations are similar to the previous case, with the only difference being that $j$ is now in $\{1,\dots,B\}$ instead of in $\{1,2,\dots\}$. The resulting relevant expressions are:
 \begin{align*}
 \pushQED{\qed} 
 \pi_0 &=\left(1+\sum\limits_{l=1}^k\sum\limits_{j=1}^{B}\prod\limits_{i=1}^j\frac{\mu_l}{\gamma-\mu_l+i\alpha}\right)^{-1},\\
 C &=\pi_0\sum\limits_{l=1}^k\sum\limits_{j=1}^{B}(\gamma-\mu_l)\prod\limits_{i=1}^j \frac{\mu_l}{\gamma-\mu_l+i\alpha},\\
 E[Q] &=\pi_0\sum\limits_{j=1}^{B}j\sum\limits_{l=1}^k\prod\limits_{i=1}^j \frac{\mu_l}{\gamma-\mu_l+i\alpha}.
 \qedhere
\popQED
 \end{align*}
\section{CTMC Derivations for $n$-Partite Entanglements}
%%!TEX PS-program = pdflatexmk
%%!TEX root = quantum_jrnl_short.tex
\subsection{Capacity and Expected Number of Qubits}
\label{app:npartite}
This section contains the proof of Proposition 1. Recall that $k\geq n$.
We assume that the irreducible and aperiodic DTMC $X$ is positive recurrent (\emph{i.e.} stable), with $\pi$ its stationary distribution. Assume that it is in steady state at time $t=1$ (which implies that it is in steady state for $t>1$).
For every mapping
$V:\{0,1,\ldots\}^{n-1}\to [0,\infty)$ such that ${E[V(\vect{Q}_1)]<\infty}$, we have
\begin{align}
&0=E[V(\vect{Q}_{t+1})-V(\vect{Q}_t)]=\nonumber\\
&\sum_{\vect{i}\geq \vect{0}} \pi(\vect{i}) E[V(\vect{Q}_{t+1})-V(\vect{i})\,|\, \vect{Q}_{t}=\vect{i}]=\nonumber\\
&\sum_{\vect{i}\geq \vect{1}}  \pi(\vect{i})\Biggl[ \frac{(k-(n-1))}{k}(V(\vect{i}-\vect{1})-V(\vect{i}))\nonumber\\
&+ \sum_{l=1}^{n-1}\frac{1}{k} (V(\vect{i}+\vect{e}_l)-V(\vect{i}))\Biggr]\nonumber\\ 
&+\sum_{j=1}^{n-2} \sum_{\vect{i}\in S_j} \pi(\vect{i})  \Biggl[ \sum_{l:i_l=0} \frac{(k-(n-1-j))}{kj} (V(\vect{i}+\vect{e}_l)-V(\vect{i}))\nonumber\\
&+ \sum_{l:i_l\neq0}^{n-1} \frac{1}{k}  (V(\vect{i}+\vect{e}_l)-V(\vect{i})) \Biggr]
+ \pi(\vect{0})\sum_{l=1}^{n-1} \frac{(V(\vect{e}_l)-V(\vect{0}))}{(n-1)}.
\label{eq:40}
\end{align}
Take $V(\vect{i})=\sum\limits_{l=1}^{n-1}i_l$.  Multiplying both sides of (\ref{eq:40}) by $k$ yields
\begin{align}
&0%= (-(k-(n-1))(n-1)+(n-1)) \sum_{\vect{i}\geq \vect{1}}  \pi(\vect{i})\nonumber\\
%&+k \sum_{j=1}^{n-2} \sum_{\vect{i}\in S_j}\pi(i_1,\ldots,i_{n-1})+ k\pi(0,\ldots,0)\nonumber\\
=-(n-1)(k-n) \sum_{\vect{i}\geq \vect{1}}  \pi(\vect{i}) + k \sum_{j=1}^{n-1} \sum_{\vect{i}\in S_j}\pi(\vect{i}).
\label{eq:41}
\end{align}
From the identities
\[
1=\sum_{\vect{i}\geq \vect{0}}\pi(\vect{i})=\sum_{\vect{i}\geq \vect{1}} \pi(\vect{i})+\sum_{j=1}^{n-1}\sum_{\vect{i}\in S_j}\pi(\vect{i}),
\]
%where the latter identity holds from  (\ref{partition}), 
we deduce that
\begin{equation}
\label{eq:42}
\sum_{j=1}^{n-1}\sum_{\vect{i}\in S_j} \pi(\vect{i})= 1-\sum_{\vect{i}\geq \vect{1}} \pi(\vect{i}).
\end{equation}
Hence, cf. Eqs. (\ref{eq:41}) and (\ref{eq:42}),
\[
0=-n(k-(n-1)) \sum_{\vect{i}\geq \vect{1}}  \pi(\vect{i}) +  k,
\]
so that
%The above equation makes sense only if $k-(n-1)>0$, namely if $k>n-1$, which shows that this condition is necessary for stability. When $k>n-1$,
\begin{align}
\sum_{\vect{i}\geq \vect{1}}  \pi(\vect{i})=\frac{k}{n(k-(n-1))}.
\label{eq:piS0}
\end{align}
The capacity $C$ is then given by 
\[
C=q\mu (k-(n-1)) \sum_{\vect{i}\geq \vect{1}} \pi(\vect{i})= \frac{q\mu k}{n}.
\]
This completes the proof for Eq. (\ref{eq:Cnpart}). Now, return to Eq. (\ref{eq:40}) and take ${V(\vect{i})=\sum\limits_{l=1}^{n-1}i_l^2}$, then multiply both sides by $k$:
\begin{align}
&0=\sum_{\vect{i}\geq \vect{1}}  \pi(\vect{i})\Biggl[(k-(n-1))(-2|\vect{i}|+(n-1))+ 2|\vect{i}|+n-1\Biggr]\nonumber\\ 
&+\sum_{j=1}^{n-2} \sum_{\vect{i}\in S_j} \pi(\vect{i})  (2|\vect{i}|+k)
+ k\pi(\vect{0})=\nonumber\\
&-2(k-n)\sum_{\vect{i}\geq \vect{1}}\pi(\vect{i})|\vect{i}|+(k-n+2)(n-1)\sum_{\vect{i}\geq \vect{1}}\pi(\vect{i})\nonumber\\
&+2\sum_{j=1}^{n-2} \sum_{\vect{i}\in S_j} \pi(\vect{i}) |\vect{i}|+k\sum_{j=1}^{n-1}\sum_{\vect{i}\in S_j}\pi(\vect{i}).
\label{eq:25app}
\end{align}
Using Eqs. (\ref{eq:42}) and (\ref{eq:piS0}), we can write
\begin{align}
\sum_{j=1}^{n-1}\sum_{\vect{i}\in S_j}\pi(\vect{i}) = 1-\sum_{\vect{i}\geq \vect{1}}\pi(\vect{i}) = \frac{(k-n)(n-1)}{n(k-(n-1))}.
\label{eq:26app}
\end{align}
Hence, from Eqs. (\ref{eq:25app}) and (\ref{eq:26app}), we obtain
\begin{align}
(k-n)\sum_{\vect{i}\in S_0}\pi(\vect{i})|\vect{i}|-\sum_{j=1}^{n-2} \sum_{\vect{i}\in S_j} \pi(\vect{i}) |\vect{i}| = \frac{k(n-1)}{n}.
\label{eq:27app}
\end{align}
Next, take $V(\vect{i})=|\vect{i}|^2$. After substituting this into Eq. (\ref{eq:40}) and multiplying both sides by $k$, we get
\begin{align*}
&0=\sum_{\vect{i}\geq \vect{1}}  \pi(\vect{i})\Biggl[(k-(n-1))(-2|\vect{i}|(n-1)+(n-1)^2)+\\
& (n-1)(2|\vect{i}|+1)\Biggr]+k\sum_{j=1}^{n-2} \sum_{\vect{i}\in S_j} \pi(\vect{i})(2|\vect{i}|+1)+ k\pi(\vect{0})=\\
&-2(k-n)(n-1)\sum_{\vect{i}\geq \vect{1}}  \pi(\vect{i})|\vect{i}|\\
&+(n-1)((n-1)(k-(n-1))+1)\sum_{\vect{i}\geq \vect{1}}  \pi(\vect{i})\\
&+2k\sum_{j=1}^{n-2} \sum_{\vect{i}\in S_j} \pi(\vect{i})|\vect{i}|+k\sum_{j=1}^{n-1} \sum_{\vect{i}\in S_j} \pi(\vect{i}).
\end{align*}
Using Eqs. (\ref{eq:piS0}) and (\ref{eq:26app}) yields
\begin{align}
2(k-n)(n-1)\sum_{\vect{i}\in S_0}  \pi(\vect{i})|\vect{i}|-2k\sum_{j=1}^{n-2} \sum_{\vect{i}\in S_j} \pi(\vect{i})|\vect{i}|=k(n-1).
\label{eq:28app}
\end{align}
Let $A\coloneqq \sum_{\vect{i}\in S_0}  \pi(\vect{i})|\vect{i}|$ and $B\coloneqq \sum_{j=1}^{n-2} \sum_{\vect{i}\in S_j} \pi(\vect{i})|\vect{i}|$. Note that $E[|\vect{Q}|] = A+B$. From Eqs. (\ref{eq:27app}) and (\ref{eq:28app}), define the following set of linear equations in the unknowns $A$ and $B$:
\begin{align*}
(k-n)A-B &= \frac{k(n-1)}{n}\\
2(k-n)(n-1)A-2kB &= k(n-1).
\end{align*}
%When $k\neq n-1$ and $k\neq n$, there is a unique solution:
Since $k\geq n$, this system of equations has a unique solution, given by
\begin{align*}
A &= \frac{k(n-1)(2k-n)}{2n(k-n)(k-(n-1))},\quad
B &= \frac{k(n-1)(n-2)}{2n(k-(n-1))}.
\end{align*}
We find $E[|\vect{Q}|]=A+B=\frac{(n-1)k}{2(k-n)}$.\qed
%%!TEX PS-program = pdflatexmk
%%!TEX root = quantum_jrnl_short.tex
\subsection{Stability for Tripartite Switching}
\label{app:tripartiteStab}
In this section, we prove that the DTMC $X$ is ergodic for $k>3$ when $n=3$.
Recall that $k\geq 3$ when $n=3$.  From Section \ref{sec:nPartite}, we see that its non-zero transition probabilities are
\begin{align*}
p_{i,j;i-1,j-1}&=\frac{k-2}{2}, \quad 
%p_{i,j;i+1,j}=\frac{1}{k},\quad p_{i,j;i,j+1}=\frac{1}{k}\quad  i,j\geq 1,\\
p_{i,j;i+1,j}=p_{i,j;i,j+1}=\frac{1}{k}\quad  i,j\geq 1,\\
p_{i,0;i+1,0}&=\frac{1}{k},\quad p_{i,0;i,1}=\frac{k-1}{k}\quad i\geq 1,\\
p_{0,j;0,j+1}&=\frac{1}{k},\quad p_{0,j;1,j}=\frac{k-1}{k}\quad j\geq 1,\\
p_{0,0;1,0}&=\frac{1}{2},\quad p_{0,0;0,1}=\frac{1}{2}.
\end{align*}
To prove that this chain is ergodic for $k>3$, we use \emph{Theorem 1.2.1} from \cite{fayolle1999random}, henceforth referred to as Malyshev's result, which defines
expected jumps\footnote{
%\begin{itemize}
%\item[-] 
$M_x$ (resp. $M_y$) is the expected horizontal (resp. vertical) jump size when leaving state $(i,j)$ with $i\geq 1$, $j\geq 1$;\\
%\item[-] 
$M'_x$  (resp. $M'_y$) is the expected horizontal  (resp. vertical) jump size when leaving state $(i,0)$ with $i\geq 1$;\\
%\textcolor{blue}{in passing, this shows why the initial definition of $M'_y$ does not make sense.}
%\item[-] 
$M''_x$  (resp. $M''_y$) is the expected horizontal  (resp. vertical) jump size when leaving state $(0,j)$ with $j\geq 1$.
%\textcolor{blue}{in passing, this shows why the initial definition of $M''_x$ does not make sense.}
%\end{itemize}
} by
\begin{align*}
M_x&=\hspace{-0.5em}\sum_{i',j'\geq 0} (i'-i)p_{i,j;i',j'}, ~ M_y=\hspace{-0.5em}\sum_{i',j'\geq 0} (j'-j)p_{i,j;i',j'}, i,j\geq 1,\\
M'_x&=\hspace{-0.5em}\sum_{i',j'\geq 0} (i'-i)p_{i,0;i',j'},~ M'_y=\hspace{-0.5em}\sum_{i',j'\geq 0} j' p_{i,0;i',j'},~ i\geq 1,\\
M''_x&=\hspace{-0.5em}\sum_{i',j'\geq 0} i'p_{0,j;i',j'},~ M''_y=\hspace{-0.5em}\sum_{i',j'\geq 0} (j'-j)p_{0,j;i',j'},~ j\geq 1.
\end{align*}
The theorem states that if $(M_x,M_y)\not=(0,0)$, then  the irreducible and aperiodic DTMC is ergodic if and only if one of the three following conditions holds:
\begin{align*}
&1.~ M_x<0,~ M_y<0, ~M_xM'_y-M_yM'_x<0,\text{ and }\\
& \quad M_y M''_x-M_xM''_y<0,\\
&2.~M_x<0,~ M_y\geq 0,~ M_y M''_x-M_xM''_y<0,\\
&3.~M_x\geq 0,~ M_y<0, ~  M_xM'_y-M_yM'_x<0.
\end{align*}
For all  $i,j\geq 1$,
\begin{align*}
&M_x=\sum_{i',j'\geq 1} (i'-i)p_{i,j;i',j'}-\sum_{j'\geq 1}ip_{i,j;0,j'}\\&\qquad+\sum_{i'\geq 1} (i'-i)p_{i,j;i',0}-ip_{i,j;0,0}\\
&=-{\bf 1}_{\{i\geq 2, j\geq 2\}} \,p_{i,j;i-1,j-1}+ p_{i,j;i+1,j}-{\bf 1}_{\{i=1,j\geq 2\}}p_{1,j;0,j-1}\\&\quad-{\bf 1}_{\{i\geq 2, j=1\}} p_{i,1;i-1,0} -{\bf 1}_{\{i=j=1\}} p_{1,1;0,0}\\
&=-{\bf 1}_{\{i\geq 2, j\geq 2\}} \,\frac{k-2}{k} + \frac{1}{k} - {\bf 1}_{\{i=1,j\geq 2\}} \frac{k-2}{k}\\&\quad-{\bf 1}_{\{ i\geq 2,j=1\}}\frac{k-2}{k} -{\bf 1}_{\{i=j=1\}} \frac{k-2}{k}\\
&=-\frac{k-2}{k}+\frac{1}{k}=-\frac{k-3}{k}.
\end{align*}
Similarly, we find that $M_y=-\frac{k-3}{k}$.  Therefore, $(M_x,M_y)\not=(0,0)$ iff $k\not=3$. From now on we will assume that $k>3$ so that we can use Malyshev's result. When $k>3$,
$M_x<0$ and $M_y<0$. Hence, by Malyshev's result, the DTMC is ergodic when $k>3$ iff $M_x M'_y-M_y M'_x<0$ and ${M_y M''_x-M_x M''_y<0}$.
For $i\geq 1$,
\begin{align*}
M'_x&= \sum_{i',j'\geq 1} (i'-i)p_{i,0;i',j'}-\sum_{j'\geq 1} ip_{i,0;0,j'}\\&\quad+\sum_{i'\geq 1}(i'-i)p_{i,0;i',0}-ip_{i,0;0,0}=p_{i,0;i+1,0}=\frac{1}{k},\\
M'_y&=\sum_{i',j'\geq 1} j'p_{i,0;i',j'}+\sum_{j'\geq 1} j' p_{i,0;0,j'}=p_{i,0;i,1}=\frac{k-1}{k}.
\end{align*}
Since $M_x=M_y<0$, we have 
$\hbox{sgn}(M_x M'_y-M_y M'_x)=-\hbox{sgn}(M'_y-M'_x)= -\hbox{sgn}\left(\frac{k-2}{k}\right)<0.$
\if{false}
\begin{align*}
\hbox{sgn}(M_x M'_y-M_y M'_x)&=-\hbox{sgn}(M'_y-M'_x)
%&= -\hbox{sign}\left(\frac{k-1}{k}-\frac{1}{k}\right)
= -\hbox{sgn}\left(\frac{k-2}{k}\right)<0.
\end{align*}
\fi
For $j\geq 1$,
\begin{align*}
M''_x&=\hspace{-0.5em}\sum_{i',j'\geq 1} i'p_{0,j;i',j'}+\sum_{i'\geq 1} i'p(0,j;i',0)=p_{0,j;1,j}=\frac{k-1}{k},\\
M''_y&=\sum_{i',j'\geq 1} (j'-j)p_{0,j;i',j'}- \sum_{i'\geq 1}j p_{0,j;i',0}\\
&\quad+\sum_{j'\geq 1}( j'-j) p_{0,j;0,j'}-jp_{0,j;0,0}=p_{0,j;0,j+1}=\frac{1}{k}.
\end{align*}
Since $M_x=M_y<0$, we have 
$\hbox{sgn}(M_y M''_x-M_x M''_y)=-\hbox{sgn}(M''_x-M''_y)=-\hbox{sgn}\left(\frac{k-2}{k}\right)<0,$
\if{false}
\begin{align*}
\hbox{sgn}(M_y M''_x-M_x M''_y)&=-\hbox{sgn}(M''_x-M''_y)\\
%=-\hbox{sign}\left(\frac{k-1}{k}-\frac{1}{k}\right)
&=-\hbox{sgn}\left(\frac{k-2}{k}\right)<0,
\end{align*}
\fi
for all $x\geq 1$, $y\geq 1$ when $k>3$.\\
This proves that  the Markov chain is ergodic when $k>3$.\qed
\section*{Acknowledgment}
The work was supported in part by the National  Science Foundation under grant CNS-1617437.

\ifCLASSOPTIONcaptionsoff
  \newpage
\fi



\bibliographystyle{IEEEtran}
\bibliography{qnet}
%\begin{thebibliography}{1}


%\end{thebibliography}

\if{false}
\begin{IEEEbiographynophoto}{Gayane Vardoyan}
\input{Gayane-bio}
\end{IEEEbiographynophoto}

\begin{IEEEbiographynophoto}{Saikat Guha}
\input{Saikat-bio}
\end{IEEEbiographynophoto}
%\newpage
\begin{IEEEbiographynophoto}{Philippe Nain}
\input{Bio-Nain-ToN}
\end{IEEEbiographynophoto}

\begin{IEEEbiographynophoto}{Don Towsley}
\input{Don-ieee-bio}
\end{IEEEbiographynophoto}
\fi

%\vfill


\end{document}